%% file: MES_under_dependence_uncertainty_v4.tex
\documentclass[11pt,a4paper]{article}

\usepackage[utf8]{inputenc}
\usepackage{bold-extra}
\usepackage{chngcntr}
\usepackage{tikz}
\usepackage{setspace}
\usepackage{url}
\usepackage{amsfonts}
\usepackage{amssymb}
\usepackage{epsfig}
\usepackage{amsmath}
\usepackage{commath}
\usepackage{natbib}
\usepackage{graphicx}
\usepackage{bbm,float}
\usepackage{dsfont}
\usepackage{mathtools}
\usepackage{color}
\usepackage{adjustbox}
\usepackage{amsmath,amssymb,amstext,amsfonts,amsthm,latexsym}
\usepackage{diagbox}
\usepackage{bm}
\usepackage{caption}
\usepackage{subcaption}
\usepackage{multirow}
\usepackage{color}
\usepackage{listings}
\usepackage{titlesec}
\usepackage{enumerate}
\usepackage{environ}
\usepackage{booktabs}
\usepackage{appendix}
\usepackage{bbm}
\usepackage{pgfplots}
\usepackage{makecell}
\usepackage{tabularx}
\newcolumntype{Y}{>{\centering\arraybackslash}X}
\numberwithin{equation}{section}
\theoremstyle{plain}
\newtheorem{theorem}{Theorem}[section]
\newtheorem{corollary}{Corollary}[section]
\newtheorem{proposition}{Proposition}[section]

\theoremstyle{definition}

\newtheorem{definition}{Definition}[section]

\newtheorem{example}{Example}[section]

\theoremstyle{remark}

\newtheorem{remark}{Remark}[section]
\pgfplotsset{compat=1.17}

\definecolor{darkblue}{rgb}{0.0, 0.0, 0.55}

\begin{document}
\title{\textbf{Static marginal expected shortfall: Systemic risk measurement under dependence uncertainty}
}
\date{\today}
\author{Jinghui Chen\thanks{Corresponding author: Jinghui Chen, Department of Mathematics and Statistics at York University, and RISC Foundation.  (email: \texttt{jh8chen@yorku.ca}).}, Edward Furman\thanks{Edward Furman, Department of Mathematics and Statistics at York University, and RISC Foundation.  (email: \texttt{efurman@yorku.ca}).}, and X.\ Sheldon Lin\thanks{X.\ Sheldon Lin, Department of Statistical Sciences at University of Toronto, and RISC Foundation. (email: \texttt{sheldon.lin@utoronto.ca}).}}
\maketitle
\begin{abstract}
Measuring the contribution of a bank or an insurance company to overall systemic risk is a key concern, particularly in the aftermath of the 2007--2009 financial crisis and the 2020 downturn. In this paper, we derive worst-case and best-case bounds for the marginal expected shortfall (MES)---a key measure of systemic risk contribution---under the assumption that individual firms' risk distributions are known but their dependence structure is not. We further derive tighter MES bounds when partial information on companies' risk exposures, and thus their dependence, is available. To represent this partial information, we employ three standard factor models: additive, minimum-based, and multiplicative background risk models. Additionally, we propose an alternative set of improved MES bounds based on a linear relationship between firm-specific and market-wide risks, consistent with the Capital Asset Pricing Model in finance and the Weighted Insurance Pricing Model in insurance. Finally, empirical analyses demonstrate the practical relevance of the theoretical bounds for industry practitioners and policymakers.
\end{abstract}


\section{Introduction}
Understanding and measuring systemic risk, that is, the risk that the failure of a single component could destabilize the entire system, has become a central concern in finance, insurance, and risk management, especially in the aftermath of the 2007--2009 financial crisis and the global economic disruption caused by the COVID-19 pandemic. For example, Canada’s Office of the Superintendent of Financial Institutions (OSFI) explicitly highlights the importance of addressing systemic risk arising from `concentration' in its 2024–2025 Annual Risk Outlook \citep{osfi2024riskoutlook}.

A key aspect of systemic risk measurement is assessing the contribution of individual institutions to overall market risk. In the insurance sector, this poses a significant challenge for industry-funded guarantee schemes---also called Policyholder Protection Schemes (PPSs)---such as Canada’s Property and Casualty Insurance Compensation Corporation (PACICC), State Guaranty Associations (SGAs) in the United States, and National Guarantee Schemes (NGSs) across the European Union. More specifically, PPSs charge premiums to their member insurers and provide protection to policyholders by covering claims and unearned premiums in the event of a member insurer’s insolvency. At present, virtually all PPSs calculate the said premiums primarily based on each member insurer’s direct written premiums in covered lines of business---a practice followed by, e.g., PACICC and various SGAs in the United States. However, this approach is limited in that it does not account for an individual member insurer’s contribution to the aggregate systemic risk. Recognizing this gap, the International Association of Insurance Supervisors has encouraged the development of risk-based premium assessment frameworks for PPSs, as recently discussed in \citet{iais2023pps}.

The marginal expected shortfall (MES) has emerged as a widely used measure of systemic risk contribution. Initially proposed by \citet{acharya2017measuring}---and long beforehand recognized in actuarial science \citep{denault2001coherent,panjer2001solvency} as a risk capital allocation (RCA) rule based on the expected shortfall (ES) risk measure---MES has received considerable attention in both research and applications. Speaking more formally, let the random variables (RVs) $X_1,X_2\ldots,X_d,\ d\in\mathbb{N}$ denote the individual risks associated with insurers labeled $\mathcal{N}:=\{1,2,\ldots,d\}$, and let the RV $S=X_1+X_2+\cdots+X_d$ represent the aggregate risk in the market. Furthermore, we assume that all the aforementioned RVs have cumulative distribution functions with finite means. The ES risk measure of the aggregate risk, for a prudence level $p\in(0,\ 1)$ and the Value-at-Risk (VaR) risk measure $\mathrm{VaR}_p(S)$, is defined as 
\begin{equation*}
\label{ES-intro-def}
\mathrm{ES}_p(S)=\mathbb{E}[S\vert S> \mathrm{VaR}_p(S)],
\end{equation*}
\citep[e.g.,][]{denuit2006actuarial}, and the corresponding MES risk measure is defined as
\begin{equation*}
\label{MES-intro-def}
\mathrm{MES}_p(X_j,S)=\mathbb{E}[X_j\vert S> \mathrm{VaR}_p(S)],\ j\in\mathcal{N},
\end{equation*}
as introduced in the systemic risk literature \citep[e.g.,][]{acharya2017measuring} for dynamic, time-varying frameworks, and in \citet{qin2021systemic} for static, tail-asymptotic, cross-sectional formulations.

	The theoretical properties of both the ES and MES functionals have been thoroughly studied. In particular, the former functional has been shown to be coherent \citep{artzner1999coherent,acerbi2002expected}, comonotonic additive \citep{kusuoka2001law}, convex \citep{follmer2002convex}, elicitable \citep{fissler2016higher}, and uniquely characterized by a set of `economic' axioms, justifying its role as a leading regulatory risk measure \citep{wang2021axiomatic}. Furthermore, it is `distorted' in the sense of \citet{wang2000distortion} and `weighted' in the sense of \citet{furman2008weightedb}. Finally, the Solvency III regulatory framework---based on Basel III, which is known as the Fundamental Review of the Trading Book \citep{basel2016es}---is shifting from the VaR risk measure at the prudence level $p=0.99$ to the ES risk measure at the prudence level $p=0.975$ for calculating market risk capital. The latter functional, that is, the MES risk measure, inherits much of the appeal of the ES risk measure on which it is based \citep[e.g.,][and references therein]{tsanakas2003risk,furman2008weighted,dhaene2012optimal}.

On a more application-oriented front, a substantial body of literature has derived explicit formulas for the ES and MES functionals under specific assumptions about the joint cumulative distribution function of the risk RVs $X_1,X_2,\ldots,X_d,\ d \in \mathbb{N}$. In particular, \citet{panjer2001solvency} provided formulations for the multivariate normal distribution, while \citet{landsman2003tail} and \cite{dhaene2008some} extended these results to the broader class of elliptical distributions. \citet{furman2005risk} and \citet{zhou2018approximation} derived results for a class of multivariate gamma distributions based on additive background risk models, and \citet{furman2010multivariate} further extended these findings to a more general class of multivariate Tweedie distributions. The multivariate phase-type and multivariate Pareto distributions were addressed in \citet{cai2005conditional} and \citet{vernic2006multivariate,vernic2011tail}, respectively. More recently, \citet{landsman2016tail} and \citet{ignatieva2019conditional} derived formulations for risk variables following multivariate log-elliptical and skewed generalized hyperbolic distributions, and \citet{cossette2018dependent} and \citet{marri2022risk} considered the case of dependent risks modeled via Archimedean and mixed Bernstein copulas. It should be noted that this overview is far from exhaustive. A considerable body of additional literature has explored ES and MES in diverse modeling frameworks beyond those mentioned here.

Specific assumptions on the joint cumulative distribution of risks are more of a mathematical nicety than a reflection of reality. In practice, the form of dependence that governs the joint behaviour of the risk RVs $X_1,X_2,\ldots,X_d,\ d\in\mathbb{N}$---and hence the distribution of the aggregate risk RV $S$---is rarely known. A plethora of authors have addressed this dependence uncertainty in the context of the aggregate risk $S$. In the case of complete uncertainty, \cite{embrechts2014academic} provided a comprehensive discussion and numerical illustrations of such problems in risk management. \citet{bernard2014risk} derived a lower bound for $\mathrm{ES}_p(S)$ for homogeneous risks by leveraging properties of complete mixability \citep{wang2011complete} and coherent risk measures \citep{artzner1999coherent}. \cite{jakobsons2016general} further discovered the minimum $\mathrm{ES}_p(S)$ for heterogeneous risks under the assumptions of joint mixability; see \cite{wang2016joint} for the detailed information of joint mixability. Moreover, \cite{wang2021axiomatic} proposed the concept of $p$-concentration, which refers to the absence of diversification benefits in portfolio construction, and illustrated the upper bound of $\mathrm{ES}_p(S)$ for any risks is obtained via comonotonicity, as discussed more generally by \cite{dhaene2002concept} and \cite{deelstra2011overview} in the contexts of finance and insurance. More recently, \cite{bernard2023coskewness} studied sharp lower and upper bounds on the expectation of the product of $d\geq 3$ RVs for some specific marginal distributions. In the case of partial uncertainty, related analyses for $\text{VaR}_p(S)$ and convex risk measures when partial information known from the factor models have been previously studied by \cite{bernard2017risk}. Additionally, \cite{bernard2017value} discovered the improved bounds on VaR of $S$ by imposing its variance constraints.

However, the impact of dependence uncertainty on the systemic risk contribution, as measured by MES, remains largely unexplored. Understanding the lower and upper bounds of $\mathrm{MES}_p(X_j, S)$---as well as assessing how close a firm in the market lies to either of these bounds---is critically important for PPSs and financial regulators, particularly given the growing prominence of systemic risks, such as cyber threats, climate-related disasters, global pandemics, and geopolitical instability, in today’s interconnected risk landscape. Quantifying these bounds provides key insights into worst-case scenarios and supports the design of more resilient financial and insurance systems. 

In this paper, we address this gap by studying the effect of dependence uncertainty on the marginal expected shortfall. Our main contributions are as follows. 

First, we derive sharp lower and upper bounds for $\mathrm{MES}_p(X_j,S)$, together with the dependence structures (copulas) that attain these bounds, under the assumption that only marginal information is available (Section~\ref{no information}). In this setting, the distributions of $X_i$, $i\in\mathcal{N}$, are known, but the dependence among the risk components $(X_1,X_2,\dots,X_d)$, $d\in\mathbb{N}$, is completely unspecified. These bounds characterize the maximum and minimum possible systemic risk contributions of individual firms under extreme dependence scenarios, thus capturing the full range of feasible MES values consistent with observed marginals. 

We then extend our analysis to constrained risk bounds for MES when partial dependence information is available through factor-model structures. In these models, the risks $X_i$, $i\in\mathcal{N}$, are represented as
\begin{equation}\label{factor model}
	X_i = f_i(Y, Z_i),
\end{equation}
where $Y$ denotes a common risk factor and $Z_i$ represents idiosyncratic risk factor. Factor models are widely used across disciplines such as statistics, economics, and insurance \citep{bernard2017risk}, with applications in asset pricing \citep{fama1993common}, portfolio optimization \citep{santos2013comparing}, and risk management \citep{gordy2000comparative}. Incorporating factor structures provides partial information on the dependence among risks, thereby reducing the dependence-uncertainty spread relative to the unconstrained case. In Section~\ref{factor models sec}, we show analytically how the inclusion of such structure tightens the feasible range of MES values. Representative examples include the additive \citep[][]{gollier1996risk,furman2005risk}, multiplicative \citep[][]{franke2006multiplicative,asimit2016background}, and minimum-based \citep[][]{marshall1967multivariate,pai2020livestock} background risk models, all of which are extensively applied in insurance and quantitative risk management. 

In Section~\ref{linear condition}, we further examine MES bounds under the assumption of a linear relationship between $\mathbb{E}[X_i \mid S]$ and the aggregate risk $S$, an idea central to both the Capital Asset Pricing Model (CAPM) \citep[][]{levy2011capm} and the Weighted Insurance Pricing Model (WIPM) \citep[][]{furman2017beyond}. Under this assumption, we obtain explicit constrained bounds in several settings, including non-negative risks, non-positive risks, and bivariate normal risks. As with factor models, these structural constraints further shrink the uncertainty interval for MES, providing interpretable benchmarks for assessing systemic importance. 

Finally, in Section~\ref{empirical study}, we complement our theoretical results with an empirical analysis of unconstrained and constrained MES bounds using loss data from constituent stocks of the S\&P 500 over a specified period. The constrained bounds are computed under the Fama--French five-factor model \citep{fama2015five}, illustrating how dependence information reduces systemic risk uncertainty in practice. The empirical findings confirm the practical relevance of our theoretical framework for practitioners and regulators. In addition, we propose a Systemic Risk Criticality Index (SRCI), which quantifies how close a firm's MES is to its theoretical upper bound---the closer this value is to one, the more systemically critical the firm is. This index provides an interpretable metric for identifying firms whose distress would pose significant systemic threats. While the upper MES bounds can be used by PPPs and financial supervisors to allocate capital or price participation based on systemic risk contribution, the SRCI serves as a practical tool for assessing the relative systemic position of firms within the market. During periods of extreme market stress---such as the global financial crisis---an institution’s MES may approach its theoretical upper bound, indicating systemic criticality. Conversely, the lower bound can be observed in more diversified market phases, as seen during the pandemic recovery period.

\section{MES bounds with fixed marginal distributions}\label{no information}
In this section, we derive lower and upper bounds for the MES when the marginal distributions of the risk variables are known, but the dependence structure among them is unspecified. Additionally, it is important to recognize that risk bounds under the assumption of completely uncertain dependence are often of limited practical use. Typically, the gap between the lower and upper bounds is substantial; see \citet{bernard2017risk}. Consequently, we also derive tighter and more informative bounds for the MES by incorporating partial dependence information from the structure of factor models.

\subsection{Preliminaries}
Let $(\Omega, \mathcal{F}, \mathbb{P})$ denote a standard atomless probability space, and let $\mathcal{X} := L^0(\Omega, \mathcal{F}, \mathbb{P})$ be the space of all real-valued random variables (RVs) defined on this space. 
We denote generic elements of $\mathcal{X}$ by $X,\ Y,\ U,\ X_i,$ and $Z_i$, where $i \in \mathcal{N} := \{1, 2, \ldots, d\}$ and $d \in \mathbb{N}$. 
Each $X_i,\ i \in \mathcal{N},$ is interpreted as a risk component, and $U \sim \mathrm{Unif}[0,1]$ is assumed to be independent of all other RVs unless stated otherwise. 
Throughout the paper, the notation $X \sim F_X$ indicates that the RV $X$ has cumulative distribution function (CDF) $F_X(x) = \mathbb{P}(X \le x)$, $x \in \mathbb{R}$. 
The quantile function of $X$, for $p \in (0,1)$, is defined by  
\[
F_X^{-1}(p) = \inf\{x \in \mathbb{R} : F_X(x) \ge p\}.
\]
This quantity is also known as the Value-at-Risk (VaR) of $X$ at prudence level $p$, denoted by $\mathrm{VaR}_p(X)$. 
Unless otherwise specified, all RVs are assumed to have finite expectations. 

For the random vector $\mathbf{X} = (X_1, \ldots, X_d)'$ of risk components $X_i \sim F_i$, we denote its joint CDF by  
\[
F(\mathbf{x}) = \mathbb{P}(\mathbf{X} \le \mathbf{x}), \quad \mathbf{x} = (x_1, \ldots, x_d) \in \mathbb{R}^d,
\]
where inequalities are understood componentwise. 
Further, let $Y \sim H$ be a random variable interpreted as a risk factor. 
For a given realization $Y = y$, we denote by $\mathbf{X}_{|y}$ and $\mathbf{X}_{|Y}$ the random vectors of conditional risk components $X_{i|y}:=X_i|\ (Y=y) \sim F_{i|y}$ and $X_{i|Y}:= X_i|\ Y \sim F_{i|Y}$, $i \in \mathcal{N}$. 
The associated aggregate risks are defined by  
\begin{equation}\label{sumrvs}
	S := \sum_{i=1}^d X_i, \quad 
	S_{|y} := \sum_{i=1}^d X_{i|y}, \quad 
	S_{|Y} := \sum_{i=1}^d X_{i|Y}.
\end{equation}

In what follows, we frequently consider the comonotonic and antimonotonic copies of the random vector $\mathbf{X}$, denoted respectively by  
\[
\mathbf{X}^c = (X_1^c, \ldots, X_d^c)' 
\quad \textnormal{and} \quad
\mathbf{X}^a = (X_1^a, \ldots, X_d^a)'.
\]
Each of these vectors has coordinates with the same marginal distributions as those of $\mathbf{X}$ but differing dependence structures. 
Specifically, $\mathbf{X}^c$ exhibits perfect positive dependence (comonotonicity), given by $X_i^c = F_i^{-1}(U),\ i \in \mathcal{N}$. 
In contrast, $\mathbf{X}^a$ represents the conceptual opposite---perfect negative dependence (antimonotonicity in the multivariate case and countermonotonicity when $d=2$)---and is constructed as  
\[
X_j^c = F_j^{-1}(U), \quad
X_i^a = F_i^{-1}(1 - U), \quad i \in \mathcal{N} \setminus \{j\}, \quad j \in \mathcal{N}.
\]
Analogously to earlier, we write $\mathbf{X}_{|y}^c$ ($\mathbf{X}_{|y}^a$) and $\mathbf{X}_{|Y}^c$ ($\mathbf{X}_{|Y}^a$) for the comonotonic (antimonotonic) copies of the conditional random vectors $\mathbf{X}_{|y}$ and $\mathbf{X}_{|Y}$, respectively. 
The (conditional) aggregate risk RVs in the comonotonic case are constructed as $S^c:=F_S^{-1}(U)\ S^c_{|y}:=F_{S_{|y}}^{-1}(U),$ and $S^c_{|Y}:=F_{S_{|Y}}^{-1}(U)$, whereas (conditional) aggregate risk RVs in the antimonotonic case are constructed as
\[
S^a := F_S^{-1}(1-U), \quad 
S^a_{|y} := F_{S_{|y}}^{-1}(1-U), \quad 
S^a_{|Y} := F_{S_{|Y}}^{-1}(1-U).
\]

\citet{wang2016joint} and \citet{wang2021axiomatic} introduced the concepts of joint mixability and risk concentration, respectively, and established their key properties. We recall these two definitions below and refer the reader to the cited references for further details.
\begin{definition}[Joint mixability]
A $d$-tuple of probability distributions $(F_1,F_2,\dots, F_d)$ on $\mathbb{R}$ is said to be jointly mixable if there exist RVs $X_1\sim F_1$, $X_2\sim F_2,\dots,X_d\sim F_d$ such that $X_1+X_2+\cdots+X_d=c$ a.s., for some constant $c\in \mathbb{R}$, termed the joint center of $(F_1,F_2,\dots, F_d)$. The random vector $(X_1,X_2,\dots, X_d)'$ is called \textit{jointly mixable}.
\end{definition}
\begin{definition}[Risk concentration]
	For $p\in (0, 1),$ a random vector $(X_1,X_2,\dots, X_d)'$ is \textit{$p$-concentrated} if its coordinates share a common tail event of probability $1-p$. A tail event of probability $1-p$ is called a $p$-tail event.
\end{definition}
Before presenting the main results of this section, we introduce the left expected shortfall (LES) as a measure of left-tail risk. 
\begin{definition}[LES]
	The LES risk measure of a RV $X\in\mathcal{X}$ at the prudence level $p\in(0,\ 1)$, denoted by $\mathrm{LES}_p(X)$, is
	\begin{equation*}
		\mathrm{LES}_p(X)=\mathbb{E}[X\vert X\leq\text{VaR}_p(X)].
	\end{equation*}
\end{definition}

\subsection{Risk bounds}
Let $m_j(p)$ $(M_j(p))$ and $m^f_j(p)$ $(M^f_j(p))$ denote the unconstrained and constrained lower (upper) bounds of the functional $\mathrm{MES}_p(X_j,S),\ j\in\mathcal{N}$. That are
\begin{equation*}
	\begin{aligned}
		m_j(p) &= \inf\Bigl\{\mathrm{MES}_p(X_j,S):\; S=\sum_{i=1}^d X_i,\; X_i\sim F_i\Bigr\},\\[2pt]
		M_j(p) &= \sup\Bigl\{\mathrm{MES}_p(X_j,S):\; S=\sum_{i=1}^d X_i,\; X_i\sim F_i\Bigr\},\\[6pt]
		m^{f}_j(p) &= \inf\Bigl\{\mathrm{MES}_p(X_j,S):\; S=\sum_{i=1}^d X_i,\; X_i\sim F_i,\; (X_i,Y)\sim H_i\Bigr\},\\[2pt]
		M^{f}_j(p) &= \sup\Bigl\{\mathrm{MES}_p(X_j,S):\; S=\sum_{i=1}^d X_i,\; X_i\sim F_i,\; (X_i,Y)\sim H_i\Bigr\}.
	\end{aligned}
\end{equation*}

\begin{theorem}[Unconstrained and constrained risk bounds]\label{upper bound}
	Let $X_i=f_i(Y, Z_i)\sim F_i$, where $i\in\mathcal{N},$ $Y\sim H$, and $Z_i$ are independent of $Y$, and let $S=\sum_{i=1}^{d}X_i$. Then, for $p\in (0, 1)$ and $j\in\mathcal{N}$, we have
	\begin{equation}\label{unique risk bounds}
		\begin{aligned}
			\mathrm{MES}_p(X_j^c,S^a)\leq m_j(p)&\leq \mathrm{MES}_p(X_{j\vert Y}^c,S^a_Y) \leq m^f_j(p)\\
			&\leq \mathrm{MES}_p(X_j,S)\\
			&\leq M^f_j(p)\coloneqq \mathrm{MES}_p(X_{j\vert Y}^c,S^c_Y) \leq M_j(p)\coloneqq\mathrm{MES}_p(X_{j}^c,S^c).
		\end{aligned}
	\end{equation}
\end{theorem}
The proofs of this and all subsequent theorems and propositions are relegated to the Appendix. The (improved) upper bound of $\mathrm{MES}_p(X_j, S)$ is always sharp due to (conditionally) comonotonicity. The comonotonic random vector $\mathbf{X}^c$ yields sharp upper bounds for various risk measures, including law-invariant convex risk measures; see, e.g., \citet{meilijson1979convex}, \citet{puccetti2012bounds} and \citet{bernard2017risk}.
\begin{proposition}\label{cont case}
	Let $X_i\sim F_i$, where $i\in\mathcal{N}$ and $X_i$ have continuous CDFs, and let $S=\sum_{i=1}^{d}X_i$. Then the unconstrained risk bounds in~\eqref{unique risk bounds} reduce to
	\begin{equation}\label{cont bounds}
			\mathrm{LES}_{1-p}(X_j)\leq m_j(p) \leq \mathrm{MES}_p(X_j,S)\leq M_j(p)= \mathrm{ES}_p\left(X_j\right),\ p\in (0, 1),\ j\in\mathcal{N}.
	\end{equation}
	Furthermore, the upper bound is obtained when the random vector $(X_1,X_2,\dots, X_d)'$ is \textit{$p$-concentrated}. 
\end{proposition}

The upper bound $\mathrm{ES}_p\left(X_j\right)$ is well-known due to a property of risk capital allocation rules known as `no-undercut'. That is, no individual risk component $X_j,\ j\in\mathcal{N}$ is allocated more risk capital than it would be allocated if considered alone \citep{furman2008weighted}. In this case $M_j(p)$ is independent of the dimension of the random vector and can be computed explicitly from the marginal distribution of the risk component \(X_j\) (the worst-case MES of \(X_j\) at prudence level \(p\) coincides with ES of the marginal risk). The lower bound $m_j(p)$ in Inequality~\eqref{cont bounds} implies that the minimum systemic risk contribution of an individual risk component $X_j$ cannot be smaller than its conditional expected value, given that the loss does not exceed its VaR at the prudence level \(1 - p\).

\begin{figure}
	\centering
	\input{lognorm_bounds_of_mes.tex}
	\caption{Risk bounds of $\mathrm{MES}_p(X_j,S)$, where $X_i\sim \text{Lognorm}(0,1)$ and $S=\sum_{i=1}^{2}X_i$, as functions of the prudence level $p$. Under the assumptions of factor models, $X_1=\exp(b_1Y+\sqrt{1-b_1^2}Z_1)$ and $X_2=\exp(b_2Y+\sqrt{1-b_2^2}Z_2)$, where $b_1=0.1$, $b_2=0.9$, and $Y\overset{d}{=}Z_i\sim N(0,1)$.}
	\label{risk bounds figure}
\end{figure}

Figure~\ref{risk bounds figure} presents the unconstrained and constrained risk bounds of  $\mathrm{MES}_p(X_j,S)$ as a function of the prudence level $p$ for the case $X_i=\exp\left(b_iY+\sqrt{1-b_i^2}Z_i\right)\sim \text{Lognorm}(0,1)$, where $i=1,2$, $b_1=0.1$, $b_2=0.9$, $Y\overset{d}{=}Z_i\sim N(0,1)$. The vertical dashed line illustrates the full range of possible MES values, from the minimum (LES) to the maximum (ES), while the solid one shows its improved range of values. As shown, \begin{equation*}
	\lim\limits_{p\rightarrow 0}\mathrm{MES}_p(X_j,S)=\lim\limits_{p\rightarrow 0}\mathrm{LES}_{1-p}(X_j)=\lim\limits_{p\rightarrow 0}\mathrm{ES}_p(X_j)=\lim\limits_{p\rightarrow 0}m^f_j(p)=\lim\limits_{p\rightarrow 0}M^f_j(p)=\mu_j.
\end{equation*}

In contrast to the (improved) upper bound of $\mathrm{MES}_p(X_j, S)$ in Theorem~\ref{unique risk bounds}, which is attained under (conditional) comonotonicity, the (improved) lower bound $m_j(p)$ (or $m^f_j(p))$ is not necessarily sharp. In what follows, we examine the conditions under which this lower bound becomes sharp, with particular attention to the role of the (conditional) marginal distributions.

\subsection{On the sharpness of the lower bounds}
In this subsection, we first identify the conditions under which $m_j(p)$ and $m^f_j(p)$ in Theorem~\ref{upper bound} are attained and subsequently demonstrate sharpness through examples and counterexamples.

\begin{corollary}[Sharpness of the lower bounds]\label{sharpness condition}
	Let $X_i=f_i(Y, Z_i)\sim F_i$, $i\in\mathcal{N},$ be $d$ risks, and $S=\sum_{i=1}^{d}X_i$ denote the aggregate risk. The lower bound of $m_j(p)$ in Theorem~\ref{upper bound} is attained if and only if $X_j$ and $S$ are antimonotonic, while the lower bound of $m^f_j(p)$ is attained when they are conditionally antimonotonic.
\end{corollary}
The proof of Corollary~\ref{sharpness condition} follows from Theorem~\ref{unique risk bounds} and is therefore omitted.
\begin{remark}\label{discrete point}
	The (conditional) antimonotonicity condition in Theorem~\ref{sharpness condition} may appear to hold universally since it involves only two random variables, $X_j$ and $S$. However, as $S=X_j+\sum_{i=1,i\neq j}^{d}X_i$ represents an aggregate risk, the validity of this condition depends on the properties of the (conditional) marginal distributions $F_i$ ($F_{i\vert Y}$).
\end{remark}
The following examples illustrate the point of Remark~\ref{discrete point}.

\subsubsection{The unconstrained lower bound}
To demonstrate that $m_j(p)>\mathrm{MES}_p(X_j^c,S^a)$ in \eqref{unique risk bounds}, we provide several illustrative examples.
\begin{example}
	Let $X_1\sim F_1$ and $X_2\sim F_2$ denote two discrete RVs having respective mass points $\{-2,-1,0,1,2\}$ and $\{-1,1,3,5,7\}$ with equal probability and $p=0.6$. Under an antimonotonic dependence between $X_1$ and $X_2$, $0=\mathrm{MES}_p(X_2^c,S^a)<m_2(p)=M_2(p)=6$ since $S$ and $X_2$ are comonotonic.
\end{example} 
An example with normal marginal distributions is provided to further illustrate this result.
\begin{example}
	Let $X_i\sim N(\mu_i, \sigma_i^2)$, $i\in\mathcal{N},$ such that $\sum_{i=1\&i\neq j}^{d}\sigma_i<\sigma_j$, and $S=\sum_{i=1}^{d}X_i$. Then the lower bound of $m_j(p)$ is $$\mathrm{LES}_{1-p}(X_j)=\frac{\mu_j-\sigma_j\phi\left(\Phi^{-1}(1-p)\right)}{1-p}.$$ However, $m_j(p)=M_j(p)>\mathrm{LES}_{1-p}(X_j)$ is obtained when $X_j=\mu_j+\sigma_j\Phi^{-1}(1-U)$ and $X_i=\mu_i+\sigma_i\Phi^{-1}(U)$ for $i\in\mathcal{N}/j$. In this case, $S=\sum_{i=1}^{d}\mu_i+\left(\sigma_j-\sum_{i=1\&i\neq j}^{d}\sigma_i\right)\Phi^{-1}(1-U)$ is comonotonic to $X_j$.
\end{example}

We next provide propositions and examples illustrating when the lower bound of $m_j(p)$ is sharp.
\begin{proposition}[Symmetric marginal]\label{symmetric margin}
	Let $X_i\sim F$, $i\in\mathcal{N},$ be $d\ge 3$ continuous risks such that $F$ is symmetric, and $S=\sum_{i=1}^{d}X_i$ denote the aggregate risk. There exists a random vector $(X_1,X_2,\dots, X_d)$ such that $X_j$ and $S$ are antimonotonic. Hence, $m_j(p)=\mathrm{LES}_{1-p}(X_j)$ is obtained by such a random vector. Furthermore, $X_j=F^{-1}(1-U)$ and $X_i=F^{-1}(U)$ for $i\in\mathcal{N}/j$.
\end{proposition}
\begin{example}
	Let $U_i\sim \mathrm{Unif}[0,1]$, $i\in\mathcal{N},$ be $d\ge 3$ risks, and $S=\sum_{i=1}^{d}U_i$. Hence, $$m_j(p)=\mathrm{LES}_{1-p}(U_j)=\frac{\int_{0}^{1-p}udu}{1-p}=\frac{1-p}{2}$$ is obtained when $U_j=1-U$ and $U_i=U$ for $i\in\mathcal{N}/j$.
\end{example}
The assumption of continuous symmetric and homogeneous risks in Proposition~\ref{symmetric margin} may, however, be overly restrictive. We now propose a proposition under a weaker set of assumptions.

%\begin{proposition}\label{max margin}
%	Let $X_i$, $i\in\mathcal{N},$ be $d$ risks, and $S=\sum_{i=1}^{j}X_i$ such that $X_j$, $j\in\mathcal{N},$ and $S$ are antimonotonic. Hence, $m(p)=\mathrm{LES}_{1-p}(X_j)$.
%\end{proposition}

\begin{proposition}\label{jointly mixable margin}
	Let $X_i\sim F_i$ for $i\in\mathcal{N},$ be $d$ risks such that the tuple of distributions $(F_i, F_j)$ is jointly mixable for some $j\in\mathcal{N}/i$. Then, there exists a random vector $(X_1,X_2,\dots, X_d)$ such that $X_j$ and $S=\sum_{i=1}^{d}X_i$ are antimonotonic. In this case, the lower bound of $m_j(p)$ in Theorem~\ref{unique risk bounds} is obtained. Furthermore, it holds that $X_j=F^{-1}_j(1-U)$ and $X_i=F^{-1}_i(U)$ for all $i\in\mathcal{N}/j$.
\end{proposition}
The proof follows directly, as under this construction, we have $$S=c+\sum_{k=1, k\neq i,j}^{d}F_k^{-1}(U)=F_S^{-1}(U),$$ where $c\in\mathbb{R}$ is the joint center of $(F_i, F_j)$.

\subsubsection{The constrained lower bound}
To illustrate that $m^f_j(p)>\mathrm{MES}_p(X_{j\vert Y}^c,S^a_Y)$ and $m^f_j(p)=\mathrm{MES}_p(X_{j\vert Y}^c,S^a_Y)$ in \eqref{unique risk bounds}, we consider the following examples.

\begin{example}\label{cond com not upper}
	\begin{enumerate}[(1)]
		\item[]
		\item Let $X_i=Y+Z_i$, $i=1,2$ be two risks such that $Z_i\sim \mathrm{Unif}[0,1]$ are independent of $Y\sim \mathrm{Unif}[0,1]$. In this case, 
		\begin{equation*}
			\begin{aligned}
				m^f_j(p)&=\mathbb{E}[Y+Z_j\vert 2Y+U+1-U>\mathrm{VaR}_p(2Y+U+1-U)]\\
				&=\mathbb{E}[Y+Z_j\vert Y>\mathrm{VaR}_p(Y)]=\mathbb{E}[Y\vert Y>\mathrm{VaR}_p(Y)]+\mathbb{E}Z_j\\
				&=\frac{1+p}{2}+\frac{1}{2}>\mathrm{MES}_p(X_{j\vert Y}^c,S^a_Y).
			\end{aligned}
		\end{equation*} The lower bound of $m^f_j(p)$ is not obtained since $X_j$ and $S$ cannot be conditionally antimonotonic. 
		\item Let $\bm{X}=(X_1,X_2,X_3)=Y(-1,1,1)+(Z_1,Z_2,Z_3)$ be a three-dimensional random vector such that $Y\sim \text{N}(0, 1)$ and $Z_i\sim \text{N}(0, 1)$ are independent of $Y$. Hence, $X_i\sim\text{N}(0, 2)$. The lower bound of $\mathrm{MES}_{p}(X_2, S)$ is
		\begin{equation*}
			m^f_j(p)=\mathbb{E}[Y+Z_2\vert Y-Z_2>\mathrm{VaR}_p(Y-Z_2)]=0
		\end{equation*}since $Y+Z_2$ and $Y-Z_2$ are independent, and it is obtained when $Z_2=\Phi^{-1}(U)$ and $Z_1=Z_3=\Phi^{-1}(1-U)$. In this case, $X_2$ and $S$ are conditionally antimonotonic.
	\end{enumerate}
\end{example}
 
\section{Dependence uncertainty spread}\label{factor models sec}
In this section, we analyze how the dependence-uncertainty spread can be reduced by incorporating factor information. To capture partial dependence, we employ three widely used factor models: the additive background risk model (Section~\ref{ABRM}), the multiplicative background risk model (Section~\ref{MBRM}), and the minimum-based background risk model (Section~\ref{MBBRM}).

To quantify the improvement in the dependence-uncertainty spread achieved by introducing a factor model, we define the following measure, denoted by $\delta_{j,p}$:
\begin{equation*}
	\delta_{j,p} = 1-\frac{M^f_j(p)-m^f_j(p)}{M_j(p)-m_j(p)},
\end{equation*} where we conventionally set $\delta_{j,p}=1$ whenever $M_j(p)=m_j(p)$; see a similar definition for general risk measures in \citet{bernard2017risk}. Here, $M_j(p)-m_j(p)$ and $M^f_j(p)-m^f_j(p)$ are referred to as the dependence-uncertainty spreads of $\mathrm{MES}_p(X_j,S)$ in the unconstrained and constrained setting, respectively.

\subsection{Additive background risk models}\label{ABRM}
In this subsection, we study improved risk bounds in the context of the additive background risk models (ABRMs). Specifically, we consider risk vectors of the form
\begin{equation*}
	\bm{X}=\left(\mu_1+b_1Y+\sigma_1Z_1,\mu_2+b_2Y+\sigma_2Z_2,\dots,\mu_d+b_dY+\sigma_dZ_d\right)',
\end{equation*}where $\bm{\mu}=(\mu_1,\mu_2,\dots,\mu_d)'$, $\bm{b}=(b_1,b_2,\dots,b_d)'$ and $\bm{\sigma}=(\sigma_1,\sigma_2,\dots,\sigma_d)'$, in which $\sigma_i>0$, are parameter vectors. In this model, the functional form is $f_i(Y,Z_i)=\mu_i+b_iY+\sigma_iZ_i$ for each $i\in\mathcal{N}$.

The ABRM framework has been widely applied across economics, finance, and insurance; see, for instance, \citet{fama1993common} and \citet{fama2004capital} for the Capital Asset Pricing Model, and \citet{gollier1996risk}, \citet{furman2018weighted}, and \citet{zhou2018approximation} for applications in insurance risk management.

Without loss of generality, we assume $\bm{\mu}=(0,0,\dots,0)'$ because \begin{equation*}
		\mathrm{MES}_p(X_j, S)=\mu_j+\mathbb{E}\left[b_jY+\sigma_jZ_j\left\vert \sum_{i=1}^{d}b_iY+\sigma_iZ_i>\text{VaR}_{p}\left(\sum_{i=1}^{d}b_iY+\sigma_iZ_i\right)\right.\right].
\end{equation*} The general case can be recovered by incorporating the parameter $\mu_j$ as needed. The following proposition directly follows from Theorem~\ref{unique risk bounds}.

\begin{proposition}[Risk bounds for ABRMs]\label{risk bounds for ABRM}
	Let $X_i=b_iY+\sigma_iZ_i$, $i\in\mathcal{N},$ be $d\in \mathbb{N}$ risks such that $Z_i\sim G_i$ are independent of $Y$. Let $S=Y\sum_{i=1}^{d}b_i+\sum_{i=1}^{d}\sigma_i Z_i$ and $\sum_{i=1}^{d}\sigma_i Z_i\sim G$. The constrained upper bound of $\mathrm{MES}_p(X_j, S)$, $j\in\mathcal{N}$, is 
	\begin{equation}\label{ABRM upper bounds for d risk}
		M^f_j(p)=\mathbb{E}\left[b_jY+\sigma_jG_j^{-1}(U)\left\vert S^c_{Y}> \text{VaR}_p(S^c_{Y})\right.\right],
	\end{equation}where $S^c_{Y}=Y\sum_{i=1}^{d}b_i+\sum_{i=1}^{d}\sigma_i G_i^{-1}(U)$. Moreover, the constrained lower bound is 
	\begin{equation}\label{ABRM lower bounds for d risks}
		m^f_j(p)\geq \mathbb{E}\left[b_jY+\sigma_jG_j^{-1}(U)\left\vert S^a_{Y}> \text{VaR}_p(S^a_{Y})\right.\right],
		\end{equation}where $S^a_{Y}=Y\sum_{i=1}^{d}b_i+G^{-1}(1-U)$. 
\end{proposition}
We now consider a specific example in which each risk component $X_i$ follows a standard normal distribution.
\begin{example}\label{two normal example}
	Let $X_i=b_iY+\sqrt{1-b_i^2}Z_i$, $i=1,2,$ be two standard normally distributed risks such that $b_i\in(-1,1)$, $Z_i\sim N(0,1)$ are independent of $Y$, and $(X_i,Y)\sim N_2(\boldsymbol{\mu}, \bf{\Sigma})$ with the mean vector $\boldsymbol{\mu}=(0, 0)$ and the covariance matrix $\bf{\Sigma}=\begin{bmatrix}
		1 & b_i\\
		b_i & 1
	\end{bmatrix}$. Moreover, we denote the aggregate risk by $S=\sum_{i=1}^{2}b_iY+\sqrt{1-b_i^2} Z_i$. For the unconstrained bounds of $\mathrm{MES}_p(X_j, S)$, we obtain that
	$$M_j(p)=\frac{\phi\left(\Phi^{-1}(p)\right)}{1-p}, \text{ and } m_j(p)=0.$$ Under the ABRM setting, the constrained bounds of $\mathrm{MES}_p(X_j, S)$ are
	$$M^f_j(p)=\frac{b_j(b_1+b_2)+\sqrt{1-b_j^2}\sum_{i=1}^{2}\sqrt{1-b_i^2}}{\sqrt{2\left(1+b_1b_2+\sqrt{(1-b_1^2)(1-b_2^2)}\right)}}\frac{\phi\left(\Phi^{-1}(p)\right)}{1-p},$$ and $$m^f_j(p)=\frac{b_j(b_1+b_2)+\sqrt{1-b_j^2}\left(\sqrt{1-b_j^2}-\sqrt{1-b_i^2}\right)}{\sqrt{2\left(1+b_1b_2-\sqrt{(1-b_1^2)(1-b_2^2)}\right)}}\frac{\phi\left(\Phi^{-1}(p)\right)}{1-p},$$ where $i,j=1,2,$ and $i\neq j$. 
	Hence, the degree of improvement we obtain by adding the ABRM-type information is
	\begin{equation*}
		\delta_{j,p}=1-\frac{b_j(b_1+b_2)+\sqrt{1-b_j^2}\sum_{i=1}^{2}\sqrt{1-b_i^2}}{\sqrt{2\left(1+b_1b_2+\sqrt{(1-b_1^2)(1-b_2^2)}\right)}}+\frac{b_j(b_1+b_2)+\sqrt{1-b_j^2}\left(\sqrt{1-b_j^2}-\sqrt{1-b_i^2}\right)}{\sqrt{2\left(1+b_1b_2-\sqrt{(1-b_1^2)(1-b_2^2)}\right)}}.
	\end{equation*}
	
	Interestingly, under the ABRM setting with standard normal marginals, the measure of improvement \(\delta_{j,p}\) is independent of the prudence level \(p\). This property is also clearly illustrated in Panels A, B, C, and D of Table~\ref{normal distributed risks}, where the percentages of \(\delta_{j,p}\) remain unchanged for \(p=0.90\) and \(p=0.95\) when other parameters are held constant.
	
	In Panels A and B, where \(b_1 = b_2\), the upper bounds remain unchanged whether or not factor model information is incorporated, whereas the lower bounds show substantial improvement, with the extent of improvement increasing with \(b_1\). In contrast, when 
	\(b_1 = -b_2\), as considered in Panels C and D, the constrained lower bound coincides with the unconstrained bound, i.e., \(m^f_j(p) = m_j(p)\), while the constrained upper bound \(M^f_j(p)\) decreases as \(b_1\) increases. Across all cases, the improvement in the dependence-uncertainty spread becomes more pronounced as the magnitude of \(b_1\) increases. 
	
	Moreover, for two standard normally distributed risks, it can be analytically shown---and is further confirmed by Table~\ref{normal distributed risks}---that \(M^f_j(p) \leq M_j(p) = \mathrm{ES}_{p}(X_j)\) and \(m^f_j(p) \geq m_j(p) \geq \mathrm{LES}_{1-p}(X_j)\). While the dependence information provided by the ABRMs improves the bounds on $\mathrm{MES}_p(X_j, S)$, the improvement is invariant with respect to \(p\). However, the extent of improvement crucially depends on the correlation coefficients between each \(X_i\) and the common factor \(Y\), represented by the parameters \(b_i\).   
	
	To further explore the impact of \(b_i\), Figure~\ref{normal distributed risks for b} plots the unconstrained and constrained risk bounds for \(\mathrm{MES}_{0.95}(X_j, S)\) as a function of \(b_2 \in (-1,1)\), given \(b_1=0.3\). As \(b_2\) increases, the improvement in both the lower and upper bounds initially declines and subsequently increases. In particular, the minimum degree of improvement occurs at \(b_2=-0.3\). Notably, these findings for \(\mathrm{MES}_p(X_j, S)\) are consistent with those reported for \(\mathrm{ES}_{p}(S)\) in Example~3.7 of \cite{bernard2017risk}. 
	
	\begin{table}[!h]
		\centering 
		\begin{tabularx}{\textwidth}{XXXXXXX}
			\toprule
			$b_1$  & $\text{MES}_p$ & $m_j(p)$ & $M_j(p)$ & $m^f_j(p)$ & $M^f_j(p)$ & $\delta_{j,p}$\\
			\midrule
			\multicolumn{7}{l}{Panel A: $b_1=b_2$ and $p=0.9$} \\
			0.0 &  1.241 &     0.000 &  1.755 &  0.000 &  1.755 &    0.00\% \\
			0.3 &  1.296 &     0.000 &  1.755 &  0.526 &  1.755 &   30.00\% \\
			0.6 &  1.447 &     0.000 &  1.755 &  1.053 &  1.755 &   60.00\% \\
			0.9 &  1.670 &     0.000 &  1.755 &  1.579 &  1.755 &   90.00\% \\
			1.0 &  1.755 &     0.000 &  1.755 &  1.755 &  1.755 &  100.00\% \\
			\midrule
			\multicolumn{7}{l}{Panel B: $b_1=b_2$ and $p=0.95$}\\
			0.0 &  1.459 &    0.000 &  2.063 &  0.000 &  2.063 &    0.00\% \\
			0.3 &  1.523 &     0.000 &  2.063 &  0.619 &  2.063 &   30.00\% \\
			0.6 &  1.701 &     0.000 &  2.063 &  1.238 &  2.063 &   60.00\% \\
			0.9 &  1.962 &     0.000 &  2.063 &  1.856 &  2.063 &   90.00\% \\
			1.0 &  2.063 &     0.000 &  2.063 &  2.063 &  2.063 &  100.00\% \\
			\midrule
			\multicolumn{7}{l}{Panel C: $b_1=-b_2$ and $p=0.9$}\\
			0.0 &  1.241 &    0.000 &  1.755 &    0.000 &  1.755 &    0.00\% \\
			0.3 &  1.296 &     0.000 &  1.755 &    0.000 &  1.674 &    4.60\% \\
			0.6 &  1.447 &     0.000 &  1.755 &    0.000 &  1.404 &   20.00\% \\
			0.9 &  1.670 &     0.000 &  1.755 &    0.000 &  0.765 &   56.40\% \\
			1.0 &  1.755 &     0.000 &  1.755 &    0.000 &  0.000 &  100.00\% \\
			\midrule
			\multicolumn{7}{l}{Panel D: $b_1=-b_2$ and $p=0.95$}\\
			0.0 &  1.459 &     0.000 &  2.063 &    0.000 &  2.063 &    0.00\% \\
			0.3 &  1.523 &     0.000 &  2.063 &    0.000 &  1.968 &    4.60\% \\
			0.6 &  1.701 &     0.000 &  2.063 &    0.000 &  1.650 &   20.00\% \\
			0.9 &  1.962 &     0.000 &  2.063 &    0.000 &  0.899 &   56.40\% \\
			1.0 &  2.063 &     0.000 &  2.063 &    0.000 &  0.000 &  100.00\% \\
			\bottomrule
		\end{tabularx} 
		\caption{$\mathrm{MES}_{p}(X_1, S)$ bounds for the standard normal case with different $p$ and correlation coefficients of $X_1$ and $Y$, $b_1$. The column $\text{MES}_p$ lists the $\mathrm{MES}_{p}(X_1,S)$ when $(X_1, X_2)$ is a bivariate Gaussian random vector with correlation $b_1^2$.}
		\label{normal distributed risks}
	\end{table}
	
	\begin{figure}
		\input{additive_bounds_of_mes}
		\caption{$\mathrm{MES}_{0.95}(X_1, S)$ bounds for the standard normal case with $b_1=0.3$ and $b_2\in(-1, 1)$ (left panel), and the degree of improvement (right panel).}
		\label{normal distributed risks for b}
	\end{figure}
\end{example}

Through Example~\ref{two normal example}, we demonstrate that incorporating partial dependence information via the ABRMs can substantially affect the dependence-uncertainty spread when evaluating \(\mathrm{MES}_p(X_j, S)\). In other words, the ABRMs can improve the precision of risk assessment, although the degree of this improvement is influenced by both the dependence parameters and the marginal distributions of the underlying risks.  

\subsection{Multiplicative background risk model}\label{MBRM}
In this subsection, we derive risk bounds under the setting of the multiplicative background risk models (MBRMs). Specifically, the risk vector is given by
\begin{equation*}
	\bm{X}=\left(\mu_1+\sigma_1\frac{Z_1}{Y},\mu_2+\sigma_2\frac{Z_2}{Y},\dots,\mu_d+\sigma_d\frac{Z_d}{Y}\right),
\end{equation*}where $\bm{\mu}=(\mu_1,\mu_2,\dots,\mu_d)$ and $\bm{\sigma}=(\sigma_1,\sigma_2,\dots,\sigma_d)$, in which $\sigma_i>0$, are parameter vectors. In this model, the functional form is $f_i(Y,Z_i)=\mu_i+\sigma_i\frac{Z_i}{Y}$ for each $i\in\mathcal{N}$. 

The MBRMs have received considerable attention in the literature; see, for example, \citet{franke2011risk}, \citet{asimit2016background}, \citet{furman2021multiplicative}, and \citet{marri2022risk}.

As in the previous subsection, we set $\bm{\mu}=(0,0,\dots,0)$ for simplicity. Consequently, the risk vector reduces to
\begin{equation*}
	\bm{Z}=\left(\frac{\sigma_1Z_1}{Y},\frac{\sigma_2Z_2}{Y},\dots,\frac{\sigma_dZ_d}{Y}\right).
\end{equation*}
\begin{proposition}[Risk bounds for MBRMs]\label{risk bound for MBRM}
	Let $X_i=\frac{\sigma_i Z_i}{Y}$, $i\in\mathcal{N},$ be $d$ risks such that $Z_i$ are independent of $Y$. Let $S=\frac{\sum_{i=1}^{d}\sigma_i Z_i}{Y}$ and $\sum_{i=1}^{d}\sigma_i Z_i\sim G$. The constrained upper bound of $\mathrm{MES}_p(X_j, S)$, $j\in\mathcal{N}$, is 
	\begin{equation}\label{MBRM upper bounds for d risk}
		M^f_j(p)=\mathbb{E}\left[\left.\frac{\sigma_j G_j^{-1}(U)}{Y}\right\vert \frac{\sum_{i=1}^{d}\sigma_i G_i^{-1}(U)}{Y}> \text{VaR}_p\left(\frac{\sum_{i=1}^{d}\sigma_i G_i^{-1}(U)}{Y}\right)\right].
		\end{equation}The constrained lower bound is 
	\begin{equation}\label{MBRM lower bounds for d risks}
			m^f_j(p)\geq\mathbb{E}\left[\frac{\sigma_j G_j^{-1}(U)}{Y}\left\vert \frac{G^{-1}(1-U)}{Y}> \text{VaR}_p\left(\frac{G^{-1}(1-U)}{Y}\right)\right.\right].
	\end{equation}
\end{proposition}
The proof of Proposition~\ref{risk bound for MBRM} is omitted, as it follows directly from Theorem~\ref{unique risk bounds}. The following example illustrates the bounds for $\mathrm{MES}_p(X_j, S)$ when the individual risks are Lomax distributed.

\begin{example}\label{pareto example}
	Let $X_i=\frac{\sigma_i Z_i}{Y}$, $i\in\mathcal{N},$ be $d$ risks such that $Z_i$ follow the exponential distribution with mean 1, and $Y$, which is independent of $Z_i$, is the gamma RV with shape parameter $\alpha>0$ and rate parameter $\beta=1$. In this setting, each $X_i$ follows a Lomax distribution with shape parameter $\alpha$ and scale parameter $\sigma_i>0$. 
	
	Under the condition $\sum_{i=1\& i\neq j}^{d}\sigma_i\geq \sigma_j(\frac{1}{p}-1)^{\frac{1}{\alpha}+1}$, the unconstrained risk bounds for $\mathrm{MES}_p(X_j, S)$ are given by
	$$M_j(p)=\frac{\sigma_j\alpha(1-p)^{-\frac{1}{\alpha}}}{(\alpha-1)}-\sigma_j \text{ and } m_j(p)=\sigma_j\frac{\alpha p^{1-\frac{1}{\alpha}}+(1-\alpha)p-1}{(1-\alpha)(1-p)}.$$ The parameters constraint ensures that the lower bound of $m_j(p)$ is sharp in \eqref{unique risk bounds} of Theorem~\ref{upper bound}.

	Under the MBRMs, assuming $\alpha>1$ and $\sum_{i=1\& i\neq j}^{d}\sigma_i\geq \sigma_j(\frac{1}{p}-1)$, the constrained bounds for $\mathrm{MES}_p(X_j, S)$ are given by
	$$M^f_j(p)=\frac{\sigma_j\alpha(1-p)^{-\frac{1}{\alpha}}}{(\alpha-1)}-\sigma_j \text{ and } m^f_j(p)=\sigma_j\frac{\alpha p^{1-\frac{1}{\alpha}}+(1-\alpha)p-1}{(1-\alpha)(1-p)}.$$
	Here, the lower bound of $m^f_j(p)$ is obtained when $Z_j=G^{-1}_j(U)$ and $Z_i=G^{-1}_i(1-U)$ for $i\in\mathcal{N}/j$. In this case, $X_{j}$ and $S$ are conditionally antimonotonic.
\end{example} 
From Example~\ref{pareto example}, we observe that incorporating partial dependence information through the MBRMs may not necessarily reduce the dependence-uncertainty spread. Thus, the MBRMs do not always improve the assessment of risk exposure. It is worth to mention that this is not for general MBRMs.

\subsection{Minimum-based background risk model}\label{MBBRM}
This subsection is dedicated to study improved risk bounds within the framework of the minimum-based background risk models (MBRMs), a structure studied in, e.g., \citet{asimit2010multivariate} and \citet{pai2020livestock}. The risk vector in this model is defined as
\begin{equation*}
	\bm{X}=\left(\min(Y, Z_1),\min(Y, Z_2),\dots,\min(Y, Z_d)\right),
\end{equation*}where the functional form is $f_i(Y,Z_i)=\min(Y, Z_i)$ for each $i\in\mathcal{N}$.

The improved risk bounds for MBRMs are derived in the proposition below.
\begin{proposition}[Risk bounds for MBRMs]\label{risk bound for MBBRM}
	Let $X_i=\min(Y, Z_i)$, $i\in\mathcal{N},$ be $d$ risks such that $Z_i\sim G_i$ are independent of $Y$. Let $S=\sum_{i=1}^{d}\min(Y, Z_i)$ have conditional distribution $F_{S_y}$ given that $Y=y$. The constrained upper bound of $\mathrm{MES}_p(X_j, S)$, $j\in\mathcal{N}$, is 
	\begin{equation}\label{MBBRM upper bounds for d risk}
		M^f_j(p)=\mathbb{E}\left[\min(Y, G_j^{-1}(U))\left\vert \sum_{i=1}^{d}\min(Y, G_i^{-1}(U))> \text{VaR}_p\left(\sum_{i=1}^{d}\min(Y, G_i^{-1}(U))\right)\right.\right].
	\end{equation} The constrained lower bound is 
	\begin{equation}\label{MBBRM lower bounds for d risks}
		m^f_j(p)\geq\mathbb{E}\left[\min(Y, G_j^{-1}(U))\left\vert F_{S_Y}^{-1}(1-U)> \text{VaR}_p\left(F_{S_Y}^{-1}(1-U)\right)\right.\right].
	\end{equation}
\end{proposition}
The subsequent example illustrates the bounds when the individual risks $X_i$ for $i=1,2$ are exponentially distributed.
\begin{example}\label{exponential example}
	Let $X_i=\min(Y, Z_i)$, $i=1,2,$ be two risks such that $Z_i\sim \text{Expo}(\lambda)$ follow the exponential distribution with rate parameter $\lambda$, and $Y\sim  \text{Expo}(\lambda_0)$ is independent of $Z_i$. Under this setting, it follows that $X_i\sim\text{Expo}(\lambda_0+\lambda)$. For the unconstrained bounds of $\mathrm{MES}_{p}(X_1, S)$, we obtain
	$$M_j(p)=\frac{1-\ln(1-p)}{\lambda_0+\lambda}\text{ and } m_j(p)=-\frac{1}{\lambda_0+\lambda}\mathbb{E}\left[\ln(1-U)\vert -\ln(U-U^2)>u_p\right],$$
	where $u_p=\text{VaR}_p(-\ln(U-U^2))$. Under the MBRM setting, the constrained bounds for $\mathrm{MES}_{p}(X_1, S)$ are given by
	$$M^f_j(p)=\frac{1-\ln(1-p)}{\lambda_0+\lambda},$$ and $$ m^f_j(p)=\mathbb{E}\left[\min\left(\frac{f(U_1)}{\lambda_0}, \frac{f(U)}{\lambda}\right)\left\vert \min\left(\frac{f(U_1)}{\lambda_0}, \frac{f(U)}{\lambda}\right)+\min\left(\frac{f(U_1)}{\lambda_0}, -\frac{\ln(U)}{\lambda}\right)>v_p\right.\right],$$ where $U\perp U_1\sim \mathrm{Unif}[0,1]$, $f(x)=-\ln(1-x),$ and $$v_p=\text{VaR}_p\left(\min\left(\frac{f(U_1)}{\lambda_0}, \frac{f(U)}{\lambda}\right)+\min\left(\frac{f(U_1)}{\lambda_0}, -\frac{\ln(U)}{\lambda}\right)\right).$$
	The upper bound \( M^f_j(p) \) is attained when the \( Z_i \) are comonotonic, whereas the lower bound \( m^f_j(p) \) corresponds to the antimonotonic case. Notably, under the MBRM framework, the unconstrained and constrained upper bounds coincide. While explicit closed-form solutions for \( m_j(p) \) and \( m^f_j(p) \) are unavailable, they can be approximated via simple numerical methods by simulating \( U \) and \( U_1 \). 
	
	In Figure~\ref{exponential distributed risks}, we depict the (improved) risk bounds for \( \mathrm{MES}_{0.95}(X_1, S) \) when \( X_i \), $i=1,2,$ (along with \( Y \) and \( Z_i \)) follow an exponential distribution with rate parameter \( \lambda_0+\lambda \) (parameters \( \lambda_0\) and \( \lambda \), respectively). The left panel presents the results for \( \lambda_0=1 \) and \( \lambda \in (0.1, 2.1) \), while the right panel shows the degree of improvement \( \delta_{0.95} \). The (improved) lower bound is computed numerically using a discretized version of \( U = \left[\frac{1}{n+1}, \frac{2}{n+1}, \dots, \frac{n}{n+1} \right] \) with \( n=10^5 \), while \( U_1 \) is obtained by randomly shuffling the elements of \( U \). The results indicate that as \( \lambda \) increases, the degree of improvement decreases.
	
	\begin{figure}
		\input{min_bounds_of_mes}
		\caption{$\mathrm{MES}_{0.95}(X_1, S)$ bounds for the exponential case with $\lambda=1$ and $\lambda_0\in(0.1, 2.1)$ (left panel), and the degree of improvement (right panel).}
		\label{exponential distributed risks}
	\end{figure}
\end{example}

Example~\ref{exponential example} demonstrates that the MBRMs can reduce the spread of dependence uncertainty by incorporating additional information about the dependence structure. However, the marginal distribution parameters significantly influence the degree of improvement, and under specific parameter choices, the effect may even vanish.

\section{Alternative constrained risk bounds}\label{linear condition}
In this section, we derive a different form of constrained risk bounds compared to those for factor models discussed previously. Specifically, we assume that the conditional expectation of each individual RV has a common linear relationship with the aggregate sum of all risks. For $\alpha, \beta\in \mathbb{R},$ we formulate the corresponding problems as 
\begin{align}
	m^l_j(p)&=\inf \left\{\mathrm{MES}_p(X_j,S): S=\sum_{i=1}^{d}X_i, X_i\sim F_i, \mathbb{E}[X_i|S]=\alpha+\beta S \right\}, \label{m for linear} \\
	M^l_j(p)&=\sup \left\{\mathrm{MES}_p(X_j,S): S=\sum_{i=1}^{d}X_i, X_i\sim F_i, \mathbb{E}[X_i|S]=\alpha+\beta S \right\}.\label{M for linear}
\end{align}

We first recall the following theorem regarding the Weighted Insurance Pricing Model (WIPM) from \citet{furman2008weighted, furman2017adaptation} and references therein.

\begin{theorem}[WIPM]\label{WIPM}
	Let $X_i\sim F_i$, $i\in\mathcal{N},$ be $d$ risks and $S=\sum_{i=1}^{d}X_i$. Let $\Pi_{w}(X_j, S)=\frac{\mathbb{E}[X_jw(S)]}{\mathbb{E}[w(S)]}$ and $\pi_{w}(S)=\Pi_{w}(S, S)$, where $w(\cdot)$ is a weight function. If there exist constants $\alpha, \beta\in \mathbb{R}$ such that $\mathbb{E}[X_i|S]=\alpha+\beta S$ for all $i$, then
	\begin{equation}\label{wipm}
		\Pi_{w}(X_j, S)=\mathbb{E}[X_j]+\beta\left(\pi_{w}(S)-\mathbb{E}[S]\right)
	\end{equation} holds for every non-decreasing weight function $w(\cdot)$ for which all quantities under consideration are well-defined and finite.
\end{theorem}

\begin{remark}The weighted risk capital allocation rule $\Pi_{w}(X_j, S)$ is a versatile risk measure enables the study of different risk problems through appropriate choices of $w$. For example, by choosing $w(s)=\mathds{1}_{s>t},$ where $t\in\mathbb{R}$, $\Pi_{w}(X_j, S)$ reduces to the systemic risk allocation measure $\mathrm{MES}_p(X_j,S)$. We refer to \cite{furman2008weighted} for further details on different weighted allocation rules corresponding to various choices of $w$.
\end{remark}
\begin{proposition}\label{bivariate normal}
		Let $(X,S)\sim N_2(\boldsymbol{\mu}, \bf{\Sigma})$, where the mean vector $\boldsymbol{\mu}=(\mu_X, \mu_S)$ and covariance matrix $\bf{\Sigma}=\begin{bmatrix}
			\sigma_X^2 & \rho\sigma_X\sigma_S \\
			\rho\sigma_X\sigma_S & \sigma_S^2
		\end{bmatrix}.$  The constrained lower and upper bounds of $\mathrm{MES}_{p}(X, S)$ are 
	\begin{equation}\label{risk bounds for normal risk}
		\mu_X-\frac{\sigma_X}{\sigma_S}\left(\mathrm{ES}_{p}(S)-\mu_S\right)\leq\mathrm{MES}_{p}(X, S)\leq \mu_X+\frac{\sigma_X}{\sigma_S}\left(\mathrm{ES}_{p}\left(S\right)-\mu_S\right).
	\end{equation}
\end{proposition}

\begin{remark}
	Proposition~\ref{bivariate normal} can be further extended to the case where $(X,S)$ follows a bivariate elliptical distribution. In this generalization, the intercept $\alpha$ and slope $\beta$ retain the same forms as in the normal case \citep{fang2018symmetric}, although $\sigma_S^2$, $\sigma_X^2$ and $\rho\sigma_{X}\sigma_{S}$ may not represent the actual variances and covariance.
\end{remark}

\begin{theorem}\label{non-positive or negative}
	Let $X_i\sim F_i$, $i\in\mathcal{N},$ be $d$ non-negative (or non-positive) and continuous risks with $\mu_i=\mathbb{E}[X_i]$, and $S=\sum_{i=1}^{d}X_i$. Suppose there exist constants $\alpha, \beta\in \mathbb{R}$ such that $\mathbb{E}[X_i|S]=\alpha+\beta S$. Then, the constrained lower and upper bounds for $\mathrm{MES}_p(X_j, S)$ are given by
	\begin{equation}\label{risk bounds for nonnegative risk}
		\mu_j\leq m^l_j(p) \leq\mathrm{MES}_p(X_j, S)\leq r_j\sum_{i=1}^{d}\mathrm{ES}_{p}\left(X_i\right)=M^l_j(p),
	\end{equation} where $r_j=\frac{\mu_j}{\sum_{i=1}^{d}\mu_i}$. Furthermore, the lower bound $\mu_j$ is sharp if there exists $Y_i\overset{d}{=}X_i$ such that $(Y_1,Y_2,\dots,Y_d)$ is jointly mixable. The upper bound $r_j\sum_{i=1}^{d}\mathrm{ES}_{p}\left(X_i\right)$ is obtained when $X_i=F_i^{-1}(U)$.
\end{theorem}

\begin{remark}
	To achieve the sharp lower bound in \eqref{risk bounds for nonnegative risk}, joint mixability of $(F_1, F_2, \dots, F_d)$ must be established, which in general is a nontrivial problem. Necessary and sufficient conditions for joint mixability for various classes of distributions are provided in \cite{wang2016joint}; see also the initial concept introduced by \cite{wang2011complete}. Therefore, the sharpness of the lower bound is affected by the nature of the distribution functions. The upper bound, however, is always sharp and corresponds to the comonotonic case.
\end{remark}
The following example illustrates a case where the linear conditional expectation assumption is satisfied.
\begin{example}\label{unif example}
	Let $X_i\sim \mathrm{Unif}[0,1]$, $i=1,2,3,$ where $\mu_i=\frac{1}{2}$, and $S=X_1+X_2+X_3$ denote the aggregate risk. The unconstrained bounds for $\mathrm{MES}_p(X_j, S)$ are
	$$M_j(p)=\frac{1}{2}(1+p) \text{ and } m_j(p)=\frac{1}{2}(1-p).$$
	Assuming $\mathbb{E}[X_i|S]=\frac{1}{3}S$, the constrained bounds for $\mathrm{MES}_p(X_j, S)$ are
	$$M^l_j(p)=\frac{1}{2}(1+p)\text{ and }m^l_j(p)=\frac{1}{2}.$$
    Furthermore, the lower bound $\frac{1}{2}$ is obtained when 
	\begin{equation*}
		\begin{aligned}
			Y_1&=U\sim \mathrm{Unif}[0,1]; \\
			Y_2&=\left\{
			\begin{aligned}
				&-2U+1,  &\text{if } 0\le U \le \frac{1}{2}; \\
				&-2U+2,    &\text{if } \frac{1}{2}< U \le 1;
			\end{aligned}
			\right. \\
			Y_3&=\left\{
			\begin{aligned}
				&U+\frac{1}{2},  &\text{if } 0\le U \le \frac{1}{2}; \\
				&U-\frac{1}{2},    &\text{if } \frac{1}{2}< U \le 1.
			\end{aligned}
			\right. \\
		\end{aligned}
	\end{equation*} Note that $(Y_1,Y_2,Y_3)$ is jointly mixable with joint center $K=\frac{3}{2}$, i.e., $\sum_{i=1}^{3}Y_i=\frac{3}{2}$ almost surely; see \citet{ruschendorf2002variance} for the concept of the mixing copula. The upper bound $\frac{1}{2}(1+p)$ is obtained when $H_i=U$ for each $i$.
	
	The degree of improvement is $\delta_{j,p}=\frac{1}{2},$ which is independent of $p$. Importantly, while an improvement in the lower bound exists: $$m_j(p)=\frac{1}{2}(1-p)\leq \frac{1}{2}=m^l_j(p),$$ the upper bound remains unchanged. These findings are illustrated in Table~\ref{unif distributed risks}, where the gap between improved bounds constitutes exactly 50\% of the original spread between unconstrained bounds. 
	
	\begin{table}[!h]
		\centering 
		\captionsetup{justification=raggedright,singlelinecheck=false}
		\begin{tabularx}{\textwidth}{XXXXXX}
			\toprule
			p  & $m_j(p)$ & $M_j(p)$ & $m^l_j(p)$ & $M^l_j(p)$ & $\delta_{j,p}$\\
			\midrule
			0.55 &  0.225 &  0.775 &   0.500 &   0.775 &  50\% \\
			0.65 &  0.175 &  0.825 &   0.500 &   0.825 &  50\% \\
			0.75 &  0.125 &  0.875 &   0.500 &   0.875 &  50\% \\
			0.85 &  0.075 &  0.925 &   0.500 &   0.925 &  50\% \\
			0.95 &  0.025 &  0.975 &   0.500 &   0.975 &  50\% \\
			\bottomrule
		\end{tabularx} 
		\caption{$\mathrm{MES}_p(X_j, S)$ bounds for the standard uniform case with various $p$.}
		\label{unif distributed risks}
	\end{table}
\end{example}

\section{An empirical study with real data}\label{empirical study}
In this section, we conduct empirical analyses on risk bounds under full dependence uncertainty, as well as on improved risk bounds within the framework of the Fama–French five-factor model (FFM). The FFM is one of the additive background risk models introduced in Section~\ref{ABRM}. For a detailed introduction to the FFM, we refer the reader to \cite{fama2015five}.

We consider a vector of risk components $\bm{X}=\left(X_1,X_2,\dots,X_d\right)'$ with
\begin{equation}\label{ffm}
	X_i=X_f+\mu_i+\bm{b_i}'\bm{Y}+Z_i,
\end{equation}where $X_f$ denotes the risk-free loss (i.e., the negative of the risk-free return), $\mu_i$ represents abnormal losses, $\bm{Y}=(Y_1,Y_2,Y_3,Y_4,Y_5)'$ contains the Fama–French five factors, and $\bm{b_i}=(b_{i1},b_{i2},b_{i3},b_{i4}, b_{i5})'$ are the corresponding factor exposures. The $Z_i$ terms are idiosyncratic components, assumed to be independent in the standard FFM. However, as discussed earlier, the independence assumption is often unrealistic and lacks empirical support. We therefore evaluate the sensitivity of the FFM to this assumption in the context of MES using real data. 

For the empirical analysis, we use daily losses of stocks that were continuously included in the S\&P 500 index during the sample period to represent the random losses $X_i$ and the aggregate loss $S=\sum_{i=1}^dX_i$. In these experiments, the individual risk is represented by the loss of a selected constituent $X_j$, while the market risk corresponds to the sum of losses of all constituents in the index ($S$). Historical returns and constituent data for the S\&P 500 are obtained from the CRSP dataset in Wharton Research Data Services. Losses are computed as the negative of returns. The daily Fama–French five factors are collected from the Kenneth R.\ French Data Library\footnote{\url{https://mba.tuck.dartmouth.edu/pages/faculty/ken.french/data_library.html}.}. The abnormal losses $\mu_i$ and factor exposures $\bm{b_i}$ in the FFM are estimated using generalized linear models (GLMs) via the Python package \texttt{statsmodels}, fitted on daily losses and factor data. For simplicity, we do not discuss model fit in detail, as it is not the primary focus of this paper. Nevertheless, we note that the GLMs exhibit strong explanatory performance.

Throughout the experiments, daily losses, factors, and risk-free losses are treated as sampled values for the loss vector $\bm{X}$, factor vector $\bm{Y}$, and $X_f$, respectively. The idiosyncratic risks $Z_i$ are computed from~\eqref{ffm} using the estimated parameters $\mu_i$ and $\bm{b_i}$. We then calculate the unconstrained risk bounds for MES, $M_j(p)$ and $m_j(p)$, and the constrained bounds, $M^f_j(p)$ and $m^f_j(p)$ (Theorem~\ref{upper bound}). For comparison, we also compute the empirical MES directly from the sampled values of $\bm{X}$. The prudence level $p$ is taken from linearly spaced values in $[0,0.99]$. Note that when $p=0$, all five quantities reduce to the sample mean of $X_j$.

\subsection{Long-term empirical experiment}\label{long term case}
In the first study, we analyze daily losses and factor data over a twenty-year period, from 1 January 2005 to 31 December 2024. The dataset contains 213 stocks that were continuously included in the S\&P 500 during this period, each with 5033 daily losses. Here, $X_j$ represents the random loss of Apple Inc., one of the most heavily weighted constituents in the index.

Figure~\ref{es with max and min m} presents the empirical unconstrained and constrained risk bounds, along with $\mathrm{MES}_p(X_j,S)$, as functions of the prudence level $p$. We can conclude the following findings from the plot. First, the unconstrained risk bounds $M_j(p)$ and $m_j(p)$ indicate that the loss distribution of Apple is slightly left-skewed, since for symmetric distributions we would expect $M_j(p)=-m_j(p)$. Second, adding partial dependence information via the FFM structure does not improve the upper bound; i.e., $M_j(p)\approx M^f_j(p)$. This is consistent with the results in Panels A and B of Table~\ref{normal distributed risks} for standard normal marginals and additive background risk models with a single common factor. For certain values of $p$, we have $M_j(p)< M^f_j(p)$ due to the fitting error introduced by the FFM. Essentially, the FFM is a multiple linear regression model built under specific assumptions. Third, the lower bound improves substantially when partial dependence is incorporated: $m^f_j(p)$ is much larger than $m_j(p)$, which assumes fully dependence uncertainty. Finally, the contribution of Apple's risk to systemic market risk, measured by $\mathrm{MES}_p(X_j, S)$, lies between $M_j(p)$ and $m_j(p)$. The constrained bounds $M^f_j(p)$ and $m^f_j(p)$ provide more realistic and practically useful ranges for MES.

The following subsections illustrate the relevance of these bounds in extreme market scenarios, specifically the 2007--2009 financial crisis and the COVID-19 pandemic.

\begin{figure}
	\centering
	\input{empirical_bounds_of_mes.tex}
	\caption{Unconstrained and constrained risk bounds, together with the empirical values of $\mathrm{MES}_p(X_j, S)$, as functions of the prudence level $p$. $X_i$ denotes the daily loss of the $i$-th S\&P 500 constituent that remained continuously in the index from 1 January 2005 to 31 December 2024 $(i=1,2,\dots,213)$. The aggregate loss is given by $S=\sum_{i=1}^{213} X_i$, and $X_j$ corresponds to Apple Inc.}
	\label{es with max and min m}
\end{figure}

\subsection{The financial crisis}\label{extrema case}
For this experiment, we use daily losses and factor data over a one-year period from 18 September 2007 to 17 September 2008. This period corresponds to the peak of the financial crisis, culminating in the bankruptcy and delisting of Lehman Brothers Holdings Inc.\ on 17 September 2008. The dataset contains 453 stocks that remained continuously in the S\&P 500 during this period, each with 253 daily losses. Here, $X_j$ represents the random loss of Lehman Brothers.

Figure~\ref{es with max and min m for fin crisis} shows the empirical lower and upper bounds under fully uncertain and partially known dependence structures. In contrast to Apple’s long-term case, the loss distribution of Lehman Brothers during the crisis is right-skewed, as indicated by $M_j(p)$ and $m_j(p)$. This indicates that its stock price exhibited a downward trend during the financial crisis. Moreover, the attainable MES during the crisis is ten times magnitude larger than in the long-term study, as seen from the scales of Figures~\ref{es with max and min m} and \ref{es with max and min m for fin crisis}.

Crucially, the contribution of the individual risk induced by Lehman Brothers to market systemic risk, represented by the aggregated loss, nearly coincides with the constrained and unconstrained upper bounds, $M_j(p)$ and $M^f_j(p)$. This highlights the practical value of both the general upper bound and the constrained bound incorporating partial dependence information in Section~\ref{no information}. In worst-case scenarios, the collapse of a single institution, such as Lehman Brothers, can destabilize the entire financial system. It is evident that empirical $\mathrm{MES}_p(X_j, S)$ is quite close to both the unconstrained and constrained upper bounds. The constrained lower bound is substantially improved compared to the unconstrained one. Additionally, the degree of improvement for dependence-uncertainty spread through FFM is observed to be higher than the long-term case of Apple.

\begin{figure}
	\centering
	\input{empirical_bounds_fin_cris.tex}
	\caption{Unconstrained and constrained risk bounds, together with the empirical values of $\mathrm{MES}_p(X_j, S)$, as functions of the prudence level $p$. $X_i$ denotes the daily loss of the $i$-th S\&P 500 constituent that remained continuously in the index from 18 September 2007 and 17 September 2008 $(i=1,2,\dots,453)$. The aggregate loss is given by $S=\sum_{i=1}^{453} X_i$, and $X_j$ corresponds to Lehman Brothers Holdings Inc.}
	\label{es with max and min m for fin crisis}
\end{figure}

\subsection{The pandemic downturn}
The most recent major downturn was triggered by the pandemic, began with an outbreak of COVID-19 in December 2019. To further examine the practical usefulness of the bounds, we use daily losses and factors from 18 April 2019 to 23 March 2020. The start date corresponds to the listing of Zoom Communications Inc., while the end date marks the last day of the pandemic-induced downturn. The dataset includes 476 stocks that remained continuously in the S\&P 500 (excluding Zoom) during this period, each with 234 daily losses. Unlike most firms, Zoom benefited from the pandemic due to increased demand for its services, making it an ideal case of an individual risk that contributes minimally---or even negatively---to systemic risk.

In Figure~\ref{es with max and min m for covid crisis}, we present the empirical lower and upper bounds when the dependence information is completely unknown or partially known from the structure of FFM. The unconstrained bounds reveal that the loss distribution of Zoom during the pandemic is left-skewed, reflecting an increase in its stock price. Although the pandemic produced a narrower attainable range of MES than the financial crisis, it resulted in a wider range than in the long-term case. 

Interestingly, unlike in previous subsections, the unconstrained and constrained upper bounds no longer coincide, while the lower bounds tend to close with each other. The empirical $\mathrm{MES}_p(X_j, S)$ demonstrates the contribution of the individual risk produced by Zoom to the systemic risk is negligible and, in some cases ($p>0.8$), even reduces systemic risk. The lower bounds are thus attainable in practice, and the constrained upper bound provides a significant improvement over the unconstrained one. The empirical $\mathrm{MES}_p(X_j, S)$ decreases for large $p$ because Zoom's stock return tends to rise when the market declines during the pandemic downtown.

\begin{figure}
	\centering
	\input{empirical_bounds_pandemic.tex}
	\caption{Unconstrained and constrained risk bounds, together with the empirical values of $\mathrm{MES}_p(X_j, S)$, as functions of the prudence level $p$. $X_i$ denotes the daily loss of the $i$-th S\&P 500 constituent that remained continuously in the index from 18 April 2019 and 23 March 2020 $(i=1,2,\dots,476)$. The aggregate loss is given by $S=\sum_{i=1}^{476} X_i$, and $X_j$ corresponds to Zoom Communications Inc.}
	\label{es with max and min m for covid crisis}
\end{figure}

\subsection{Systemic Risk Criticality Index}

To compare the systemic risk exposures of different companies, we define two Systemic Risk Criticality Indices (SRCIs):
\begin{equation*}
	\beta_{j,p} = 1-\frac{M_j(p)-\mathrm{MES}_p(X_j,S)}{M_j(p)-m_j(p)},
\end{equation*}and
\begin{equation*}
	\beta_{j,p}^f = 1-\frac{M^f_j(p)-\mathrm{MES}_p(X_j,S)}{M^f_j(p)-m^f_j(p)}.
\end{equation*}

SRCIs serve as measures of each firm’s contribution to systemic risk. Higher values indicate a greater potential for the firm to trigger or amplify market collapse, providing valuable information for policymakers concerned with financial stability.

Table~\ref{index of systemic importance} reports the values of $\beta_{j,p}$ and $\beta_{j,p}^f$ for $p \in \{0.99, 0.993, 0.995\}$, covering a selection of companies that are currently or were previously included in the S\&P 500 index. SRCIs for Apple, Coca-Cola, Microsoft, and BNY Mellon are computed using daily loss data from 1 January 2005 to 31 December 2024, corresponding to the long-term analysis described in Section~\ref{long term case}. The results for Lehman Brothers and Zoom, by contrast, are based on the periods from 18 September 2007 to 17 September 2008 and from 18 April 2019 to 23 March 2020, respectively, as discussed in Section~\ref{extrema case}.

As expected, Lehman Brothers during the global financial crisis exhibits the highest systemic risk importance across all three prudence levels, as measured by both $\beta_{j,p}$ and $\beta_{j,p}^f$, while Zoom during the COVID-19 pandemic shows the lowest systemic risk importance. Moreover, financial institutions such as BNY Mellon display higher systemic risk importance than firms in other industry sectors. For instance, $\beta_{j,0.995}$ equals 0.823 for Coca-Cola but rises to 0.927 for BNY Mellon.

\begin{table}[!h]
	\centering 
	\begin{tabularx}{\textwidth}{l|lXX|XXX}
		\hline
		\multirow{2}{*}{Individual risk}  & \multicolumn{3}{c|}{Estimated $\beta_{j,p}$} & \multicolumn{3}{c}{Estimated $\beta_{j,p}^f$} \\
		\cline{2-7}
		& $p=0.990$ & $0.993$ & $0.995$ & $0.990$ & $0.993$ & $0.995$ \\
		\hline
		Apple           & 0.832  &  0.839  & 0.823 & 0.737 & 0.768 &  0.737  \\
		Lehman Brothers & 1.000  &  1.000  & 1.000 & 1.000 & 1.000 &  1.000   \\
		Zoom            & 0.661  &  0.666  & 0.666 & 0.600 & 0.250 &  0.250   \\
		Coco Cola       & 0.859  &  0.839  & 0.823 & 0.857 & 0.846 &  0.840  \\
		Microsoft       & 0.838  &  0.864  & 0.865 & 0.647 & 0.718 &  0.751  \\
		BNY Mellon      & 0.921  &  0.932  & 0.927 & 0.965 & 0.970 &  0.951  \\
		\hline
	\end{tabularx} 
	\caption{SRCIs $\beta_{j,p}$ and $\beta_{j,p}^f$. The indices for Apple, Coco Cola, Microsoft and BNY Mellon are computed using daily loss data from 1 January 2005 to 31 December 2024, whereas those for Lehman Brothers and Zoom are based on the periods from 18 September 2007 to 17 September 2008 and from 18 April 2019 to 23 March 2020, respectively.}
	\label{index of systemic importance}
\end{table}

The empirical studies in this section confirm the theoretical properties of the unconstrained and constrained risk bounds developed in Section~\ref{no information}. For practitioners and policymakers, these results are particularly valuable in managing systemic risk in worst-case scenarios. For instance, global regulators could employ such bounds to determine appropriate reserve requirements for banks and insurers, thereby reducing the likelihood of market collapse triggered by individual institutions. Moreover, PPPs could use the SRCIs to assess the relative systemic position of firms within the market.

\section{Conclusion} \label{conclusion}
In this paper, we studied the static systemic risk contribution measure known as MES, which plays a central role in financial and insurance risk management. We first derived risk bounds for MES under the assumption of fixed marginal distributions, with either completely unknown or partially specified dependence structures. Partial dependence information was incorporated through structural factor models, resulting in more practical and tighter bounds. Our analytical results further demonstrate that incorporating partial dependence information may improve the bounds. Specifically, the dependence-uncertainty spread---the gap between the upper and lower bounds---reduces compared with the case of complete dependence uncertainty. In addition, we derived constrained risk bounds for MES under the assumption that the conditional expectations of individual risks, given the aggregate risk, follow a linear relationship. In this framework, the partial dependence information is implicit, while for non-negative risks we provide explicit bounds. Finally, we conducted empirical analyses based on financial loss data to obtain both unconstrained and constrained risk bounds. These empirical results illustrate the practical relevance of our theoretical findings and underscore the importance of incorporating dependence information in systemic risk assessment.

\paragraph{Acknowledgments:}
The authors thank the three reviewers and the editor for their valuable comments and suggestions, which have helped improve the paper. Financial support from the project “New Order of Risk Management: Theory and Applications in the Era of Systemic Risk”, provided through the Natural Sciences and Engineering Research Council of Canada (NSERC) grant ALLRP 580632 and the Mathematics of Information Technology and Complex Systems (Mitacs) grant IT33381, is gratefully acknowledged. The authors also thank Yang Yang for his helpful remarks on the paper.  

\titleformat{\section}[block]{\Large\bfseries}{Appendix \Alph{section}:}{1ex}{}[]
\begin{appendices}
	\section{Proof of Theorem~\ref{upper bound}}\label{stoch order proof}
	Inequalities~\eqref{unique risk bounds} is a direct consequence, if we prove 
	\begin{equation}\label{fsd}
		\begin{aligned}
			X_j^c\vert S^a>\text{VaR}_{p}(S^a) &\leq_{st} X_{j\vert Y}^c\vert S^a_Y> \text{VaR}_p(S^a_Y) \\ &\leq_{st} X_j\vert S> \text{VaR}_p(S) \\
			&\leq_{st} X_{j\vert Y}^c\vert S^c_Y> \text{VaR}_p(S^c_Y) \leq_{st} X_j^c\vert S^c>\text{VaR}_{p}(S^c),   
		\end{aligned}
	\end{equation}where $\leq_{st}$ denotes the usual stochastic order or first order stochastic dominance.
	For random variables $X_1\sim F_1$ and $X_2\sim F_2$, $X_1\leq_{st} X_2$ if and only if $F_1(t)\geq F_2(t)$ for all $t\in\mathbb{R}$ \citep[see, e.g., Theorem~2.2 of][]{bauerle2006stochastic}. Thus, proving \eqref{fsd} is equivalent to proving 
	\begin{equation*}
		\begin{aligned}
			F_{X_j^c\vert S^a>\text{VaR}_{p}(S^a)}(t) \geq F_{X^c_{j\vert Y}\vert S^a_Y>  \text{VaR}_{p}(S^a_Y)}(t) &\geq F_{X_j\vert S> \text{VaR}_{p}(S)}(t) \\
			&\geq F_{X_{j\vert Y}^c\vert S^c_Y> \text{VaR}_{p}(S^a_Y)}(t) \geq F_{X_j^c\vert S^c>\text{VaR}_{p}(S^c)}(t),
		\end{aligned}
	\end{equation*}for all $t$. Let $s_p=\text{VaR}_{p}(S)$. First, we have
	\begin{equation}\label{prob of cond Xj}
		\begin{aligned}
			F_{X_j\vert S>s_p}(t)=\mathbb{P}(X_j\leq t\vert S> s_p)=\frac{\mathbb{P}(X_j\leq t, S> s_p)}{\mathbb{P}(S> s_p)}=\frac{\mathbb{P}(X_j\leq t)-\mathbb{P}(X_j\leq t, S\leq s_p)}{\mathbb{P}(S> s_p)}.
		\end{aligned}
	\end{equation}Under the assumptions of factor models in~\eqref{factor model}, 
	\begin{equation}
			\mathbb{P}(X_j\leq t, S\leq s_p)=\int\mathbb{P}\left(X_{j\vert y}\leq t, S_{y}\leq s_p\right)dH(y)
	\end{equation}By the Fréchet-Hoeffding bounds for joint distribution functions \citep[see, e.g.,][]{nelsen2006introduction}, 
	\begin{equation}\label{frech-hoef}
			\max(F_{j\vert y}(t)+F_{S_y}(s_p)-1, 0)\leq \mathbb{P}\left(X_{j\vert y}\leq t, S_{y}\leq s_p\right)\leq \min(F_{j\vert y}(t), F_{S_y}(s_p)),
	\end{equation}where the upper bound is obtained by $S_y^c$ and $X_{j\vert y}^c$, and the lower is obtained by $S_y^a$ and $X_{j\vert y}^c$. Therefore, by \eqref{prob of cond Xj}-\eqref{frech-hoef}, for every $t$,
	\begin{equation*}
		\begin{aligned}
			F_{X_{j\vert Y}^c\vert S_Y^a>\text{VaR}_{p}(S_Y^a)}(t)&=\frac{F_j(t)-\int\max(F_{j\vert y}(t)+F_{S_y}(s_p)-1, 0)dH(y)}{\mathbb{P}(S> s_p)}\\&\geq F_{X_j\vert S>s_p}(t)\\
			&\geq \frac{F_j(t)-\int\min(F_{j\vert y}(t), F_{S_y}(s_p))dH(y)}{\mathbb{P}(S> s_p)}=F_{X_{j\vert Y}^c\vert S_Y^c>\text{VaR}_{p}(S_Y^c)}(t).
		\end{aligned}
		\end{equation*}Furthermore, 
	$\min(F_{j\vert y}(t), F_{S_y}(s_p))\leq F_{j\vert y}(t)$ and $\min(F_{j\vert y}(t), F_{S_y}(s_p))\leq F_{S_y}(s_p)$. Integrating both sides of these two inequalities, we have
	$$\int\min(F_{j\vert y}(t), F_{S_y}(s_p))dH(y)\leq \min(F_{j}(t), F_{S}(s_p)).$$ 
	Thus, 
	$$F_{X_{j\vert Y}^c\vert S_Y^c>\text{VaR}_{p}(S_Y^c)}(t)\geq \frac{F_j(t)-\min(F_{j}(t), F_{S}(s_p))}{\mathbb{P}(S> s_p)}=F_{X_j^c\vert S^c>\text{VaR}_{p}(S^c)}(t).$$ 
	Similarly, we can prove that 
	$$\int\max(F_{j\vert y}(t)+F_{S_y}(s_p)-1, 0)dH(y)\geq \max(F_{j}(t)+F_{S}(s_p)-1, 0),$$ 
	and, then, 
	$$F_{X_{j\vert Y}^c\vert S_Y^a>\text{VaR}_{p}(S_Y^a)}(t)\leq \frac{F_j(t)-\max(F_{j}(t)+F_{S}(s_p)-1, 0)}{\mathbb{P}(S> s_p)}=F_{X_j^c\vert S^a>\text{VaR}_{p}(S^a)}(t).$$

	\section{Proof of Proposition~\ref{cont case}}
	For continuous RVs $X_i$,
	\begin{equation*}
		\mathbb{E}[X_j^c\vert S^c>\text{VaR}_{p}(S^c)]=\mathbb{E}[F_j^{-1}(U)\vert U>p]=\frac{\int_p^1 \mathrm{VaR}_q(X_j)dq}{1-p}=\mathrm{ES}_p(X_j),
	\end{equation*}where the components of $(X_1^c, X_2^c,\dots, X_d^c)$ share a common tail event of probability $1-p$ (i.e., $X_i^c\mathds{1}_{X_i^c>\mathrm{VaR}_p(X_i^c)}=F_i^{-1}(U\mathds{1}_{U>p})$), and
				\begin{equation*}
					\mathbb{E}[X_j^c\vert S^a>\text{VaR}_{p}(S^a)]=\mathbb{E}[F_j^{-1}(U)\vert U<1-p]=\frac{\int_0^{1-p} \mathrm{VaR}_q(X_j)dq}{1-p}=\mathrm{LES}_{1-p}(X_j).
				\end{equation*}

	\section{Proof of Proposition~\ref{symmetric margin}}
		When $X_j=F^{-1}(U)$ and $X_i=F^{-1}(1-U)$ for $i\in\mathcal{N}/j$, we have $$S=(d-2)F^{-1}(1-U)+2\mu,$$ where $\mu$ is the mean of $F$. This result follows from the property $F^{-1}(U)+F^{-1}(1-U)=2\mu$, as discussed in the concept of complete mixability; see \citet{wang2011complete}. Thus $X_j$ and $S$ are antimonotonic.
	
	\section{Proof of Proposition~\ref{bivariate normal}}
	Under the bivariate normal setting, it is well-known that $\mathbb{E}[X|S]=\alpha+\beta S$, where $\alpha=\mu_X-\beta\mu_S$ and $\beta=\rho\frac{\sigma_X}{\sigma_S}$; see, e.g., \cite{furman2017beyond}. An application of Theorem~\ref{WIPM} then yields
		\begin{equation*}
			\mathrm{MES}_{p}(X, S)=\mu_X+\rho \frac{\sigma_X}{\sigma_S}\left(\mathrm{ES}_{p}(S)-\mu_S\right),
		\end{equation*} Since $\rho\in [-1,1]$ and all other terms are constants, the corresponding risk bounds follow as shown in Equation~\eqref{risk bounds for normal risk}. 
	
	\section{Proof of Theorem~\ref{non-positive or negative}}
		We first prove the case of $d$ non-negative risks. Since $\sum_{i=1}^{d}\mathbb{E}[X_i|S=0]=\mathbb{E}[S|S=0]=0$ and $\mathbb{E}[X_i|S=0]\geq 0$ for all $i$, it follows that $\alpha=0$ if $\mathbb{E}[X_j|S]=\alpha+\beta S$. In addition,\begin{equation*}
			\beta=\frac{\mu_j}{\sum_{i=1}^{d}\mu_i},
		\end{equation*}because $\mathbb{E}[\mathbb{E}[X_j|S]]=\mathbb{E}[X_j]=\beta \mathbb{E}[S]$. Since $\Pi_{w}(X_j, S)=\mathrm{MES}_p(X_j, S)$ for $w(s)=\mathds{1}_{s>t}$, Theorem~\ref{WIPM} applies. Then,
		\begin{equation}\label{connection between x and s}
			\mathrm{MES}_p(X_j, S)=\mu_j+\frac{\mu_j}{\sum_{i=1}^{d}\mu_i}\left(\mathrm{ES}_{p}(S)-\sum_{i=1}^{d}\mu_i\right)=\frac{\mu_j}{\sum_{i=1}^{d}\mu_i}\mathrm{ES}_{p}(S).
		\end{equation}In this case, solving for risk bounds on $\mathrm{MES}_p(X_j, S)$ reduces to solving for bounds on $\mathrm{ES}_{p}(S)$. By Propositions~3.2 and 3.5 of \cite{bernard2017risk},
		\begin{equation*}
			\sum_{i=1}^{d}\mu_i=\mathrm{ES}_{p}\left(\sum_{i=1}^{d}\mu_i\right)	\leq \mathrm{ES}_{p}(S)\leq \mathrm{ES}_{p}\left(\sum_{i=1}^{d}F_i^{-1}(U)\right)=\sum_{i=1}^{d}\mathrm{ES}_{p}\left(X_i\right).
		\end{equation*} The last equation holds because ES is comonotonic additive and objective. The risk bounds on $\mathrm{MES}_p(X_j, S)$ are thus obtained by Equation~\eqref{connection between x and s} and $\beta\geq 0$. 
		
		The proof for non-positive risks is similar, with the crucial observation that $\beta\geq 0$ in both cases.
\end{appendices}
\bibliographystyle{chicago}
\bibliography{Bibliography}
\end{document}

%% file: lognorm_bounds_of_mes.tex
\begin{tikzpicture}
	\begin{axis}[
		xlabel={$p$},
		ylabel={$\mathrm{MES}_p(X_j,S)$},
		xtick={1},
		extra x ticks={0, 0.9402985074626866},
		extra x tick labels={0, q},
		extra y ticks={0.14569054019826758, 1.6482035551275955, 5.60425680837996, 6.908718691599656, 7.962863824636424},
		extra y tick labels={$\mathrm{LES}_{1-q}(X_j)$, $\mu_j$, $m^f_j(q)$, $M^f_j(q)$, $\mathrm{ES}_q(X_j)$},
		ytick=\empty,
		xmin=0, xmax=1.05,
		ymin=-1, ymax=16,
		axis lines=middle,
		width=0.75\textwidth,
		height=0.5\textwidth,
		xlabel style={at={(current axis.south)}, yshift=-3mm, xshift=5.4cm},
		ylabel style={at={(current axis.west)}, yshift=3.6cm,xshift=-13mm},
		%				axis x line=middle, % Add this line
		axis x line shift=1, % Add this line
		legend style={
			at={(1.25,0.5)}, % position relative to axis
			anchor=east          % anchor legend at its right side
		}
		]
		\addplot[] coordinates {
			(0, 1.6482035551275955)
			(0.004975124378109453, 1.65616464441183)
			(0.009950248756218905, 1.6640556436460785)
			(0.014925373134328358, 1.6719164793629757)
			(0.01990049751243781, 1.679796179411962)
			(0.024875621890547265, 1.687672742386679)
			(0.029850746268656716, 1.6955842908125223)
			(0.03482587064676617, 1.703503751858132)
			(0.03980099502487562, 1.711466680925085)
			(0.04477611940298507, 1.719444108624817)
			(0.04975124378109453, 1.727470695427991)
			(0.05472636815920398, 1.7355164275410644)
			(0.05970149253731343, 1.74361566428693)
			(0.06467661691542288, 1.7517376693574305)
			(0.06965174129353234, 1.7599167777778064)
			(0.07462686567164178, 1.7681216746483968)
			(0.07960199004975124, 1.7763868240460785)
			(0.0845771144278607, 1.784680402504114)
			(0.08955223880597014, 1.7930370993700624)
			(0.0945273631840796, 1.80142461737973)
			(0.09950248756218906, 1.8098779376957375)
			(0.1044776119402985, 1.8183643054451823)
			(0.10945273631840796, 1.826919043411986)
			(0.11442786069651742, 1.8355089447276944)
			(0.11940298507462686, 1.8441697135822415)
			(0.12437810945273632, 1.8528676903117092)
			(0.12935323383084577, 1.8616389940378033)
			(0.13432835820895522, 1.8704495073894543)
			(0.13930348258706468, 1.879335793929335)
			(0.14427860696517414, 1.8882632708904061)
			(0.14925373134328357, 1.897268973177072)
			(0.15422885572139303, 1.9063178429637175)
			(0.15920398009950248, 1.9154474117144282)
			(0.16417910447761194, 1.9246221354058743)
			(0.1691542288557214, 1.9338800662311195)
			(0.17412935323383086, 1.9431851616671967)
			(0.1791044776119403, 1.9525760182084013)
			(0.18407960199004975, 1.9620160815673133)
			(0.1890547263681592, 1.9715445158485307)
			(0.19402985074626866, 1.9811242408994718)
			(0.19900497512437812, 1.990795011736981)
			(0.20398009950248755, 2.0005192074857856)
			(0.208955223880597, 2.0103371975735023)
			(0.21393034825870647, 2.0202108048344933)
			(0.21890547263681592, 2.030181036970368)
			(0.22388059701492538, 2.0402091442715244)
			(0.22885572139303484, 2.0503367970852824)
			(0.23383084577114427, 2.0605246562267996)
			(0.23880597014925373, 2.0708150796966787)
			(0.24378109452736318, 2.081168121222422)
			(0.24875621890547264, 2.091626852217856)
			(0.2537313432835821, 2.102150701017019)
			(0.25870646766169153, 2.1127834790690554)
			(0.263681592039801, 2.123483970296184)
			(0.26865671641791045, 2.1342967537733144)
			(0.2736318407960199, 2.1451799492552763)
			(0.27860696517412936, 2.1561789321065983)
			(0.2835820895522388, 2.167251137392599)
			(0.2885572139303483, 2.1784427666106083)
			(0.2935323383084577, 2.189710548814414)
			(0.29850746268656714, 2.201101542765148)
			(0.3034825870646766, 2.2125717493461483)
			(0.30845771144278605, 2.224169117113721)
			(0.31343283582089554, 2.2358488957445966)
			(0.31840796019900497, 2.247659957640017)
			(0.32338308457711445, 2.259556777313121)
			(0.3283582089552239, 2.2715891867030713)
			(0.3333333333333333, 2.283735365511175)
			(0.3383084577114428, 2.295972626854814)
			(0.34328358208955223, 2.308352313349196)
			(0.3482587064676617, 2.320826849885114)
			(0.35323383084577115, 2.333448635954346)
			(0.3582089552238806, 2.346169229466332)
			(0.36318407960199006, 2.3590421293477526)
			(0.3681592039800995, 2.3720179978020286)
			(0.373134328358209, 2.3851514780751972)
			(0.3781094527363184, 2.3983923067231316)
			(0.38308457711442784, 2.411796318709605)
			(0.3880597014925373, 2.4253122937201805)
			(0.39303482587064675, 2.4389973085534504)
			(0.39800995024875624, 2.452799153452875)
			(0.40298507462686567, 2.4667762001116307)
			(0.4079601990049751, 2.480875215339001)
			(0.4129353233830846, 2.4951559219707815)
			(0.417910447761194, 2.5095640278949944)
			(0.4228855721393035, 2.524160666796341)
			(0.42786069651741293, 2.5388904505812557)
			(0.43283582089552236, 2.553815987245433)
			(0.43781094527363185, 2.5688807539985046)
			(0.4427860696517413, 2.5841489006937075)
			(0.44776119402985076, 2.599562729388832)
			(0.4527363184079602, 2.615188003789604)
			(0.4577114427860697, 2.6309658085190786)
			(0.4626865671641791, 2.646963597982762)
			(0.46766169154228854, 2.6631211951674456)
			(0.472636815920398, 2.6795078273274724)
			(0.47761194029850745, 2.6960620096001704)
			(0.48258706467661694, 2.712854830040796)
			(0.48756218905472637, 2.729823447617726)
			(0.4925373134328358, 2.747040905502681)
			(0.4975124378109453, 2.764442955974051)
			(0.5024875621890548, 2.782104698627303)
			(0.5074626865671642, 2.7999604262336537)
			(0.5124378109452736, 2.8180874038174593)
			(0.5174129353233831, 2.836418409437115)
			(0.5223880597014925, 2.8550329910446997)
			(0.527363184079602, 2.8738623543039563)
			(0.5323383084577115, 2.8929884569866533)
			(0.5373134328358209, 2.912340872123609)
			(0.5422885572139303, 2.9320041046109835)
			(0.5472636815920398, 2.9519060319829795)
			(0.5522388059701493, 2.972133855136883)
			(0.5572139303482587, 2.9926136905684615)
			(0.5621890547263682, 3.013435596947267)
			(0.5671641791044776, 3.0345238614806322)
			(0.572139303482587, 3.0559715767893696)
			(0.5771144278606966, 3.0777011298351353)
			(0.582089552238806, 3.0998088395150103)
			(0.5870646766169154, 3.122215118923207)
			(0.5920398009950248, 3.1450197237076356)
			(0.5970149253731343, 3.168141016907226)
			(0.6019900497512438, 3.1916824218631437)
			(0.6069651741293532, 3.2155601729774292)
			(0.6119402985074627, 3.2398816153881067)
			(0.6169154228855721, 3.264560774154664)
			(0.6218905472636815, 3.289709196619006)
			(0.6268656716417911, 3.3152386160660927)
			(0.6318407960199005, 3.3412650924344476)
			(0.6368159203980099, 3.3676979836420795)
			(0.6417910447761194, 3.394658206938007)
			(0.6467661691542289, 3.4220526609068584)
			(0.6517412935323383, 3.450007504272814)
			(0.6567164179104478, 3.4784270930227543)
			(0.6616915422885572, 3.507443257072593)
			(0.6666666666666666, 3.537017687181317)
			(0.6716417910447762, 3.5671084915960036)
			(0.6766169154228856, 3.5978569340619044)
			(0.681592039800995, 3.629160667545599)
			(0.6865671641791045, 3.6611679524915126)
			(0.6915422885572139, 3.6937736393558)
			(0.6965174129353234, 3.7271336425797847)
			(0.7014925373134329, 3.7611399569026904)
			(0.7064676616915423, 3.795956900793788)
			(0.7114427860696517, 3.8314735778932287)
			(0.7164179104477612, 3.8678635711075904)
			(0.7213930348258707, 3.9050130826545053)
			(0.7263681592039801, 3.9431059525634646)
			(0.7313432835820896, 3.9820255060698244)
			(0.736318407960199, 4.0219669926886805)
			(0.7412935323383084, 4.062810932947765)
			(0.746268656716418, 4.104765332478103)
			(0.7512437810945274, 4.147708044900827)
			(0.7562189054726368, 4.1918614169433175)
			(0.7611940298507462, 4.237100862740219)
			(0.7661691542288557, 4.283664950200481)
			(0.7711442786069652, 4.331427004055585)
			(0.7761194029850746, 4.380644062463952)
			(0.7810945273631841, 4.431187879196796)
			(0.7860696517412935, 4.483336678031136)
			(0.7910447761194029, 4.536961392371045)
			(0.7960199004975125, 4.59236474314533)
			(0.8009950248756219, 4.6494179161394715)
			(0.8059701492537313, 4.708452212875661)
			(0.8109452736318408, 4.769340594956169)
			(0.8159203980099502, 4.832448041366214)
			(0.8208955223880597, 4.897651449703718)
			(0.8258706467661692, 4.96535592427255)
			(0.8308457711442786, 5.035445369064366)
			(0.835820895522388, 5.108373293182549)
			(0.8407960199004975, 5.184034996586361)
			(0.845771144278607, 5.262943202899134)
			(0.8507462686567164, 5.345010940383215)
			(0.8557213930348259, 5.430824526167687)
			(0.8606965174129353, 5.520323964024207)
			(0.8656716417910447, 5.614188698453729)
			(0.8706467661691543, 5.71239942491851)
			(0.8756218905472637, 5.815755895207952)
			(0.8805970149253731, 5.924300239073558)
			(0.8855721393034826, 6.038991387356083)
			(0.8905472636815921, 6.159965024785064)
			(0.8955223880597015, 6.2883966508627)
			(0.900497512437811, 6.424566703563415)
			(0.9054726368159204, 6.569955197851533)
			(0.9104477611940298, 6.725071866348183)
			(0.9154228855721394, 6.891842119199647)
			(0.9203980099502488, 7.071150799623545)
			(0.9253731343283582, 7.26560680697733)
			(0.9303482587064676, 7.476735680641075)
			(0.9353233830845771, 7.708259750621973)
			(0.9402985074626866, 7.962863824636424)
			(0.945273631840796, 8.246228189022677)
			(0.9502487562189055, 8.563306319309117)
			(0.9552238805970149, 8.923589921901106)
			(0.9601990049751243, 9.336990392552693)
			(0.9651741293532339, 9.821555301702325)
			(0.9701492537313433, 10.399969995157921)
			(0.9751243781094527, 11.114133521813642)
			(0.9800995024875622, 12.029880571380126)
			(0.9850746268656716, 13.286188007649281)
			(0.9900497512437811, 15.199527601302604)
			(0.9950248756218906, 18.90041806745566)
		};
		\addlegendentry{$M_j(p)$}
		\addplot[dashdotted] coordinates {
			(0, 1.6482035551275955)
			(0.004975124378109453, 1.6563126848059104)
			(0.009950248756218905, 1.6639848652810354)
			(0.014925373134328358, 1.6716001242472436)
			(0.01990049751243781, 1.6791889954487957)
			(0.024875621890547265, 1.686742631756359)
			(0.029850746268656716, 1.6943204272714432)
			(0.03482587064676617, 1.7019131930219589)
			(0.03980099502487562, 1.709509067881219)
			(0.04477611940298507, 1.71716329763316)
			(0.04975124378109453, 1.7248228772333651)
			(0.05472636815920398, 1.7325104355003327)
			(0.05970149253731343, 1.7402396697215754)
			(0.06467661691542288, 1.7479611415450256)
			(0.06965174129353234, 1.7557420093980483)
			(0.07462686567164178, 1.7635134396010135)
			(0.07960199004975124, 1.7713402165868743)
			(0.0845771144278607, 1.7792125691445058)
			(0.08955223880597014, 1.7871227242196301)
			(0.0945273631840796, 1.7950097000535858)
			(0.09950248756218906, 1.803042668440956)
			(0.1044776119402985, 1.8110158858743424)
			(0.10945273631840796, 1.8190743483633125)
			(0.11442786069651742, 1.8272014209435308)
			(0.11940298507462686, 1.8353851652610744)
			(0.12437810945273632, 1.8435974138303841)
			(0.12935323383084577, 1.851834341496373)
			(0.13432835820895522, 1.860090275905252)
			(0.13930348258706468, 1.868479794639543)
			(0.14427860696517414, 1.8768910144331679)
			(0.14925373134328357, 1.8852958191652698)
			(0.15422885572139303, 1.8937514050959603)
			(0.15920398009950248, 1.902277772993379)
			(0.16417910447761194, 1.9108905807149121)
			(0.1691542288557214, 1.919522771432712)
			(0.17412935323383086, 1.9281517747250798)
			(0.1791044776119403, 1.9368874830690215)
			(0.18407960199004975, 1.9455746868473358)
			(0.1890547263681592, 1.9544345701107695)
			(0.19402985074626866, 1.9632543220527123)
			(0.19900497512437812, 1.9722446045560325)
			(0.20398009950248755, 1.9811378908679538)
			(0.208955223880597, 1.990171765461749)
			(0.21393034825870647, 1.9992847057903211)
			(0.21890547263681592, 2.0085283873764377)
			(0.22388059701492538, 2.0176561743438013)
			(0.22885572139303484, 2.0269918699045886)
			(0.23383084577114427, 2.0363971437095425)
			(0.23880597014925373, 2.0458154112596207)
			(0.24378109452736318, 2.0554759135740324)
			(0.24875621890547264, 2.065015366533497)
			(0.2537313432835821, 2.0746626023257075)
			(0.25870646766169153, 2.084361606854696)
			(0.263681592039801, 2.0941975432947344)
			(0.26865671641791045, 2.104133389333585)
			(0.2736318407960199, 2.114106414887896)
			(0.27860696517412936, 2.124260458270355)
			(0.2835820895522388, 2.1344358982554628)
			(0.2885572139303483, 2.144704147367433)
			(0.2935323383084577, 2.155127855382398)
			(0.29850746268656714, 2.165660508232908)
			(0.3034825870646766, 2.176124086548324)
			(0.30845771144278605, 2.1866047808004563)
			(0.31343283582089554, 2.197378592112454)
			(0.31840796019900497, 2.207861384435343)
			(0.32338308457711445, 2.2185857385252166)
			(0.3283582089552239, 2.2295817345541153)
			(0.3333333333333333, 2.24062526934185)
			(0.3383084577114428, 2.2516046405100862)
			(0.34328358208955223, 2.2629670096923618)
			(0.3482587064676617, 2.2743208306977847)
			(0.35323383084577115, 2.2858270320787035)
			(0.3582089552238806, 2.2975816581753294)
			(0.36318407960199006, 2.309228149762525)
			(0.3681592039800995, 2.3207997569155085)
			(0.373134328358209, 2.332477896155605)
			(0.3781094527363184, 2.3443460133659784)
			(0.38308457711442784, 2.3564969300107776)
			(0.3880597014925373, 2.3686822239983547)
			(0.39303482587064675, 2.381248840121692)
			(0.39800995024875624, 2.393612903559977)
			(0.40298507462686567, 2.406017608945554)
			(0.4079601990049751, 2.4188673386364123)
			(0.4129353233830846, 2.431479146050859)
			(0.417910447761194, 2.4442715368036665)
			(0.4228855721393035, 2.4573333441189193)
			(0.42786069651741293, 2.47045432284502)
			(0.43283582089552236, 2.4835530337482403)
			(0.43781094527363185, 2.496981163835929)
			(0.4427860696517413, 2.5106646655480684)
			(0.44776119402985076, 2.5246947775114945)
			(0.4527363184079602, 2.5386522832476004)
			(0.4577114427860697, 2.5526315992623383)
			(0.4626865671641791, 2.5668230410191066)
			(0.46766169154228854, 2.5808818774276276)
			(0.472636815920398, 2.595622252774166)
			(0.47761194029850745, 2.610197989763689)
			(0.48258706467661694, 2.624881600648767)
			(0.48756218905472637, 2.639867114567536)
			(0.4925373134328358, 2.65525220626262)
			(0.4975124378109453, 2.670402459238764)
			(0.5024875621890548, 2.6859255308783276)
			(0.5074626865671642, 2.702069526361567)
			(0.5124378109452736, 2.718012268566301)
			(0.5174129353233831, 2.734150475756375)
			(0.5223880597014925, 2.7505641512738492)
			(0.527363184079602, 2.7671843646664427)
			(0.5323383084577115, 2.7834296182612843)
			(0.5373134328358209, 2.8000072796699187)
			(0.5422885572139303, 2.8175785681786634)
			(0.5472636815920398, 2.834958222290345)
			(0.5522388059701493, 2.8525809045372275)
			(0.5572139303482587, 2.87000859942843)
			(0.5621890547263682, 2.8880517588346444)
			(0.5671641791044776, 2.9065449680092446)
			(0.572139303482587, 2.9245608051433853)
			(0.5771144278606966, 2.943498286681858)
			(0.582089552238806, 2.962586578792332)
			(0.5870646766169154, 2.981530400050134)
			(0.5920398009950248, 3.0010779608492015)
			(0.5970149253731343, 3.0213736377697766)
			(0.6019900497512438, 3.041860317399742)
			(0.6069651741293532, 3.062563158815998)
			(0.6119402985074627, 3.0836970146586915)
			(0.6169154228855721, 3.105248880631012)
			(0.6218905472636815, 3.1272797903694527)
			(0.6268656716417911, 3.149055035797179)
			(0.6318407960199005, 3.1710296636208968)
			(0.6368159203980099, 3.1941137076405153)
			(0.6417910447761194, 3.2173991824846855)
			(0.6467661691542289, 3.2404546501435556)
			(0.6517412935323383, 3.2642647047169047)
			(0.6567164179104478, 3.2895143260594053)
			(0.6616915422885572, 3.314005056276192)
			(0.6666666666666666, 3.3391335688772807)
			(0.6716417910447762, 3.3648771970118867)
			(0.6766169154228856, 3.3912727921978543)
			(0.681592039800995, 3.417756171216258)
			(0.6865671641791045, 3.445245666174406)
			(0.6915422885572139, 3.473081584599152)
			(0.6965174129353234, 3.502067772066079)
			(0.7014925373134329, 3.530835083808501)
			(0.7064676616915423, 3.5598328441818023)
			(0.7114427860696517, 3.5889907886152055)
			(0.7164179104477612, 3.61969026989432)
			(0.7213930348258707, 3.6510716137153256)
			(0.7263681592039801, 3.6828009480045916)
			(0.7313432835820896, 3.7150699587113807)
			(0.736318407960199, 3.7484939759875373)
			(0.7412935323383084, 3.783133787601001)
			(0.746268656716418, 3.817960420518385)
			(0.7512437810945274, 3.8535862380206836)
			(0.7562189054726368, 3.8906269183140094)
			(0.7611940298507462, 3.9280216204472356)
			(0.7661691542288557, 3.9676309382115833)
			(0.7711442786069652, 4.006676912356246)
			(0.7761194029850746, 4.048426270216005)
			(0.7810945273631841, 4.0895476233735195)
			(0.7860696517412935, 4.131840560789625)
			(0.7910447761194029, 4.175576387543237)
			(0.7960199004975125, 4.218282924290762)
			(0.8009950248756219, 4.266820498373536)
			(0.8059701492537313, 4.31373412472119)
			(0.8109452736318408, 4.361866725428203)
			(0.8159203980099502, 4.411695103237047)
			(0.8208955223880597, 4.466149034305494)
			(0.8258706467661692, 4.5235713162117435)
			(0.8308457711442786, 4.579827334412641)
			(0.835820895522388, 4.641855241746181)
			(0.8407960199004975, 4.70259840231471)
			(0.845771144278607, 4.766174907985859)
			(0.8507462686567164, 4.833285508287397)
			(0.8557213930348259, 4.9029702709284475)
			(0.8606965174129353, 4.975359651120665)
			(0.8656716417910447, 5.0512507183736)
			(0.8706467661691543, 5.131014347078688)
			(0.8756218905472637, 5.212338277788525)
			(0.8805970149253731, 5.29756195644101)
			(0.8855721393034826, 5.3924103313769916)
			(0.8905472636815921, 5.489614391606363)
			(0.8955223880597015, 5.592223036339595)
			(0.900497512437811, 5.7010759945742775)
			(0.9054726368159204, 5.819107143871815)
			(0.9104477611940298, 5.934139359292336)
			(0.9154228855721394, 6.074622490860403)
			(0.9203980099502488, 6.215631003164875)
			(0.9253731343283582, 6.35608967866232)
			(0.9303482587064676, 6.532999588276467)
			(0.9353233830845771, 6.707737476994331)
			(0.9402985074626866, 6.908718691599656)
			(0.945273631840796, 7.11590406898024)
			(0.9502487562189055, 7.365458110168125)
			(0.9552238805970149, 7.641921208328522)
			(0.9601990049751243, 7.9552276517129705)
			(0.9651741293532339, 8.31776483902309)
			(0.9701492537313433, 8.742371387255522)
			(0.9751243781094527, 9.248461009422648)
			(0.9800995024875622, 9.97235955406369)
			(0.9850746268656716, 10.91412890337495)
			(0.9900497512437811, 12.279427084879895)
			(0.9950248756218906, 15.050621135290607)
		};
		\addlegendentry{$M^f_j(p)$}
		\addplot[dotted] coordinates {
			(0, 1.6482035551275955)
			(0.004975124378109453, 1.655616800475805)
			(0.009950248756218905, 1.6625134181577617)
			(0.014925373134328358, 1.6692986272544528)
			(0.01990049751243781, 1.676007978870056)
			(0.024875621890547265, 1.6826531825439492)
			(0.029850746268656716, 1.6893276579906384)
			(0.03482587064676617, 1.6959288982934972)
			(0.03980099502487562, 1.7025369902998284)
			(0.04477611940298507, 1.7090667841669653)
			(0.04975124378109453, 1.715691739307781)
			(0.05472636815920398, 1.7222503057434624)
			(0.05970149253731343, 1.7287635375571078)
			(0.06467661691542288, 1.7352288422933655)
			(0.06965174129353234, 1.7418545928164186)
			(0.07462686567164178, 1.748455835679024)
			(0.07960199004975124, 1.7550276225346486)
			(0.0845771144278607, 1.761625867510884)
			(0.08955223880597014, 1.7682190696033913)
			(0.0945273631840796, 1.774853595557883)
			(0.09950248756218906, 1.7814855671412342)
			(0.1044776119402985, 1.7880952918960353)
			(0.10945273631840796, 1.7947576656185285)
			(0.11442786069651742, 1.801455036919762)
			(0.11940298507462686, 1.8082922249680131)
			(0.12437810945273632, 1.815087200255847)
			(0.12935323383084577, 1.8218522318542267)
			(0.13432835820895522, 1.8287397341555929)
			(0.13930348258706468, 1.8356982400275097)
			(0.14427860696517414, 1.8426655817978677)
			(0.14925373134328357, 1.8496799678709863)
			(0.15422885572139303, 1.8566383452557398)
			(0.15920398009950248, 1.8634762811865162)
			(0.16417910447761194, 1.8705200719953488)
			(0.1691542288557214, 1.8773886392306571)
			(0.17412935323383086, 1.8845069124876128)
			(0.1791044776119403, 1.8915610201493525)
			(0.18407960199004975, 1.898702796761128)
			(0.1890547263681592, 1.9058766264973757)
			(0.19402985074626866, 1.9131079087358278)
			(0.19900497512437812, 1.9203583620737554)
			(0.20398009950248755, 1.927593524057098)
			(0.208955223880597, 1.9348402704000265)
			(0.21393034825870647, 1.9422440191670935)
			(0.21890547263681592, 1.9496665882840598)
			(0.22388059701492538, 1.9571466876117567)
			(0.22885572139303484, 1.9647060848274305)
			(0.23383084577114427, 1.972309405288504)
			(0.23880597014925373, 1.979723626365568)
			(0.24378109452736318, 1.987472016950495)
			(0.24875621890547264, 1.9953702445879413)
			(0.2537313432835821, 2.0028207365577617)
			(0.25870646766169153, 2.010837119880017)
			(0.263681592039801, 2.018426217484265)
			(0.26865671641791045, 2.026184535027385)
			(0.2736318407960199, 2.0342068829661635)
			(0.27860696517412936, 2.042224855229395)
			(0.2835820895522388, 2.0504725987995283)
			(0.2885572139303483, 2.058507089220637)
			(0.2935323383084577, 2.0664674520137116)
			(0.29850746268656714, 2.0744747655889615)
			(0.3034825870646766, 2.082847699141516)
			(0.30845771144278605, 2.0910503311804707)
			(0.31343283582089554, 2.0994386128474796)
			(0.31840796019900497, 2.1078216813319743)
			(0.32338308457711445, 2.1162749754783308)
			(0.3283582089552239, 2.1248354192135728)
			(0.3333333333333333, 2.1334482152662817)
			(0.3383084577114428, 2.1421629486346174)
			(0.34328358208955223, 2.150876403029792)
			(0.3482587064676617, 2.1599313978050807)
			(0.35323383084577115, 2.169008024653913)
			(0.3582089552238806, 2.177848072419131)
			(0.36318407960199006, 2.1869674481924855)
			(0.3681592039800995, 2.196098833774304)
			(0.373134328358209, 2.2056858884470527)
			(0.3781094527363184, 2.2148564811811844)
			(0.38308457711442784, 2.224328975447704)
			(0.3880597014925373, 2.2340154873965146)
			(0.39303482587064675, 2.2436003056099563)
			(0.39800995024875624, 2.252908506154879)
			(0.40298507462686567, 2.2622843331123406)
			(0.4079601990049751, 2.2721369382561982)
			(0.4129353233830846, 2.2820358151226654)
			(0.417910447761194, 2.291821405935899)
			(0.4228855721393035, 2.301653884372472)
			(0.42786069651741293, 2.3118108194416793)
			(0.43283582089552236, 2.322182634018187)
			(0.43781094527363185, 2.332477706666981)
			(0.4427860696517413, 2.343085413272412)
			(0.44776119402985076, 2.35338634875335)
			(0.4527363184079602, 2.364209759635522)
			(0.4577114427860697, 2.374650458957806)
			(0.4626865671641791, 2.385394714409742)
			(0.46766169154228854, 2.396320504140443)
			(0.472636815920398, 2.407629127572389)
			(0.47761194029850745, 2.418509360884972)
			(0.48258706467661694, 2.4295601019651447)
			(0.48756218905472637, 2.4409596842283143)
			(0.4925373134328358, 2.45275780295144)
			(0.4975124378109453, 2.4645967016506205)
			(0.5024875621890548, 2.476149575777299)
			(0.5074626865671642, 2.4883860182930095)
			(0.5124378109452736, 2.500626236671753)
			(0.5174129353233831, 2.5129588602388617)
			(0.5223880597014925, 2.5251461568834093)
			(0.527363184079602, 2.537843897089055)
			(0.5323383084577115, 2.5508343695673816)
			(0.5373134328358209, 2.563774121519215)
			(0.5422885572139303, 2.576576008826038)
			(0.5472636815920398, 2.5897529376808213)
			(0.5522388059701493, 2.6037635937091994)
			(0.5572139303482587, 2.6180452026681884)
			(0.5621890547263682, 2.6316143582820515)
			(0.5671641791044776, 2.645807751984791)
			(0.572139303482587, 2.659253827283682)
			(0.5771144278606966, 2.673763926983124)
			(0.582089552238806, 2.6880426700463045)
			(0.5870646766169154, 2.7036861218818546)
			(0.5920398009950248, 2.717481622727415)
			(0.5970149253731343, 2.732910918077307)
			(0.6019900497512438, 2.7473232096163067)
			(0.6069651741293532, 2.761937538399819)
			(0.6119402985074627, 2.778626599689773)
			(0.6169154228855721, 2.794207948236281)
			(0.6218905472636815, 2.810757560723972)
			(0.6268656716417911, 2.8273412830017266)
			(0.6318407960199005, 2.8441836010441293)
			(0.6368159203980099, 2.8606435181096406)
			(0.6417910447761194, 2.8779948575832734)
			(0.6467661691542289, 2.8964247861008556)
			(0.6517412935323383, 2.9147710648849796)
			(0.6567164179104478, 2.9325901508503014)
			(0.6616915422885572, 2.9491659398180614)
			(0.6666666666666666, 2.967653022100594)
			(0.6716417910447762, 2.986816989093643)
			(0.6766169154228856, 3.006849263344343)
			(0.681592039800995, 3.026318359448995)
			(0.6865671641791045, 3.0468771238275636)
			(0.6915422885572139, 3.068455007253499)
			(0.6965174129353234, 3.0902688670115652)
			(0.7014925373134329, 3.1113037872459244)
			(0.7064676616915423, 3.1348090184172532)
			(0.7114427860696517, 3.1571583135092958)
			(0.7164179104477612, 3.1784015411884576)
			(0.7213930348258707, 3.200989296607074)
			(0.7263681592039801, 3.2238449676387706)
			(0.7313432835820896, 3.2458451441261977)
			(0.736318407960199, 3.271095804626054)
			(0.7412935323383084, 3.2966564681316175)
			(0.746268656716418, 3.324255812972197)
			(0.7512437810945274, 3.349156550457468)
			(0.7562189054726368, 3.3775351152191213)
			(0.7611940298507462, 3.404965496768451)
			(0.7661691542288557, 3.4348035165927056)
			(0.7711442786069652, 3.4629089624962885)
			(0.7761194029850746, 3.4922883367899953)
			(0.7810945273631841, 3.5257895816924454)
			(0.7860696517412935, 3.5565103177790065)
			(0.7910447761194029, 3.5875008611051284)
			(0.7960199004975125, 3.621365525758431)
			(0.8009950248756219, 3.656616319574654)
			(0.8059701492537313, 3.694367837038037)
			(0.8109452736318408, 3.7285669211927663)
			(0.8159203980099502, 3.7664341574862377)
			(0.8208955223880597, 3.8047907640447582)
			(0.8258706467661692, 3.8471566884232242)
			(0.8308457711442786, 3.8883628897286373)
			(0.835820895522388, 3.930199157218499)
			(0.8407960199004975, 3.977547082685738)
			(0.845771144278607, 4.022536384944323)
			(0.8507462686567164, 4.070291416432101)
			(0.8557213930348259, 4.122754366228391)
			(0.8606965174129353, 4.176688380266617)
			(0.8656716417910447, 4.235871628672414)
			(0.8706467661691543, 4.290130930727484)
			(0.8756218905472637, 4.355062646262541)
			(0.8805970149253731, 4.422554408298671)
			(0.8855721393034826, 4.494725272811929)
			(0.8905472636815921, 4.561686436736479)
			(0.8955223880597015, 4.641133823304851)
			(0.900497512437811, 4.7198278470989)
			(0.9054726368159204, 4.808243563056527)
			(0.9104477611940298, 4.891824225308885)
			(0.9154228855721394, 4.980746452929139)
			(0.9203980099502488, 5.075947780635086)
			(0.9253731343283582, 5.190596312875258)
			(0.9303482587064676, 5.316667296725895)
			(0.9353233830845771, 5.45883105524221)
			(0.9402985074626866, 5.60425680837996)
			(0.945273631840796, 5.760369782415794)
			(0.9502487562189055, 5.937953312034807)
			(0.9552238805970149, 6.133874473021606)
			(0.9601990049751243, 6.3945541583390675)
			(0.9651741293532339, 6.649330389873672)
			(0.9701492537313433, 6.9962410672381194)
			(0.9751243781094527, 7.332568066193079)
			(0.9800995024875622, 7.8027262902771595)
			(0.9850746268656716, 8.527436702088627)
			(0.9900497512437811, 9.768669663387989)
			(0.9950248756218906, 11.996131168996198)
		};
		\addlegendentry{$m^f_j(p)$}
		\addplot[dash pattern=on 8pt off 2pt] coordinates {
			(0, 1.6482035551275955)
			(0.004975124378109453, 1.5618575236192904)
			(0.009950248756218905, 1.5118745305428283)
			(0.014925373134328358, 1.4718149656099486)
			(0.01990049751243781, 1.4373053831295262)
			(0.024875621890547265, 1.4066821653795147)
			(0.029850746268656716, 1.3788323861217766)
			(0.03482587064676617, 1.3532525710178558)
			(0.03980099502487562, 1.3294222680928494)
			(0.04477611940298507, 1.307138877352719)
			(0.04975124378109453, 1.2860892329082907)
			(0.05472636815920398, 1.2661858372130552)
			(0.05970149253731343, 1.2472119837056173)
			(0.06467661691542288, 1.2291336809406048)
			(0.06965174129353234, 1.2117872472581406)
			(0.07462686567164178, 1.1951666026809573)
			(0.07960199004975124, 1.1791406251366978)
			(0.0845771144278607, 1.1637188926238116)
			(0.08955223880597014, 1.1487919234408115)
			(0.0945273631840796, 1.1343784888231419)
			(0.09950248756218906, 1.120384371237882)
			(0.1044776119402985, 1.1068338764577685)
			(0.10945273631840796, 1.0936440626076567)
			(0.11442786069651742, 1.080842588019512)
			(0.11940298507462686, 1.0683551523181303)
			(0.12437810945273632, 1.0562114123496846)
			(0.12935323383084577, 1.044343891298513)
			(0.13432835820895522, 1.0327833784979337)
			(0.13930348258706468, 1.021467919575438)
			(0.14427860696517414, 1.010428869831396)
			(0.14925373134328357, 0.9996088372076041)
			(0.15422885572139303, 0.989039366428772)
			(0.15920398009950248, 0.9786669005590948)
			(0.16417910447761194, 0.9685229210271169)
			(0.1691542288557214, 0.9585571427671605)
			(0.17412935323383086, 0.9488008137285734)
			(0.1791044776119403, 0.9392064758629814)
			(0.18407960199004975, 0.9298050295039992)
			(0.1890547263681592, 0.9205514862322518)
			(0.19402985074626866, 0.9114763232667856)
			(0.19900497512437812, 0.9025367303420512)
			(0.20398009950248755, 0.8937627130233052)
			(0.208955223880597, 0.885113398282947)
			(0.21393034825870647, 0.8766182904784496)
			(0.21890547263681592, 0.8682382523046321)
			(0.22388059701492538, 0.8600022708127171)
			(0.22885572139303484, 0.8518727740480779)
			(0.23383084577114427, 0.8438782252270611)
			(0.23880597014925373, 0.8359824723086274)
			(0.24378109452736318, 0.8282134549537883)
			(0.24875621890547264, 0.8205363157953065)
			(0.2537313432835821, 0.8129784760447646)
			(0.25870646766169153, 0.8055062643023373)
			(0.263681592039801, 0.7981465918288849)
			(0.26865671641791045, 0.7908668781508965)
			(0.2736318407960199, 0.7836935354263738)
			(0.27860696517412936, 0.7765949904617805)
			(0.2835820895522388, 0.7695971687458548)
			(0.2885572139303483, 0.7626694302973206)
			(0.2935323383084577, 0.7558372273947871)
			(0.29850746268656714, 0.7490707873135379)
			(0.3034825870646766, 0.7423951031951453)
			(0.30845771144278605, 0.735781210532877)
			(0.31343283582089554, 0.7292536577161558)
			(0.31840796019900497, 0.7227842353532205)
			(0.32338308457711445, 0.7163970615569407)
			(0.3283582089552239, 0.7100646340705323)
			(0.3333333333333333, 0.7038106551358952)
			(0.3383084577114428, 0.6976082860967627)
			(0.34328358208955223, 0.6914685604967405)
			(0.3482587064676617, 0.6854020647643931)
			(0.35323383084577115, 0.6793828826594983)
			(0.3582089552238806, 0.6734337381703243)
			(0.36318407960199006, 0.6675292857345659)
			(0.3681592039800995, 0.6616918819589289)
			(0.373134328358209, 0.6558967200470504)
			(0.3781094527363184, 0.6501658023026312)
			(0.38308457711442784, 0.6444748312666121)
			(0.3880597014925373, 0.6388454676276877)
			(0.39303482587064675, 0.6332538960401792)
			(0.39800995024875624, 0.6277214478676492)
			(0.40298507462686567, 0.6222247646172637)
			(0.4079601990049751, 0.616784860218615)
			(0.4129353233830846, 0.6113788095244895)
			(0.417910447761194, 0.6060273205275438)
			(0.4228855721393035, 0.6007078794885126)
			(0.42786069651741293, 0.5954408995743983)
			(0.43283582089552236, 0.5902042579242525)
			(0.43781094527363185, 0.5850180836160602)
			(0.4427860696517413, 0.579860625402966)
			(0.44776119402985076, 0.5747517386489662)
			(0.4527363184079602, 0.5696700255959172)
			(0.4577114427860697, 0.5646350779215967)
			(0.4626865671641791, 0.5596258342571764)
			(0.46766169154228854, 0.5546616322899227)
			(0.472636815920398, 0.5497217308657731)
			(0.47761194029850745, 0.5448252230608104)
			(0.48258706467661694, 0.5399516725949215)
			(0.48756218905472637, 0.5351199370118557)
			(0.4925373134328358, 0.5303098703158322)
			(0.4975124378109453, 0.5255401033125556)
			(0.5024875621890548, 0.5207907663769474)
			(0.5074626865671642, 0.5160802721021751)
			(0.5124378109452736, 0.511389013927246)
			(0.5174129353233831, 0.50673519450504)
			(0.5223880597014925, 0.5020994575753499)
			(0.527363184079602, 0.49749980388494464)
			(0.5323383084577115, 0.49291711519513715)
			(0.5373134328358209, 0.48836919815746616)
			(0.5422885572139303, 0.4838371607039054)
			(0.5472636815920398, 0.47933862299266067)
			(0.5522388059701493, 0.47485490765121335)
			(0.5572139303482587, 0.47040345575115766)
			(0.5621890547263682, 0.4659657934656048)
			(0.5671641791044776, 0.4615591900043342)
			(0.572139303482587, 0.45716536421267745)
			(0.5771144278606966, 0.4528014204941222)
			(0.582089552238806, 0.4484492597214679)
			(0.5870646766169154, 0.4441258283901506)
			(0.5920398009950248, 0.43981319893688964)
			(0.5970149253731343, 0.4355281667012867)
			(0.6019900497512438, 0.4312529653538845)
			(0.6069651741293532, 0.42700424569507794)
			(0.6119402985074627, 0.4227643923838289)
			(0.6169154228855721, 0.4185499181718482)
			(0.6218905472636815, 0.41434334849527293)
			(0.6268656716417911, 0.4101610644299134)
			(0.6318407960199005, 0.40598572195882804)
			(0.6368159203980099, 0.4018335767439997)
			(0.6417910447761194, 0.3976874050093482)
			(0.6467661691542289, 0.39356334315978764)
			(0.6517412935323383, 0.38944427721665326)
			(0.6567164179104478, 0.3853462303826066)
			(0.6616915422885572, 0.38125218782783193)
			(0.6666666666666666, 0.3771780655064365)
			(0.6716417910447762, 0.3731069368082138)
			(0.6766169154228856, 0.3690464730719036)
			(0.681592039800995, 0.3650042542625439)
			(0.6865671641791045, 0.3609634648126477)
			(0.6915422885572139, 0.3569397778602515)
			(0.6965174129353234, 0.3529164369085817)
			(0.7014925373134329, 0.34890902781572597)
			(0.7064676616915423, 0.3449008404819835)
			(0.7114427860696517, 0.34090737888184286)
			(0.7164179104477612, 0.3369119658536933)
			(0.7213930348258707, 0.3329300286968054)
			(0.7263681592039801, 0.32894490877111576)
			(0.7313432835820896, 0.32497196167282005)
			(0.736318407960199, 0.3209945317017204)
			(0.7412935323383084, 0.3170279074194887)
			(0.746268656716418, 0.31305541906603473)
			(0.7512437810945274, 0.3090922924403183)
			(0.7562189054726368, 0.30512182499292934)
			(0.7611940298507462, 0.3011591835068941)
			(0.7661691542288557, 0.29718761179621467)
			(0.7711442786069652, 0.29322222067312265)
			(0.7761194029850746, 0.2892461768607304)
			(0.7810945273631841, 0.28527453730031194)
			(0.7860696517412935, 0.28129036493874443)
			(0.7910447761194029, 0.2773086636631543)
			(0.7960199004975125, 0.27331236192156505)
			(0.8009950248756219, 0.2693164096095251)
			(0.8059701492537313, 0.2653035648603228)
			(0.8109452736318408, 0.26128872022314814)
			(0.8159203980099502, 0.2572544212037889)
			(0.8208955223880597, 0.2532154962899962)
			(0.8258706467661692, 0.24915422765467182)
			(0.8308457711442786, 0.245085368291161)
			(0.835820895522388, 0.24099087533595903)
			(0.8407960199004975, 0.23688540815414824)
			(0.845771144278607, 0.23275052249350173)
			(0.8507462686567164, 0.22860075631069182)
			(0.8557213930348259, 0.22441716773812898)
			(0.8606965174129353, 0.22021413143748889)
			(0.8656716417910447, 0.2159720841641241)
			(0.8706467661691543, 0.21170517438561062)
			(0.8756218905472637, 0.20739305464590102)
			(0.8805970149253731, 0.2030495523340584)
			(0.8855721393034826, 0.19865331595815558)
			(0.8905472636815921, 0.194217706993294)
			(0.8955223880597015, 0.18972006427379617)
			(0.900497512437811, 0.18517305810432114)
			(0.9054726368159204, 0.1805522778882049)
			(0.9104477611940298, 0.17586934320138675)
			(0.9154228855721394, 0.17109743517710646)
			(0.9203980099502488, 0.16624652862486597)
			(0.9253731343283582, 0.16128636015287465)
			(0.9303482587064676, 0.15622423327291757)
			(0.9353233830845771, 0.15102470971249055)
			(0.9402985074626866, 0.14569054019826758)
			(0.945273631840796, 0.1401779949909383)
			(0.9502487562189055, 0.13448154148113875)
			(0.9552238805970149, 0.1285429586198767)
			(0.9601990049751243, 0.12234018513284564)
			(0.9651741293532339, 0.11578636080053616)
			(0.9701492537313433, 0.10882154181967267)
			(0.9751243781094527, 0.10129061782545688)
			(0.9800995024875622, 0.0930244926030174)
			(0.9850746268656716, 0.08364292039580182)
			(0.9900497512437811, 0.07248049118795448)
			(0.9950248756218906, 0.05755233834066966)
		};
		\addlegendentry{$m_j(p)$}
		\addplot [dashed] coordinates {(0,7.962863824636424) (0.9402985074626866,7.962863824636424)};
		\addplot [dashed] coordinates {(0,5.60425680837996) (0.9402985074626866,5.60425680837996)};
		\addplot [dashed] coordinates {(0,6.908718691599656) (0.9402985074626866,6.908718691599656)};
		\addplot [dashed] coordinates {(0,0.14569054019826758) (0.9402985074626866,0.14569054019826758)};
%		\addplot [dashed] coordinates {(0.9402985074626866,-1) (0.9402985074626866,0.14569054019826758)};
		\addplot [] coordinates {(1,16) (1,-1)};
		\addplot [] coordinates {(0.9402985074626866,5.60425680837996) (0.9402985074626866,6.908718691599656)};
		\addplot [dash pattern=on 4pt off 2pt] coordinates {(0.9402985074626866,0.14569054019826758) (0.9402985074626866,5.60425680837996)};
		\addplot [dash pattern=on 4pt off 2pt] coordinates {(0.9402985074626866,6.908718691599656) (0.9402985074626866,7.962863824636424)};
		%				\node[anchor=south, text=blue] at (axis cs:0.4,0.53) {$f_1(U)$};
		%				\node[anchor=south, text=orange] at (axis cs:0.4,0.1) {$f_2(U)$};
	\end{axis}
\end{tikzpicture}

%% file: additive_bounds_of_mes.tex
\begin{subfigure}{0.5\textwidth}
	\centering
	\begin{tikzpicture}
		\begin{axis}[
			xlabel={$b_2$},
			ylabel={$\mathrm{MES}_{0.95}(X_1, S)$},
			xmin=-1.0, xmax=1.05,
			ymin=-0.05, ymax=2.3,
			axis x line=bottom,
			axis y line=left,
			width=\linewidth,
			height=\textwidth,
			xlabel style={at={(current axis.south)}, below=5mm},
			ylabel style={at={(axis description cs:-0.1,0.5)}, anchor=south},
			legend style={at={(0.6,0.75)}}
			]
			\addplot[dashed] coordinates {
				(-1.0, 1.2203173538835335)
				(-0.9797979797979798, 1.3812775101240606)
				(-0.9595959595959596, 1.444298290379351)
				(-0.9393939393939394, 1.4911787315422385)
				(-0.9191919191919192, 1.5296883336469291)
				(-0.898989898989899, 1.5628362257904018)
				(-0.8787878787878788, 1.5921654073987714)
				(-0.8585858585858586, 1.6185922615262425)
				(-0.8383838383838383, 1.6427136579968646)
				(-0.8181818181818181, 1.6649439978991731)
				(-0.797979797979798, 1.685584831969128)
				(-0.7777777777777778, 1.7048636453356072)
				(-0.7575757575757576, 1.72295701859953)
				(-0.7373737373737373, 1.740005231225183)
				(-0.7171717171717171, 1.75612188101335)
				(-0.696969696969697, 1.7714004530961376)
				(-0.6767676767676767, 1.78591894342335)
				(-0.6565656565656566, 1.7997431976975282)
				(-0.6363636363636364, 1.81292937666413)
				(-0.6161616161616161, 1.825525811798773)
				(-0.5959595959595959, 1.8375744259981075)
				(-0.5757575757575757, 1.8491118376775355)
				(-0.5555555555555556, 1.860170230369815)
				(-0.5353535353535352, 1.870778045880615)
				(-0.5151515151515151, 1.880960542791527)
				(-0.4949494949494949, 1.890740250876927)
				(-0.4747474747474747, 1.9001373441172476)
				(-0.4545454545454545, 1.9091699493638492)
				(-0.43434343434343425, 1.9178544036345777)
				(-0.41414141414141414, 1.9262054700265316)
				(-0.3939393939393939, 1.9342365200082807)
				(-0.3737373737373737, 1.9419596881814112)
				(-0.3535353535353535, 1.9493860043305309)
				(-0.33333333333333326, 1.9565255066056833)
				(-0.31313131313131304, 1.9633873389258376)
				(-0.2929292929292928, 1.969979835102036)
				(-0.2727272727272727, 1.9763105917139727)
				(-0.2525252525252525, 1.9823865314048117)
				(-0.23232323232323226, 1.9882139579639555)
				(-0.21212121212121204, 1.9937986043297524)
				(-0.19191919191919182, 1.9991456744513054)
				(-0.1717171717171716, 2.0042598797909914)
				(-0.1515151515151515, 2.0091454711197088)
				(-0.13131313131313127, 2.013806266149461)
				(-0.11111111111111105, 2.0182456734581757)
				(-0.09090909090909083, 2.0224667130861653)
				(-0.07070707070707061, 2.0264720341194082)
				(-0.050505050505050386, 2.0302639295197467)
				(-0.030303030303030276, 2.033844348414186)
				(-0.010101010101010055, 2.037214906013291)
				(0.010101010101010166, 2.040376891290876)
				(0.030303030303030498, 2.043331272522778)
				(0.05050505050505061, 2.046078700750375)
				(0.07070707070707072, 2.048619511203968)
				(0.09090909090909105, 2.050953722691286)
				(0.11111111111111116, 2.0530810349264064)
				(0.1313131313131315, 2.0550008237435877)
				(0.1515151515151516, 2.056712134107975)
				(0.1717171717171717, 2.0582136708000207)
				(0.19191919191919204, 2.059503786611672)
				(0.21212121212121215, 2.0605804678487734)
				(0.2323232323232325, 2.061441316884245)
				(0.2525252525252526, 2.06208353144874)
				(0.27272727272727293, 2.0625038802775357)
				(0.29292929292929304, 2.0626986746516947)
				(0.31313131313131315, 2.0626637352747648)
				(0.3333333333333335, 2.062394353809218)
				(0.3535353535353536, 2.061885248253991)
				(0.3737373737373739, 2.0611305111689227)
				(0.39393939393939403, 2.0601235495342793)
				(0.41414141414141437, 2.058857014761894)
				(0.4343434343434345, 2.057322721032561)
				(0.4545454545454546, 2.0555115497005594)
				(0.4747474747474749, 2.053413336951314)
				(0.49494949494949503, 2.0510167411820124)
				(0.5151515151515154, 2.0483090856419794)
				(0.5353535353535355, 2.0452761706419826)
				(0.5555555555555556, 2.0419020480089847)
				(0.5757575757575759, 2.0381687482665756)
				(0.595959595959596, 2.0340559480298905)
				(0.6161616161616164, 2.029540560974138)
				(0.6363636363636365, 2.024596229950455)
				(0.6565656565656568, 2.0191926895862315)
				(0.6767676767676769, 2.013294956771234)
				(0.696969696969697, 2.0068622887915017)
				(0.7171717171717173, 1.9998468222292132)
				(0.7373737373737375, 1.9921917645045042)
				(0.7575757575757578, 1.983828944307472)
				(0.7777777777777779, 1.9746754193993217)
				(0.7979797979797982, 1.9646286567744597)
				(0.8181818181818183, 1.9535594743589613)
				(0.8383838383838385, 1.9413013254898819)
				(0.8585858585858588, 1.9276333037468818)
				(0.8787878787878789, 1.9122516831267087)
				(0.8989898989898992, 1.8947188324943034)
				(0.9191919191919193, 1.8743626527765267)
				(0.9393939393939394, 1.8500512252081056)
				(0.9595959595959598, 1.8195767773152527)
				(0.9797979797979799, 1.7772383710096178)
				(1.0, 1.66301223148414)
			};
			\addlegendentry{$M^f_1(0.95)$}
			\addplot[] coordinates {
				(-1.0, 1.2203173538835335)
				(-0.9797979797979798, 1.0469994741480269)
				(-0.9595959595959596, 0.9715577582986031)
				(-0.9393939393939394, 0.9121921893774119)
				(-0.9191919191919192, 0.8611322616948622)
				(-0.898989898989899, 0.8153678139628712)
				(-0.8787878787878788, 0.7733547870377714)
				(-0.8585858585858586, 0.7341757095825557)
				(-0.8383838383838383, 0.6972325938353688)
				(-0.8181818181818181, 0.662109889971073)
				(-0.797979797979798, 0.628504866517156)
				(-0.7777777777777778, 0.5961888243461805)
				(-0.7575757575757576, 0.5649839342698758)
				(-0.7373737373737373, 0.5347486322521988)
				(-0.7171717171717171, 0.5053679984673396)
				(-0.696969696969697, 0.4767471867580368)
				(-0.6767676767676767, 0.4488067995192168)
				(-0.6565656565656566, 0.42147954704432233)
				(-0.6363636363636364, 0.3947077804218862)
				(-0.6161616161616161, 0.36844163393398943)
				(-0.5959595959595959, 0.3426376023430463)
				(-0.5757575757575757, 0.3172574346561915)
				(-0.5555555555555556, 0.29226726226500466)
				(-0.5353535353535352, 0.2676369033956897)
				(-0.5151515151515151, 0.24333930206952623)
				(-0.4949494949494949, 0.2193500709967605)
				(-0.4747474747474747, 0.19564711570998275)
				(-0.4545454545454545, 0.172210322869369)
				(-0.43434343434343425, 0.1490212997472045)
				(-0.41414141414141414, 0.12606315489035472)
				(-0.3939393939393939, 0.1033203121823117)
				(-0.3737373737373737, 0.08077835219724078)
				(-0.3535353535353535, 0.05842387600754064)
				(-0.33333333333333326, 0.036244387579669995)
				(-0.31313131313131304, 0.014228191646212711)
				(-0.2929292929292928, 0.007635695470035721)
				(-0.2727272727272727, 0.02935762414222717)
				(-0.2525252525252525, 0.050947380531882726)
				(-0.23232323232323226, 0.07241424789167235)
				(-0.21212121212121204, 0.0937670612491238)
				(-0.19191919191919182, 0.11501425645263993)
				(-0.1717171717171716, 0.13616391440859937)
				(-0.1515151515151515, 0.1572238012140182)
				(-0.13131313131313127, 0.17820140478779053)
				(-0.11111111111111105, 0.19910396852067694)
				(-0.09090909090909083, 0.21993852239642744)
				(-0.07070707070707061, 0.2407119119810712)
				(-0.050505050505050386, 0.26143082563237263)
				(-0.030303030303030276, 0.28210182024512087)
				(-0.010101010101010055, 0.3027313458189427)
				(0.010101010101010166, 0.32332576911286354)
				(0.030303030303030498, 0.3438913966339362)
				(0.05050505050505061, 0.364434497195594)
				(0.07070707070707072, 0.38496132427442187)
				(0.09090909090909105, 0.40547813839171426)
				(0.11111111111111116, 0.42599122974836207)
				(0.1313131313131315, 0.44650694134843355)
				(0.1515151515151516, 0.4670316928585568)
				(0.1717171717171717, 0.48757200546726304)
				(0.19191919191919204, 0.5081345280315072)
				(0.21212121212121215, 0.5287260648274289)
				(0.2323232323232325, 0.5493536052602831)
				(0.2525252525252526, 0.5700243559357508)
				(0.27272727272727293, 0.5907457755535251)
				(0.29292929292929304, 0.6115256131565732)
				(0.31313131313131315, 0.6323719503590319)
				(0.3333333333333335, 0.6532932482863586)
				(0.3535353535353536, 0.6742984000985126)
				(0.3737373737373739, 0.6953967901374893)
				(0.39393939393939403, 0.7165983609536271)
				(0.41414141414141437, 0.737913689732774)
				(0.4343434343434345, 0.7593540759847318)
				(0.4545454545454546, 0.7809316427839275)
				(0.4747474747474749, 0.8026594544053038)
				(0.49494949494949503, 0.8245516539120038)
				(0.5151515151515154, 0.8466236251821553)
				(0.5353535353535355, 0.8688921850875843)
				(0.5555555555555556, 0.8913758131680909)
				(0.5757575757575759, 0.9140949283394884)
				(0.595959595959596, 0.9370722251635096)
				(0.6161616161616164, 0.9603330863359861)
				(0.6363636363636365, 0.9839060938338462)
				(0.6565656565656568, 1.0078236693969063)
				(0.6767676767676769, 1.0321228869552272)
				(0.696969696969697, 1.0568465172511876)
				(0.7171717171717173, 1.0820443915483784)
				(0.7373737373737375, 1.1077752125608176)
				(0.7575757575757578, 1.1341090063630515)
				(0.7777777777777779, 1.1611305168102997)
				(0.7979797979797982, 1.188944027484378)
				(0.8181818181818183, 1.2176804219969557)
				(0.8383838383838385, 1.2475079014121422)
				(0.8585858585858588, 1.2786489812229673)
				(0.8787878787878789, 1.3114089528967965)
				(0.8989898989898992, 1.3462269710611134)
				(0.9191919191919193, 1.3837766178685236)
				(0.9393939393939394, 1.425191256235895)
				(0.9595959595959598, 1.4726800652763834)
				(0.9797979797979799, 1.531945353555617)
				(1.0, 1.66301223148414)
				
			};
			\addlegendentry{$m^f_1(0.95)$}
			\addplot [dashdotted] coordinates {(-1,2.063) (1,2.063)};
			\addlegendentry{$M_1(0.95)$}
			\addplot [dotted] coordinates {(-1,0) (1,0)};
			\addlegendentry{$m_1(0.95)$}
			\addplot [] coordinates {(1,-0.3) (1,2.2)};
			\addplot [] coordinates {(-1,2.2) (1,2.2)};
		\end{axis}
	\end{tikzpicture}
\end{subfigure}%
\begin{subfigure}{0.5\textwidth}
	\centering
	\begin{tikzpicture}
		\begin{axis}[
			xlabel={$b_2$},
			ylabel={$\delta_{1,0.95}$},
			xmin=-1, xmax=1.05,
			ymin=0, ymax=1.05,
			axis x line=bottom,
			axis y line=left,
			width=\linewidth,
			height=\textwidth,
			xlabel style={at={(current axis.south)}, below=5mm},
			ylabel style={rotate=-90,at={(current axis.west)}, xshift=-5mm}
			]
			\addplot[] coordinates {
				(-1.0, 1.0)
				(-0.9797979797979798, 0.8379425217318672)
				(-0.9595959595959596, 0.7708161163492238)
				(-0.9393939393939394, 0.7193082138931053)
				(-0.9191919191919192, 0.6758850434637348)
				(-0.898989898989899, 0.6376284623302608)
				(-0.8787878787878788, 0.6030418692409014)
				(-0.8585858585858586, 0.5712362144042669)
				(-0.8383838383838383, 0.5416322326984475)
				(-0.8181818181818181, 0.5138275651956021)
				(-0.797979797979798, 0.4875292568094618)
				(-0.7777777777777778, 0.46251615011342984)
				(-0.7575757575757576, 0.43861642778621013)
				(-0.7373737373737373, 0.41569345253185697)
				(-0.7171717171717171, 0.3936364393560843)
				(-0.696969696969697, 0.3723540855391523)
				(-0.6767676767676767, 0.3517700869274699)
				(-0.6565656565656566, 0.33181989967925074)
				(-0.6363636363636364, 0.31244834904767116)
				(-0.6161616161616161, 0.29360782918417117)
				(-0.5959595959595959, 0.27525692466052165)
				(-0.5757575757575757, 0.25735933890262164)
				(-0.5555555555555556, 0.23988304993390797)
				(-0.5353535353535352, 0.22279963713312334)
				(-0.5151515151515151, 0.2060837384818025)
				(-0.4949494949494949, 0.18971260865933726)
				(-0.4747474747474747, 0.1736657559871535)
				(-0.4545454545454545, 0.15792464167931564)
				(-0.43434343434343425, 0.14247242880853372)
				(-0.41414141414141414, 0.12729377129748765)
				(-0.3939393939393939, 0.11237463540140558)
				(-0.3737373737373737, 0.09770214776859076)
				(-0.3535353535353535, 0.08326446539689558)
				(-0.33333333333333326, 0.06905066374873892)
				(-0.31313131313131304, 0.05505064001857762)
				(-0.2929292929292928, 0.048658575983105745)
				(-0.2727272727272727, 0.05612019255146028)
				(-0.2525252525252525, 0.06364126705216289)
				(-0.23232323232323226, 0.07122324392442769)
				(-0.21212121212121204, 0.07886762705632322)
				(-0.19191919191919182, 0.08657598326766458)
				(-0.1717171717171716, 0.09434994605972702)
				(-0.1515151515151515, 0.10219121965721167)
				(-0.13131313131313127, 0.11010158337078035)
				(-0.11111111111111105, 0.1180828963118028)
				(-0.09090909090909083, 0.12613710249469667)
				(-0.07070707070707061, 0.13426623636654356)
				(-0.050505050505050386, 0.14247242880853328)
				(-0.030303030303030276, 0.15075791365940838)
				(-0.010101010101010055, 0.1591250348174782)
				(0.010101010101010166, 0.1675762539852116)
				(0.030303030303030498, 0.1761141591289005)
				(0.05050505050505061, 0.1847414737358064)
				(0.07070707070707072, 0.19346106696263532)
				(0.09090909090909105, 0.20227596478253396)
				(0.11111111111111116, 0.21118936225338447)
				(0.1313131313131315, 0.2202046370484061)
				(0.1515151515151516, 0.22932536441154783)
				(0.1717171717171717, 0.23855533372543747)
				(0.19191919191919204, 0.24789856690964474)
				(0.21212121212121215, 0.2573593389026214)
				(0.2323232323232325, 0.26694220052322126)
				(0.2525252525252526, 0.2766520040586804)
				(0.27272727272727293, 0.2864939319873242)
				(0.29292929292929304, 0.2964735293185514)
				(0.31313131313131315, 0.3065967401229781)
				(0.3333333333333335, 0.3168699489360267)
				(0.3535353535353536, 0.3273000278539827)
				(0.3737373737373739, 0.3378943903093419)
				(0.39393939393939403, 0.3486610527210703)
				(0.41414141414141437, 0.35960870547687007)
				(0.4343434343434345, 0.37074679503430763)
				(0.4545454545454546, 0.38208561934667506)
				(0.4747474747474749, 0.3936364393560845)
				(0.49494949494949503, 0.4054116099894369)
				(0.5151515151515154, 0.41742473499646504)
				(0.5353535353535355, 0.42969085115831684)
				(0.5555555555555556, 0.4422266489772828)
				(0.5757575757575759, 0.45505073908693394)
				(0.595959595959596, 0.4681839765216895)
				(0.6161616161616164, 0.4816498589882815)
				(0.6363636363636365, 0.495475020890487)
				(0.6565656565656568, 0.5096898528441001)
				(0.6767676767676769, 0.5243292879915498)
				(0.696969696969697, 0.5394338135281618)
				(0.7171717171717173, 0.5550507916853904)
				(0.7373737373737375, 0.5712362144042669)
				(0.7575757575757578, 0.5880570795644504)
				(0.7777777777777779, 0.6055946811266923)
				(0.7979797979797982, 0.6239492834548228)
				(0.8181818181818183, 0.6432469659936619)
				(0.8383838383838385, 0.6636500139269912)
				(0.8585858585858588, 0.6853733975171539)
				(0.8787878787878789, 0.7087123674982325)
				(0.8989898989898992, 0.734091988260845)
				(0.9191919191919193, 0.762164643995775)
				(0.9393939393939394, 0.7940285397822251)
				(0.9595959595959598, 0.8318250069634958)
				(0.9797979797979799, 0.8810823219978878)
				(1.0, 1.0)
			};
			\addplot [] coordinates {(-1,1) (1,1)};
			\addplot [] coordinates {(1,0) (1,1)};
		\end{axis}
	\end{tikzpicture}
\end{subfigure}

%% file: min_bounds_of_mes.tex
\begin{subfigure}{0.5\textwidth}
	\centering
	\begin{tikzpicture}
		\begin{axis}[
			xlabel={$\lambda_0$},
			ylabel={$\mathrm{MES}_{0.95}(X_1, S)$},
			xmin=0.1, xmax=2.2,
			ymin=0.55, ymax=4.15,
			axis x line=bottom,
			axis y line=left,
			width=\linewidth,
			height=\textwidth,
			xlabel style={at={(current axis.south)}, below=5mm},
			ylabel style={at={(axis description cs:-0.1,0.5)}, anchor=south},
			legend style={at={(0.93,0.93)}}
			]
			\addplot[dashed] coordinates {
				(0.1, 3.6324838850490817)
				(0.12020202020202021, 3.5669747076811995)
				(0.14040404040404042, 3.5037864931961473)
				(0.1606060606060606, 3.4427980424877718)
				(0.1808080808080808, 3.383896450657357)
				(0.20101010101010103, 3.3269764094351975)
				(0.22121212121212122, 3.2719395788407364)
				(0.24141414141414144, 3.2186940201940195)
				(0.26161616161616164, 3.1671536836016414)
				(0.28181818181818186, 3.117237943907368)
				(0.3020202020202021, 3.0688711798436383)
				(0.32222222222222224, 3.021982391763522)
				(0.3424242424242424, 2.976504853888977)
				(0.36262626262626263, 2.9323757974932914)
				(0.38282828282828285, 2.8895361218542366)
				(0.40303030303030307, 2.8479301301788698)
				(0.4232323232323233, 2.8075052880187723)
				(0.4434343434343435, 2.7682120019723233)
				(0.4636363636363636, 2.730003416713906)
				(0.48383838383838385, 2.692835228603438)
				(0.5040404040404041, 2.6566655143172935)
				(0.5242424242424243, 2.621454573106991)
				(0.5444444444444445, 2.587164781437835)
				(0.5646464646464647, 2.553760458888606)
				(0.5848484848484848, 2.5212077443074885)
				(0.6050505050505051, 2.4894744813206104)
				(0.6252525252525253, 2.4585301123793974)
				(0.6454545454545455, 2.428345580612922)
				(0.6656565656565657, 2.3988932388225894)
				(0.6858585858585858, 2.370146765020042)
				(0.706060606060606, 2.3420810839659265)
				(0.7262626262626263, 2.3146722942179343)
				(0.7464646464646465, 2.2878976002420184)
				(0.7666666666666667, 2.2617352491815037)
				(0.7868686868686869, 2.2361644719154605)
				(0.8070707070707072, 2.211165428070682)
				(0.8272727272727273, 2.1867191546812883)
				(0.8474747474747475, 2.162807518216758)
				(0.8676767676767677, 2.139413169723337)
				(0.8878787878787879, 2.1165195028456125)
				(0.9080808080808082, 2.094110614514796)
				(0.9282828282828284, 2.0721712681081454)
				(0.9484848484848485, 2.0506868589001814)
				(0.9686868686868687, 2.0296433816410726)
				(0.9888888888888889, 2.0090274001109445)
				(1.0090909090909093, 1.9888260185110356)
				(1.0292929292929294, 1.9690268545636882)
				(1.0494949494949497, 1.9496180142032775)
				(1.0696969696969698, 1.9305880677493654)
				(1.08989898989899, 1.9119260274617929)
				(1.1101010101010103, 1.8936213263850887)
				(1.1303030303030306, 1.8756637983966093)
				(1.1505050505050507, 1.8580436593792624)
				(1.1707070707070708, 1.840751489445533)
				(1.1909090909090911, 1.8237782161449743)
				(1.2111111111111112, 1.8071150985922566)
				(1.2313131313131316, 1.790753712457424)
				(1.2515151515151517, 1.7746859357642213)
				(1.2717171717171718, 1.758903935446176)
				(1.2919191919191921, 1.7434001546136844)
				(1.3121212121212122, 1.7281673004886193)
				(1.3323232323232326, 1.713198332965981)
				(1.3525252525252527, 1.698486453764899)
				(1.372727272727273, 1.6840250961338652)
				(1.3929292929292931, 1.6698079150774379)
				(1.4131313131313132, 1.6558287780738594)
				(1.4333333333333336, 1.6420817562550643)
				(1.4535353535353537, 1.6285611160224167)
				(1.473737373737374, 1.6152613110732745)
				(1.4939393939393941, 1.602176974815087)
				(1.5141414141414145, 1.5893029131452188)
				(1.5343434343434346, 1.5766340975761057)
				(1.5545454545454547, 1.5641656586866153)
				(1.574747474747475, 1.5518928798816987)
				(1.5949494949494951, 1.5398111914435382)
				(1.6151515151515154, 1.5279161648584203)
				(1.6353535353535356, 1.516203507404542)
				(1.6555555555555557, 1.504669056986858)
				(1.675757575757576, 1.493308777205908)
				(1.6959595959595961, 1.4821187526483515)
				(1.7161616161616164, 1.4710951843876718)
				(1.7363636363636366, 1.4602343856841824)
				(1.7565656565656569, 1.4495327778741112)
				(1.776767676767677, 1.438986886438141)
				(1.796969696969697, 1.4285933372403214)
				(1.8171717171717174, 1.4183488529288093)
				(1.8373737373737375, 1.4082502494903701)
				(1.8575757575757579, 1.398294432951025)
				(1.877777777777778, 1.388478396215672)
				(1.8979797979797983, 1.3787992160398919)
				(1.9181818181818184, 1.3692540501275354)
				(1.9383838383838385, 1.3598401343480404)
				(1.9585858585858589, 1.3505547800677535)
				(1.978787878787879, 1.341395371589844)
				(1.9989898989898993, 1.3323593636976927)
				(2.019191919191919, 1.3234442792969054)
				(2.0393939393939395, 1.3146477071513625)
				(2.05959595959596, 1.3059672997089633)
				(2.0797979797979798, 1.2974007710129387)
				(2.1, 1.2889458946948353)
				
			};
			\addlegendentry{$M_1(0.95)=M^f_1(0.95)$}
			\addplot[dotted] coordinates {
				(0.1, 3.415552667083784)
				(0.12020202020202021, 3.290593062498825)
				(0.14040404040404042, 3.20082488138276)
				(0.1606060606060606, 3.066305369147141)
				(0.1808080808080808, 2.9777632027868997)
				(0.20101010101010103, 2.874146422013216)
				(0.22121212121212122, 2.782861160907112)
				(0.24141414141414144, 2.66521950328885)
				(0.26161616161616164, 2.5669745243061812)
				(0.28181818181818186, 2.502495186564061)
				(0.3020202020202021, 2.381402038555184)
				(0.32222222222222224, 2.283638163836871)
				(0.3424242424242424, 2.18568192441874)
				(0.36262626262626263, 2.1237287217687624)
				(0.38282828282828285, 2.039374010610369)
				(0.40303030303030307, 1.975200973561743)
				(0.4232323232323233, 1.9126520952238104)
				(0.4434343434343435, 1.8718331613334827)
				(0.4636363636363636, 1.8299723700029813)
				(0.48383838383838385, 1.8057666520112012)
				(0.5040404040404041, 1.7505051570340264)
				(0.5242424242424243, 1.7051589871755304)
				(0.5444444444444445, 1.6739966458521767)
				(0.5646464646464647, 1.6397495157180846)
				(0.5848484848484848, 1.6560635029377169)
				(0.6050505050505051, 1.6071857390331163)
				(0.6252525252525253, 1.555185441496674)
				(0.6454545454545455, 1.5137922137420767)
				(0.6656565656565657, 1.527795352991224)
				(0.6858585858585858, 1.4765293898507017)
				(0.706060606060606, 1.4574033055736404)
				(0.7262626262626263, 1.4271750211781329)
				(0.7464646464646465, 1.4122902441712388)
				(0.7666666666666667, 1.3901861481399287)
				(0.7868686868686869, 1.37671787733869)
				(0.8070707070707072, 1.3721699978632904)
				(0.8272727272727273, 1.3192231933685692)
				(0.8474747474747475, 1.340048132975227)
				(0.8676767676767677, 1.334452345520572)
				(0.8878787878787879, 1.3546044761301095)
				(0.9080808080808082, 1.295414913179938)
				(0.9282828282828284, 1.2658373338576248)
				(0.9484848484848485, 1.2514877666382933)
				(0.9686868686868687, 1.249054788585574)
				(0.9888888888888889, 1.2144534841725914)
				(1.0090909090909093, 1.2294189161796234)
				(1.0292929292929294, 1.1905978312830798)
				(1.0494949494949497, 1.1771397654415363)
				(1.0696969696969698, 1.1800848578886278)
				(1.08989898989899, 1.151475883866779)
				(1.1101010101010103, 1.142832504653024)
				(1.1303030303030306, 1.1237682636239072)
				(1.1505050505050507, 1.139193585635389)
				(1.1707070707070708, 1.1308036178460878)
				(1.1909090909090911, 1.1079834777845667)
				(1.2111111111111112, 1.0956961986692009)
				(1.2313131313131316, 1.091732810092857)
				(1.2515151515151517, 1.0292642673919286)
				(1.2717171717171718, 1.0298170132415805)
				(1.2919191919191921, 1.0117169436473599)
				(1.3121212121212122, 1.039534487659993)
				(1.3323232323232326, 1.044585068940409)
				(1.3525252525252527, 1.0233387662990068)
				(1.372727272727273, 1.015142783210046)
				(1.3929292929292931, 1.011030683342884)
				(1.4131313131313132, 0.9718956024251489)
				(1.4333333333333336, 0.9670815572064515)
				(1.4535353535353537, 0.9805432744257913)
				(1.473737373737374, 0.9902682582534444)
				(1.4939393939393941, 0.947181157712836)
				(1.5141414141414145, 0.9648993827230812)
				(1.5343434343434346, 0.9425361771926415)
				(1.5545454545454547, 0.9160585694480098)
				(1.574747474747475, 0.9383443839258975)
				(1.5949494949494951, 0.9190466866732824)
				(1.6151515151515154, 0.9264664605437823)
				(1.6353535353535356, 0.9020675495653487)
				(1.6555555555555557, 0.8991001203616088)
				(1.675757575757576, 0.8686583488004624)
				(1.6959595959595961, 0.8885677939706554)
				(1.7161616161616164, 0.8669852127737562)
				(1.7363636363636366, 0.8841421723184336)
				(1.7565656565656569, 0.8524812353407897)
				(1.776767676767677, 0.8395611339002303)
				(1.796969696969697, 0.8394240175240544)
				(1.8171717171717174, 0.8693591291903154)
				(1.8373737373737375, 0.8452806260414615)
				(1.8575757575757579, 0.8538003136457842)
				(1.877777777777778, 0.8231401790605208)
				(1.8979797979797983, 0.8319355990901799)
				(1.9181818181818184, 0.8051450508253492)
				(1.9383838383838385, 0.8110032171721137)
				(1.9585858585858589, 0.8090873694451722)
				(1.978787878787879, 0.8117155680346112)
				(1.9989898989898993, 0.7893499861432433)
				(2.019191919191919, 0.785786988224672)
				(2.0393939393939395, 0.7805964200323229)
				(2.05959595959596, 0.7789606989727512)
				(2.0797979797979798, 0.7746327309465584)
				(2.1, 0.7571551283148598)
			};
			\addlegendentry{$m^f_1(0.95)$}
			\addplot [] coordinates {(0.1, 2.1361670368749093)
				(0.1, 2.1361670368749093)
				(0.12020202020202021, 2.0976428342261277)
				(0.14040404040404042, 2.06048352803966)
				(0.1606060606060606, 2.024617844348804)
				(0.1808080808080808, 1.9899793867893725)
				(0.20101010101010103, 1.9565062263723931)
				(0.22121212121212122, 1.9241405319741733)
				(0.24141414141414144, 1.8928282369054314)
				(0.26161616161616164, 1.8625187375154328)
				(0.28181818181818186, 1.8331646202969074)
				(0.3020202020202021, 1.8047214143962573)
				(0.32222222222222224, 1.7771473668118989)
				(0.3424242424242424, 1.7504032378907264)
				(0.36262626262626263, 1.7244521150161425)
				(0.38282828282828285, 1.6992592426273014)
				(0.40303030303030307, 1.6747918669235244)
				(0.4232323232323233, 1.651019093794731)
				(0.4434343434343435, 1.6279117586821383)
				(0.4636363636363636, 1.6054423072165465)
				(0.48383838383838385, 1.5835846856070632)
				(0.5040404040404041, 1.5623142398635164)
				(0.5242424242424243, 1.5416076230329858)
				(0.5444444444444445, 1.5214427097166618)
				(0.5646464646464647, 1.5017985172090225)
				(0.5848484848484848, 1.4826551326684356)
				(0.6050505050505051, 1.4639936457877756)
				(0.6252525252525253, 1.4457960864864985)
				(0.6454545454545455, 1.4280453671926188)
				(0.6656565656565657, 1.4107252293249093)
				(0.6858585858585858, 1.3938201936229933)
				(0.706060606060606, 1.3773155140063802)
				(0.7262626262626263, 1.361197134673362)
				(0.7464646464646465, 1.3454516501774296)
				(0.7666666666666667, 1.3300662682428677)
				(0.7868686868686869, 1.315028775102756)
				(0.8070707070707072, 1.300327503161976)
				(0.8272727272727273, 1.2859513008052936)
				(0.8474747474747475, 1.2718895041863179)
				(0.8676767676767677, 1.2581319108473636)
				(0.8878787878787879, 1.2446687550330526)
				(0.9080808080808082, 1.2314906845721418)
				(0.9282828282828284, 1.218588739212559)
				(0.9484848484848485, 1.2059543303041866)
				(0.9686868686868687, 1.1935792217325683)
				(0.9888888888888889, 1.1814555120146144)
				(1.0090909090909093, 1.1695756174744973)
				(1.0292929292929294, 1.1579322564244776)
				(1.0494949494949497, 1.1465184342813088)
				(1.0696969696969698, 1.1353274295543072)
				(1.08989898989899, 1.1243527806460973)
				(1.1101010101010103, 1.113588273411573)
				(1.1303030303030306, 1.1030279294247394)
				(1.1505050505050507, 1.0926659949068933)
				(1.1707070707070708, 1.082496930273046)
				(1.1909090909090911, 1.0725154002566968)
				(1.2111111111111112, 1.0627162645759596)
				(1.2313131313131316, 1.0530945691067342)
				(1.2515151515151517, 1.043645537531079)
				(1.2717171717171718, 1.034364563431203)
				(1.2919191919191921, 1.0252472028015758)
				(1.3121212121212122, 1.0162891669535934)
				(1.3323232323232326, 1.0074863157889888)
				(1.3525252525252527, 0.9988346514198265)
				(1.372727272727273, 0.9903303121144212)
				(1.3929292929292931, 0.9819695665499265)
				(1.4131313131313132, 0.9737488083536103)
				(1.4333333333333336, 0.9656645509160546)
				(1.4535353535353537, 0.957713422460591)
				(1.473737373737374, 0.9498921613543385)
				(1.4939393939393941, 0.9421976116471349)
				(1.5141414141414145, 0.9346267188255424)
				(1.5343434343434346, 0.9271765257699386)
				(1.5545454545454547, 0.9198441689034303)
				(1.574747474747475, 0.9126268745220776)
				(1.5949494949494951, 0.9055219552965261)
				(1.6151515151515154, 0.898526806935796)
				(1.6353535353535356, 0.8916389050045135)
				(1.6555555555555557, 0.8848558018854222)
				(1.675757575757576, 0.8781751238794925)
				(1.6959595959595961, 0.8715945684364091)
				(1.7161616161616164, 0.8651119015086557)
				(1.7363636363636366, 0.8587249550228038)
				(1.7565656565656569, 0.8524316244619917)
				(1.776767676767677, 0.846229866553938)
				(1.796969696969697, 0.8401176970591463)
				(1.8171717171717174, 0.834093188654276)
				(1.8373737373737375, 0.8281544689059365)
				(1.8575757575757579, 0.822299718330426)
				(1.877777777777778, 0.8165271685351967)
				(1.8979797979797983, 0.8108351004380536)
				(1.9181818181818184, 0.8052218425603239)
				(1.9383838383838385, 0.7996857693904349)
				(1.9585858585858589, 0.7942252998145359)
				(1.978787878787879, 0.7888388956109785)
				(1.9989898989898993, 0.7835250600056501)
				(2.019191919191919, 0.7782823362853047)
				(2.0393939393939395, 0.7731093064661932)
				(2.05959595959596, 0.7680045900154426)
				(2.0797979797979798, 0.7629668426227536)
				(2.1, 0.7579947550201289)
				
			};
			\addlegendentry{$m_1(0.95)$}
			\addplot [] coordinates {(2.1,0.55) (2.1,4)};
			\addplot [] coordinates {(0.1,4) (2.1,4)};
		\end{axis}
	\end{tikzpicture}
\end{subfigure}%
\begin{subfigure}{0.5\textwidth}
	\centering
	\begin{tikzpicture}
		\begin{axis}[
			xlabel={$\lambda_0$},
			ylabel={$\delta_{1,0.95}$},
			xmin=0.1, xmax=2.2,
			ymin=0, ymax=1.05,
			axis x line=bottom,
			axis y line=left,
			width=\linewidth,
			height=\textwidth,
			xlabel style={at={(current axis.south)}, below=5mm},
			ylabel style={rotate=-90,at={(current axis.west)}, xshift=-5mm}
			]
			\addplot[] coordinates {
				(0.1, 0.8550232069965662)
				(0.12020202020202021, 0.8118997823599412)
				(0.14040404040404042, 0.7900914644206116)
				(0.1606060606060606, 0.7345240937402102)
				(0.1808080808080808, 0.7086388721410176)
				(0.20101010101010103, 0.6695805621907283)
				(0.22121212121212122, 0.6371280874024503)
				(0.24141414141414144, 0.5825561501916366)
				(0.26161616161616164, 0.5399639101375098)
				(0.28181818181818186, 0.5212557211181524)
				(0.3020202020202021, 0.45618062030398754)
				(0.32222222222222224, 0.4068738321727856)
				(0.3424242424242424, 0.35501028695213355)
				(0.36262626262626263, 0.33054787528778606)
				(0.38282828282828285, 0.28574424482139515)
				(0.40303030303030307, 0.2560730615030937)
				(0.4232323232323233, 0.22623097684674498)
				(0.4434343434343435, 0.2139098049716639)
				(0.4636363636363636, 0.19966017043466144)
				(0.48383838383838385, 0.2002991729929351)
				(0.5040404040404041, 0.17196573126342973)
				(0.5242424242424243, 0.15145791181920354)
				(0.5444444444444445, 0.14314607924853773)
				(0.5646464646464647, 0.13113687201345592)
				(0.5848484848484848, 0.16697119464714127)
				(0.6050505050505051, 0.13963409971570928)
				(0.6252525252525253, 0.10801390317041037)
				(0.6454545454545455, 0.08572111192125587)
				(0.6656565656565657, 0.11847188184712187)
				(0.6858585858585858, 0.08471468323284281)
				(0.706060606060606, 0.08301269661874267)
				(0.7262626262626263, 0.06919727886386173)
				(0.7464646464646465, 0.07092034719787221)
				(0.7666666666666667, 0.06452922779128223)
				(0.7868686868686869, 0.06697069981045078)
				(0.8070707070707072, 0.07887516838796005)
				(0.8272727272727273, 0.03693725572033568)
				(0.8474747474747475, 0.07650381709150211)
				(0.8676767676767677, 0.08660167671164465)
				(0.8878787878787879, 0.12609465711061363)
				(0.9080808080808082, 0.07410474345526163)
				(0.9282828282828284, 0.05535328224934355)
				(0.9484848484848485, 0.053902785547735976)
				(0.9686868686868687, 0.06635324119033736)
				(0.9888888888888889, 0.03987323957303868)
				(1.0090909090909093, 0.07304640758113845)
				(1.0292929292929294, 0.04027344644329112)
				(1.0494949494949497, 0.03812893435108344)
				(1.0696969696969698, 0.056280200709924566)
				(1.08989898989899, 0.034438832616960124)
				(1.1101010101010103, 0.037491015451167176)
				(1.1303030303030306, 0.026843607748584852)
				(1.1505050505050507, 0.060790369105650055)
				(1.1707070707070708, 0.06370774430391923)
				(1.1909090909090911, 0.04721127783481904)
				(1.2111111111111112, 0.04430411841902382)
				(1.2313131313131316, 0.05237953238214499)
				(1.2515151515151517, -0.019672332984481455)
				(1.2717171717171718, -0.0062764707692497534)
				(1.2919191919191921, -0.018840358617304487)
				(1.3121212121212122, 0.032653511340443364)
				(1.3323232323232326, 0.052569252398200095)
				(1.3525252525252527, 0.03502329987151909)
				(1.372727272727273, 0.035768570943917055)
				(1.3929292929292931, 0.042249922318477795)
				(1.4131313131313132, -0.002716992157417275)
				(1.4333333333333336, 0.0020948702652924256)
				(1.4535353535353537, 0.03403134896981841)
				(1.473737373737374, 0.06068224971981573)
				(1.4939393939393941, 0.007551063478378528)
				(1.5141414141414145, 0.0462406670048503)
				(1.5343434343434346, 0.023649968973318947)
				(1.5545454545454547, -0.00587532701523652)
				(1.574747474747475, 0.040229746597196914)
				(1.5949494949494951, 0.021322656299375686)
				(1.6151515151515154, 0.04439168418767758)
				(1.6353535353535356, 0.016697463354088393)
				(1.6555555555555557, 0.022981629319714036)
				(1.675757575757576, -0.015471068811740096)
				(1.6959595959595961, 0.027801069921833066)
				(1.7161616161616164, 0.0030913579929142676)
				(1.7363636363636366, 0.04225572534695399)
				(1.7565656565656569, 8.308622168029522e-05)
				(1.776767676767677, -0.011250364702573012)
				(1.796969696969697, -0.0011787735765551588)
				(1.8171717171717174, 0.06036045979944238)
				(1.8373737373737375, 0.029522981736344933)
				(1.8575757575757579, 0.05468903536051939)
				(1.877777777777778, 0.011562193077446326)
				(1.8979797979797983, 0.03715111231942414)
				(1.9181818181818184, -0.00013614778366299873)
				(1.9383838383838385, 0.02020415887062721)
				(1.9585858585858589, 0.026714510300391958)
				(1.978787878787879, 0.04140150992368008)
				(1.9989898989898993, 0.010613269065742559)
				(2.019191919191919, 0.013765913111817407)
				(2.0393939393939395, 0.01382563740014875)
				(2.05959595959596, 0.02036592641068058)
				(2.0797979797979798, 0.021828494981493085)
				(2.1, -0.0015813634109225205)
				
			};
			\addplot [] coordinates {(0.1,1) (2.1,1)};
			\addplot [] coordinates {(2.1,0) (2.1,1)};
		\end{axis}
	\end{tikzpicture}
\end{subfigure}

%% file: empirical_bounds_of_mes.tex
\begin{tikzpicture}
	\begin{axis}[
		xlabel={$p$},
		ylabel={},
		xtick={0.2, 0.4, 0.6, 0.8, 1},
		ytick={-0.07, -0.035, 0, 0.035, 0.065},
		xmin=0, xmax=1.05,
		ymin=-0.075, ymax=0.085,
		axis lines=middle,
		width=0.75\textwidth,
		height=0.5\textwidth,
		xlabel style={below right},
		ylabel style={above left},
		%				axis x line=middle, % Add this line
		axis x line shift=0.075, % Add this line
		legend style={
			at={(-0.15,0.5)}, % position relative to axis
			anchor=east          % anchor legend at its right side
		}
		]
		\addplot[blue] coordinates {
			(0.0, -0.0012817211844197139)
			(0.0049499999999999995, -0.0008575411341853035)
			(0.009899999999999999, -0.0005416732891832229)
			(0.014849999999999999, -0.00026793525615167405)
			(0.019799999999999998, -2.8093857693087417e-05)
			(0.024749999999999998, 0.0001924578239608802)
			(0.029699999999999997, 0.00040102314151136586)
			(0.03465, 0.000601466241251544)
			(0.039599999999999996, 0.0007950926960480032)
			(0.04454999999999999, 0.0009814502911813642)
			(0.049499999999999995, 0.001161809115617813)
			(0.05445, 0.0013288867409119562)
			(0.059399999999999994, 0.001498400506970849)
			(0.06434999999999999, 0.001662739859842854)
			(0.0693, 0.0018240892399658411)
			(0.07425, 0.0019820616012019743)
			(0.07919999999999999, 0.002136335347432024)
			(0.08414999999999999, 0.002288844868735084)
			(0.08909999999999998, 0.002438579842931937)
			(0.09405, 0.0025854935292827373)
			(0.09899999999999999, 0.0027306296868107633)
			(0.10394999999999999, 0.002873921712131293)
			(0.1089, 0.003009183277591973)
			(0.11384999999999999, 0.003147497085201794)
			(0.11879999999999999, 0.0032838509582863583)
			(0.12374999999999999, 0.0034182385487528345)
			(0.12869999999999998, 0.003550703990877993)
			(0.13365, 0.0036820089449541285)
			(0.1386, 0.003812395617070358)
			(0.14354999999999998, 0.003942660788863108)
			(0.1485, 0.004072566627771296)
			(0.15344999999999998, 0.004201848591549296)
			(0.15839999999999999, 0.004330266351829988)
			(0.16335, 0.00445273877938732)
			(0.16829999999999998, 0.004579485427615862)
			(0.17325, 0.0047056676279740445)
			(0.17819999999999997, 0.004831064796905222)
			(0.18314999999999998, 0.004955376550717587)
			(0.1881, 0.0050797430249632895)
			(0.19304999999999997, 0.005203845850775671)
			(0.19799999999999998, 0.005327425173439048)
			(0.20295, 0.005450914734480179)
			(0.20789999999999997, 0.005573882338183643)
			(0.21284999999999998, 0.005696769755112346)
			(0.2178, 0.0058143014986029974)
			(0.22274999999999998, 0.0059368200408997955)
			(0.22769999999999999, 0.006059162593259584)
			(0.23264999999999997, 0.006180816416364579)
			(0.23759999999999998, 0.006301948657805577)
			(0.24255, 0.00642323111227702)
			(0.24749999999999997, 0.006544557961447056)
			(0.25244999999999995, 0.006665820574162679)
			(0.25739999999999996, 0.006786778699491571)
			(0.26234999999999997, 0.006907271551724138)
			(0.2673, 0.0070273870355302415)
			(0.27225, 0.00714239312039312)
			(0.2772, 0.007262342495876856)
			(0.28214999999999996, 0.007382482701356213)
			(0.28709999999999997, 0.0075020420847268674)
			(0.29205, 0.007621387594723547)
			(0.297, 0.0077403872244205755)
			(0.30195, 0.007858641616851694)
			(0.30689999999999995, 0.007977019208715597)
			(0.31184999999999996, 0.00809521628645683)
			(0.31679999999999997, 0.008213584642233858)
			(0.32175, 0.008332326985057134)
			(0.3267, 0.008446704632634994)
			(0.33164999999999994, 0.00856611920332937)
			(0.33659999999999995, 0.008685461515423779)
			(0.34154999999999996, 0.00880515509957755)
			(0.3465, 0.008925147765278202)
			(0.35145, 0.00904540625)
			(0.35639999999999994, 0.009166221673355974)
			(0.36134999999999995, 0.009287615743621656)
			(0.36629999999999996, 0.00940995735340232)
			(0.37124999999999997, 0.009532886852085967)
			(0.3762, 0.009656200063714558)
			(0.38115, 0.00977521926163724)
			(0.38609999999999994, 0.009899091909385115)
			(0.39104999999999995, 0.010023473735725938)
			(0.39599999999999996, 0.010148169736842107)
			(0.40095, 0.010273269651741294)
			(0.4059, 0.01039870702341137)
			(0.41084999999999994, 0.010524405059021923)
			(0.41579999999999995, 0.010650792176870746)
			(0.42074999999999996, 0.010777642195540308)
			(0.42569999999999997, 0.010904815916955017)
			(0.43065, 0.011032268760907505)
			(0.4356, 0.011155354452657517)
			(0.44054999999999994, 0.011284372514204546)
			(0.44549999999999995, 0.011413765317090648)
			(0.45044999999999996, 0.011543718365871295)
			(0.45539999999999997, 0.011674448011674571)
			(0.46035, 0.011805866347569957)
			(0.46529999999999994, 0.011937853214418433)
			(0.47024999999999995, 0.012070564516129034)
			(0.47519999999999996, 0.012204143884892087)
			(0.48014999999999997, 0.012338585626911314)
			(0.4851, 0.012479745945945945)
			(0.49004999999999993, 0.012605634592910014)
			(0.49499999999999994, 0.012743091266719119)
			(0.49994999999999995, 0.012881566547477158)
			(0.5048999999999999, 0.01302108386837881)
			(0.5098499999999999, 0.013161471422780706)
			(0.5147999999999999, 0.013302753480753481)
			(0.5197499999999999, 0.013445332230037233)
			(0.5246999999999999, 0.01358915510033445)
			(0.52965, 0.013734018166455428)
			(0.5346, 0.013880535439795045)
			(0.53955, 0.014028539922313335)
			(0.5445, 0.014171835586567814)
			(0.54945, 0.014322455908289242)
			(0.5544, 0.01447439188586714)
			(0.5593499999999999, 0.014627607754733995)
			(0.5642999999999999, 0.014782540355677153)
			(0.5692499999999999, 0.014939168819188194)
			(0.5741999999999999, 0.015097210452636494)
			(0.5791499999999999, 0.01525694192634561)
			(0.5841, 0.01541865981844243)
			(0.58905, 0.015581552707930367)
			(0.594, 0.01574627019089574)
			(0.59895, 0.01590651163942546)
			(0.6039, 0.016075884653961883)
			(0.6088499999999999, 0.016247077704418487)
			(0.6137999999999999, 0.016419919238683127)
			(0.6187499999999999, 0.01659516519020323)
			(0.6236999999999999, 0.016772523231256598)
			(0.6286499999999999, 0.0169525468164794)
			(0.6335999999999999, 0.017135082429501087)
			(0.63855, 0.01732013249037933)
			(0.6435, 0.017507751393534002)
			(0.64845, 0.017697608818541547)
			(0.6534, 0.017882889971346707)
			(0.65835, 0.01807924360465116)
			(0.6632999999999999, 0.018278622418879058)
			(0.6682499999999999, 0.01848102155688623)
			(0.6731999999999999, 0.018686787234042556)
			(0.6781499999999999, 0.01889495185185185)
			(0.6830999999999999, 0.019106866457680248)
			(0.6880499999999999, 0.019322652229299363)
			(0.693, 0.019543266019417478)
			(0.69795, 0.01976815197368421)
			(0.7029, 0.019987236631016043)
			(0.70785, 0.020219928619986404)
			(0.7127999999999999, 0.020457260719225448)
			(0.7177499999999999, 0.02070048275862069)
			(0.7226999999999999, 0.020948659742120346)
			(0.7276499999999999, 0.02120137345003647)
			(0.7325999999999999, 0.02145824219910847)
			(0.7375499999999999, 0.021720727479182435)
			(0.7424999999999999, 0.021989773148148147)
			(0.74745, 0.022264561762391822)
			(0.7524, 0.022545208667736755)
			(0.75735, 0.022820284779050737)
			(0.7623, 0.023112803675856306)
			(0.7672499999999999, 0.02341255119453925)
			(0.7721999999999999, 0.02372026852659111)
			(0.7771499999999999, 0.024033636363636368)
			(0.7820999999999999, 0.024353824065633547)
			(0.7870499999999999, 0.02468059048507463)
			(0.7919999999999999, 0.025015624641833812)
			(0.7969499999999999, 0.025359987279843444)
			(0.8019, 0.025714429287863595)
			(0.80685, 0.0260800329218107)
			(0.8118, 0.026440904008438813)
			(0.81675, 0.02682605850487541)
			(0.8216999999999999, 0.027221606904231626)
			(0.8266499999999999, 0.027626294387170673)
			(0.8315999999999999, 0.028040705188679245)
			(0.8365499999999999, 0.028467153098420416)
			(0.8414999999999999, 0.0289034335839599)
			(0.8464499999999999, 0.029354865459249674)
			(0.8513999999999999, 0.0298211229946524)
			(0.85635, 0.030303603042876905)
			(0.8613, 0.030802121776504297)
			(0.86625, 0.031293652818991095)
			(0.8712, 0.03182436363636364)
			(0.8761499999999999, 0.03238007532051282)
			(0.8810999999999999, 0.032957312186978296)
			(0.8860499999999999, 0.03355491637630662)
			(0.8909999999999999, 0.03418027322404371)
			(0.8959499999999999, 0.034831160305343514)
			(0.9008999999999999, 0.03551413026052104)
			(0.9058499999999999, 0.03623725316455697)
			(0.9107999999999999, 0.03699835189309577)
			(0.91575, 0.03780611556603773)
			(0.9207, 0.038636285)
			(0.9256499999999999, 0.039563848000000006)
			(0.9305999999999999, 0.04058012285714285)
			(0.9355499999999999, 0.041699643076923074)
			(0.9404999999999999, 0.042933723333333326)
			(0.9454499999999999, 0.044289650909090914)
			(0.9503999999999999, 0.04579364)
			(0.9553499999999999, 0.04745997333333333)
			(0.9602999999999999, 0.049324045)
			(0.9652499999999999, 0.05140320571428571)
			(0.9702, 0.05381444666666667)
			(0.97515, 0.056579984126984124)
			(0.9800999999999999, 0.06020991089108911)
			(0.9850499999999999, 0.06504493421052632)
		};
		\addlegendentry{$M_j(p)$}
		\addplot[dashdotted, red] coordinates {
			(0.0, -0.0012851315717138264)
			(0.0049499999999999995, -0.0008774819596049358)
			(0.009899999999999999, -0.0005740105763868882)
			(0.014849999999999999, -0.00030380163011653125)
			(0.019799999999999998, -4.946984683986295e-05)
			(0.024749999999999998, 0.00019325205590112163)
			(0.029699999999999997, 0.00042513464730904585)
			(0.03465, 0.0006551277656298368)
			(0.039599999999999996, 0.0008775155618620994)
			(0.04454999999999999, 0.0010990354786062442)
			(0.049499999999999995, 0.0013194583171934292)
			(0.05445, 0.0015198471078333223)
			(0.059399999999999994, 0.001729247455707733)
			(0.06434999999999999, 0.0019417758690491017)
			(0.0693, 0.0021482013265276254)
			(0.07425, 0.0023506814363119287)
			(0.07919999999999999, 0.0025571120145300576)
			(0.08414999999999999, 0.002755248630770624)
			(0.08909999999999998, 0.0029605484145777063)
			(0.09405, 0.0031723039045512447)
			(0.09899999999999999, 0.0033649696317270125)
			(0.10394999999999999, 0.003567437020655676)
			(0.1089, 0.003761358485310126)
			(0.11384999999999999, 0.0039685721922787685)
			(0.11879999999999999, 0.004174810567754187)
			(0.12374999999999999, 0.00437059430719464)
			(0.12869999999999998, 0.004571420326486231)
			(0.13365, 0.004771577974652243)
			(0.1386, 0.0049662351474726064)
			(0.14354999999999998, 0.005154635880840829)
			(0.1485, 0.005343791012410422)
			(0.15344999999999998, 0.005521630540629656)
			(0.15839999999999999, 0.005693285410668968)
			(0.16335, 0.005853234319693959)
			(0.16829999999999998, 0.006020876738931343)
			(0.17325, 0.006167089749350288)
			(0.17819999999999997, 0.00631597647156328)
			(0.18314999999999998, 0.006469236433239574)
			(0.1881, 0.0066104046945007705)
			(0.19304999999999997, 0.006768288639226787)
			(0.19799999999999998, 0.0069064338346437855)
			(0.20295, 0.007048458311918423)
			(0.20789999999999997, 0.007189120575525117)
			(0.21284999999999998, 0.0073158121206539405)
			(0.2178, 0.007446251206564496)
			(0.22274999999999998, 0.007561906805396419)
			(0.22769999999999999, 0.007696255508233764)
			(0.23264999999999997, 0.007798215876345486)
			(0.23759999999999998, 0.007919429730168308)
			(0.24255, 0.008040158326081015)
			(0.24749999999999997, 0.008147143761634217)
			(0.25244999999999995, 0.008248112886785978)
			(0.25739999999999996, 0.008362007598856209)
			(0.26234999999999997, 0.00847525831733536)
			(0.2673, 0.00859654483323939)
			(0.27225, 0.008692305259688343)
			(0.2772, 0.008797071239262098)
			(0.28214999999999996, 0.008905096594527562)
			(0.28709999999999997, 0.009030086579828423)
			(0.29205, 0.00914426404405788)
			(0.297, 0.009250039200833816)
			(0.30195, 0.00936189752195238)
			(0.30689999999999995, 0.009479053096847816)
			(0.31184999999999996, 0.009580027711826788)
			(0.31679999999999997, 0.009691900287902238)
			(0.32175, 0.009792859821388465)
			(0.3267, 0.009886334607161726)
			(0.33164999999999994, 0.0099795936502666)
			(0.33659999999999995, 0.010093879859876411)
			(0.34154999999999996, 0.010195859045005671)
			(0.3465, 0.0102962318506597)
			(0.35145, 0.010403954867958934)
			(0.35639999999999994, 0.01051454781212995)
			(0.36134999999999995, 0.010631099711060979)
			(0.36629999999999996, 0.010736923148971872)
			(0.37124999999999997, 0.010839273789931176)
			(0.3762, 0.010943670185494268)
			(0.38115, 0.011050183560099593)
			(0.38609999999999994, 0.011160804400462214)
			(0.39104999999999995, 0.011275109101448525)
			(0.39599999999999996, 0.011371742784710164)
			(0.40095, 0.01145928775806647)
			(0.4059, 0.011570458516464651)
			(0.41084999999999994, 0.011674394066080052)
			(0.41579999999999995, 0.011799065774644157)
			(0.42074999999999996, 0.011907126517822782)
			(0.42569999999999997, 0.012029118433164537)
			(0.43065, 0.01213046372097812)
			(0.4356, 0.012247199880800884)
			(0.44054999999999994, 0.012354946244838709)
			(0.44549999999999995, 0.012457200811818182)
			(0.45044999999999996, 0.01256791781209473)
			(0.45539999999999997, 0.012670706062227392)
			(0.46035, 0.01278172200022628)
			(0.46529999999999994, 0.012893496053933777)
			(0.47024999999999995, 0.01300262216402559)
			(0.47519999999999996, 0.013118816905003779)
			(0.48014999999999997, 0.013246918691862132)
			(0.4851, 0.013357689066627453)
			(0.49004999999999993, 0.01345966362005373)
			(0.49499999999999994, 0.013578769934712813)
			(0.49994999999999995, 0.013691368596707732)
			(0.5048999999999999, 0.013833722229709761)
			(0.5098499999999999, 0.013958565459632245)
			(0.5147999999999999, 0.014069261662684919)
			(0.5197499999999999, 0.014204583238498852)
			(0.5246999999999999, 0.014315661296128965)
			(0.52965, 0.014435283447097979)
			(0.5346, 0.014541774969775484)
			(0.53955, 0.014644146924162547)
			(0.5445, 0.014765308722421206)
			(0.54945, 0.014901153155542751)
			(0.5544, 0.015038711723769822)
			(0.5593499999999999, 0.015160130848945796)
			(0.5642999999999999, 0.01528210047201318)
			(0.5692499999999999, 0.015414392006185348)
			(0.5741999999999999, 0.01553951033007738)
			(0.5791499999999999, 0.0156696263468768)
			(0.5841, 0.015807655985305338)
			(0.58905, 0.015949226742694597)
			(0.594, 0.01608626905205767)
			(0.59895, 0.016196474646288372)
			(0.6039, 0.016354200234435453)
			(0.6088499999999999, 0.016500973008317276)
			(0.6137999999999999, 0.016655441069644575)
			(0.6187499999999999, 0.016808301011881417)
			(0.6236999999999999, 0.016960530149812674)
			(0.6286499999999999, 0.017135357962500864)
			(0.6335999999999999, 0.017315695380057342)
			(0.63855, 0.01749262475929688)
			(0.6435, 0.017660629843074312)
			(0.64845, 0.017856120735976127)
			(0.6534, 0.018013851495241096)
			(0.65835, 0.01818678085310015)
			(0.6632999999999999, 0.01835464281379113)
			(0.6682499999999999, 0.01852363246538495)
			(0.6731999999999999, 0.01872440470638087)
			(0.6781499999999999, 0.01890085006993857)
			(0.6830999999999999, 0.019059199770733323)
			(0.6880499999999999, 0.019220729562129918)
			(0.693, 0.019423530829334006)
			(0.69795, 0.019612679738971885)
			(0.7029, 0.01978930456270164)
			(0.70785, 0.019983169338701864)
			(0.7127999999999999, 0.020157611123849825)
			(0.7177499999999999, 0.020389322248180946)
			(0.7226999999999999, 0.020611375030223898)
			(0.7276499999999999, 0.020813223978065894)
			(0.7325999999999999, 0.02102486267327932)
			(0.7375499999999999, 0.02123825312557939)
			(0.7424999999999999, 0.021469588746937236)
			(0.74745, 0.02169248845592596)
			(0.7524, 0.021933902869740906)
			(0.75735, 0.022180311708882523)
			(0.7623, 0.022423217168761743)
			(0.7672499999999999, 0.022679566615865513)
			(0.7721999999999999, 0.02295503460809785)
			(0.7771499999999999, 0.023217908602610603)
			(0.7820999999999999, 0.0234376315114578)
			(0.7870499999999999, 0.023683202581552987)
			(0.7919999999999999, 0.02404490596969133)
			(0.7969499999999999, 0.02434219643316672)
			(0.8019, 0.024614905489376546)
			(0.80685, 0.02488173986593637)
			(0.8118, 0.025131454056903462)
			(0.81675, 0.025480014183293195)
			(0.8216999999999999, 0.025725310079625564)
			(0.8266499999999999, 0.02603028905293481)
			(0.8315999999999999, 0.02639017760742865)
			(0.8365499999999999, 0.026708191612691085)
			(0.8414999999999999, 0.02707751406247297)
			(0.8464499999999999, 0.02739959677623179)
			(0.8513999999999999, 0.027740995567340335)
			(0.85635, 0.028033827060966625)
			(0.8613, 0.02847280276242625)
			(0.86625, 0.02888708890785802)
			(0.8712, 0.029272317128475746)
			(0.8761499999999999, 0.02971223707724698)
			(0.8810999999999999, 0.03023212142993335)
			(0.8860499999999999, 0.030753107929882357)
			(0.8909999999999999, 0.03116067222184132)
			(0.8959499999999999, 0.03176291634993689)
			(0.9008999999999999, 0.03237992660412454)
			(0.9058499999999999, 0.03299374744495566)
			(0.9107999999999999, 0.03366240100186546)
			(0.91575, 0.034331132110799904)
			(0.9207, 0.035093100695897085)
			(0.9256499999999999, 0.03577331616174908)
			(0.9305999999999999, 0.0366339005416051)
			(0.9355499999999999, 0.03747659737701029)
			(0.9404999999999999, 0.03854345908147895)
			(0.9454499999999999, 0.03977875292931865)
			(0.9503999999999999, 0.041240534462205124)
			(0.9553499999999999, 0.042658917894034046)
			(0.9602999999999999, 0.04410059700151157)
			(0.9652499999999999, 0.046122118691006984)
			(0.9702, 0.048155427046334215)
			(0.97515, 0.0505873470875113)
			(0.9800999999999999, 0.05376428463602966)
			(0.9850499999999999, 0.05758296559763544)
		};
		\addlegendentry{$M^f_j(p)$}
		\addplot[dash pattern=on 8pt off 2pt] coordinates {
			(0.0, -0.0012817211844197139)
			(0.0049499999999999995, -0.0010111238019169329)
			(0.009899999999999999, -0.0008127696166967689)
			(0.014849999999999999, -0.0006671202097620008)
			(0.019799999999999998, -0.0005588801946077438)
			(0.024749999999999998, -0.00047186552567237165)
			(0.029699999999999997, -0.0003703116936309646)
			(0.03465, -0.0002845409633594071)
			(0.039599999999999996, -0.00018935092075315544)
			(0.04454999999999999, -0.0001296763727121464)
			(0.049499999999999995, -2.229291239807656e-05)
			(0.05445, 6.171086362681237e-05)
			(0.059399999999999994, 0.00016317680608365017)
			(0.06434999999999999, 0.00024483775748566575)
			(0.0693, 0.00034304654141759187)
			(0.07425, 0.0004416024898046791)
			(0.07919999999999999, 0.0005111603366422097)
			(0.08414999999999999, 0.0006125239748318507)
			(0.08909999999999998, 0.0007127253490401394)
			(0.09405, 0.0007952809826716385)
			(0.09899999999999999, 0.0008468081164534625)
			(0.10394999999999999, 0.0009272541583499668)
			(0.1089, 0.0010129433667781492)
			(0.11384999999999999, 0.0010814535874439461)
			(0.11879999999999999, 0.001136392559188275)
			(0.12374999999999999, 0.0012222639455782313)
			(0.12869999999999998, 0.001321473432155074)
			(0.13365, 0.0013994582568807336)
			(0.1386, 0.0014994297577854672)
			(0.14354999999999998, 0.0015592484918793505)
			(0.1485, 0.0016379724620770126)
			(0.15344999999999998, 0.0017081145539906103)
			(0.15839999999999999, 0.0017619090909090909)
			(0.16335, 0.0018445504630729044)
			(0.16829999999999998, 0.0019027181079789776)
			(0.17325, 0.0019667904349915883)
			(0.17819999999999997, 0.0020008232591876207)
			(0.18314999999999998, 0.0020558628071028944)
			(0.1881, 0.0021131997063142433)
			(0.19304999999999997, 0.0021472051218911596)
			(0.19799999999999998, 0.0021672594152626363)
			(0.20295, 0.00222259212166542)
			(0.20789999999999997, 0.002280811088810838)
			(0.21284999999999998, 0.002312134561979298)
			(0.2178, 0.0023695760731521466)
			(0.22274999999999998, 0.0024177185582822083)
			(0.22769999999999999, 0.002483097247234371)
			(0.23264999999999997, 0.0025191566545831176)
			(0.23759999999999998, 0.0025343776387802973)
			(0.24255, 0.0026130183630640085)
			(0.24749999999999997, 0.0026590277264325326)
			(0.25244999999999995, 0.0026984303561935143)
			(0.25739999999999996, 0.0027459531709927753)
			(0.26234999999999997, 0.00280213442887931)
			(0.2673, 0.0028365104420938437)
			(0.27225, 0.0028806188916188917)
			(0.2772, 0.002944803738317757)
			(0.28214999999999996, 0.0029926346526432325)
			(0.28709999999999997, 0.0030119885730211815)
			(0.29205, 0.003050285433623351)
			(0.297, 0.003109592990390051)
			(0.30195, 0.003173391972672929)
			(0.30689999999999995, 0.0032128514908256883)
			(0.31184999999999996, 0.0032834094715564538)
			(0.31679999999999997, 0.0033335759162303664)
			(0.32175, 0.003422998828010548)
			(0.3267, 0.0034808483328415464)
			(0.33164999999999994, 0.003556106718192628)
			(0.33659999999999995, 0.0035987532195268044)
			(0.34154999999999996, 0.0036682320458660232)
			(0.3465, 0.0037201584068105808)
			(0.35145, 0.0037444849877450977)
			(0.35639999999999994, 0.0037961673355974067)
			(0.36134999999999995, 0.003822032669570629)
			(0.36629999999999996, 0.0038630768265914073)
			(0.37124999999999997, 0.00394378982300885)
			(0.3762, 0.00397969321439949)
			(0.38115, 0.004041964686998394)
			(0.38609999999999994, 0.004146021035598706)
			(0.39104999999999995, 0.004174494616639478)
			(0.39599999999999996, 0.004236776973684211)
			(0.40095, 0.00430230447761194)
			(0.4059, 0.004348340468227425)
			(0.41084999999999994, 0.004432902866779089)
			(0.41579999999999995, 0.004513744217687075)
			(0.42074999999999996, 0.004572354030874786)
			(0.42569999999999997, 0.004593570588235294)
			(0.43065, 0.004610264921465969)
			(0.4356, 0.004669318549806406)
			(0.44054999999999994, 0.004743102982954546)
			(0.44549999999999995, 0.00479963525618058)
			(0.45044999999999996, 0.004848532899493854)
			(0.45539999999999997, 0.004926303538854433)
			(0.46035, 0.004930915316642121)
			(0.46529999999999994, 0.005021317725752508)
			(0.47024999999999995, 0.005112373593398349)
			(0.47519999999999996, 0.005160814085573646)
			(0.48014999999999997, 0.005182875)
			(0.4851, 0.00523660401389425)
			(0.49004999999999993, 0.005337636540708999)
			(0.49499999999999994, 0.005480125885129819)
			(0.49994999999999995, 0.0055449384187524825)
			(0.5048999999999999, 0.005669705457463885)
			(0.5098499999999999, 0.005699224564248075)
			(0.5147999999999999, 0.0057681433251433245)
			(0.5197499999999999, 0.00582703103020273)
			(0.5246999999999999, 0.005881726170568562)
			(0.52965, 0.005932932826362485)
			(0.5346, 0.006050624252775406)
			(0.53955, 0.0060964164868364265)
			(0.5445, 0.006160386829481029)
			(0.54945, 0.006215186507936509)
			(0.5544, 0.0062659144003566645)
			(0.5593499999999999, 0.006420132100991884)
			(0.5642999999999999, 0.006552882808937529)
			(0.5692499999999999, 0.006663593173431735)
			(0.5741999999999999, 0.006794760615958935)
			(0.5791499999999999, 0.006852374881964117)
			(0.5841, 0.006877668418537984)
			(0.58905, 0.007018806092843327)
			(0.594, 0.0070842927068037196)
			(0.59895, 0.00717345765230312)
			(0.6039, 0.007269955867602809)
			(0.6088499999999999, 0.007341422041645505)
			(0.6137999999999999, 0.00753832098765432)
			(0.6187499999999999, 0.007703097446586764)
			(0.6236999999999999, 0.0077885116156283)
			(0.6286499999999999, 0.007920521669341894)
			(0.6335999999999999, 0.00811872830802603)
			(0.63855, 0.008168346893897746)
			(0.6435, 0.008320273690078038)
			(0.64845, 0.008422994347088751)
			(0.6534, 0.008509692263610315)
			(0.65835, 0.008685226744186045)
			(0.6632999999999999, 0.008861332153392331)
			(0.6682499999999999, 0.008965643113772456)
			(0.6731999999999999, 0.009053849240121582)
			(0.6781499999999999, 0.009243377160493827)
			(0.6830999999999999, 0.009379774921630094)
			(0.6880499999999999, 0.009524388535031847)
			(0.693, 0.009723662783171522)
			(0.69795, 0.0098744125)
			(0.7029, 0.010082423128342245)
			(0.70785, 0.010256488783140722)
			(0.7127999999999999, 0.010432188105117566)
			(0.7177499999999999, 0.010577746657283603)
			(0.7226999999999999, 0.010677805157593123)
			(0.7276499999999999, 0.010860089715536104)
			(0.7325999999999999, 0.010957546062407134)
			(0.7375499999999999, 0.011117346707040121)
			(0.7424999999999999, 0.011273554012345678)
			(0.74745, 0.011448274586939419)
			(0.7524, 0.011596640449438204)
			(0.75735, 0.011872647299509003)
			(0.7623, 0.012094000835421889)
			(0.7672499999999999, 0.012271295221843004)
			(0.7721999999999999, 0.012593159546643418)
			(0.7771499999999999, 0.012781868092691624)
			(0.7820999999999999, 0.012929179580674568)
			(0.7870499999999999, 0.013185109141791043)
			(0.7919999999999999, 0.013421468958930278)
			(0.7969499999999999, 0.01350761350293542)
			(0.8019, 0.013713777331995988)
			(0.80685, 0.014056701646090535)
			(0.8118, 0.014347456751054853)
			(0.81675, 0.014573598049837487)
			(0.8216999999999999, 0.014882678173719376)
			(0.8266499999999999, 0.015053632302405499)
			(0.8315999999999999, 0.01524915212264151)
			(0.8365499999999999, 0.015626228432563793)
			(0.8414999999999999, 0.016046244360902254)
			(0.8464499999999999, 0.016242730918499357)
			(0.8513999999999999, 0.01668236229946524)
			(0.85635, 0.01695465836791148)
			(0.8613, 0.017373393982808023)
			(0.86625, 0.017781108308605343)
			(0.8712, 0.01823542681047766)
			(0.8761499999999999, 0.018647820512820514)
			(0.8810999999999999, 0.019132641068447413)
			(0.8860499999999999, 0.019839655052264808)
			(0.8909999999999999, 0.020065000000000003)
			(0.8959499999999999, 0.02032029198473282)
			(0.9008999999999999, 0.020821597194388777)
			(0.9058499999999999, 0.021307900843881858)
			(0.9107999999999999, 0.02196467706013363)
			(0.91575, 0.022613341981132073)
			(0.9207, 0.023472517499999998)
			(0.9256499999999999, 0.024149312)
			(0.9305999999999999, 0.025053037142857143)
			(0.9355499999999999, 0.025525870769230772)
			(0.9404999999999999, 0.02662403333333333)
			(0.9454499999999999, 0.027725203636363636)
			(0.9503999999999999, 0.028310188)
			(0.9553499999999999, 0.029132577777777776)
			(0.9602999999999999, 0.03002025)
			(0.9652499999999999, 0.031501811428571426)
			(0.9702, 0.03412186666666666)
			(0.97515, 0.03663338888888889)
			(0.9800999999999999, 0.0388539504950495)
			(0.9850499999999999, 0.04057242105263157)
		};
		\addlegendentry{$\mathrm{MES}_p(X_j,S)$}
		\addplot[dotted] coordinates {
			(0.0, -0.001335739232496807)
			(0.0049499999999999995, -0.001430796010806973)
			(0.009899999999999999, -0.0015834006077815447)
			(0.014849999999999999, -0.0015463946549840787)
			(0.019799999999999998, -0.0015816938775596574)
			(0.024749999999999998, -0.001652060600680196)
			(0.029699999999999997, -0.0016789789978154414)
			(0.03465, -0.0017433627538295002)
			(0.039599999999999996, -0.0017762277386532024)
			(0.04454999999999999, -0.0018349145150047856)
			(0.049499999999999995, -0.0018666148327956573)
			(0.05445, -0.001934531689148556)
			(0.059399999999999994, -0.001992228982166902)
			(0.06434999999999999, -0.0020191851630478506)
			(0.0693, -0.0020593441012653983)
			(0.07425, -0.002129965618607648)
			(0.07919999999999999, -0.0022058436553577298)
			(0.08414999999999999, -0.00225528872429345)
			(0.08909999999999998, -0.002283751132564626)
			(0.09405, -0.0023017150657027043)
			(0.09899999999999999, -0.0023469264023582223)
			(0.10394999999999999, -0.002407553806928051)
			(0.1089, -0.0024686843689012417)
			(0.11384999999999999, -0.0025264064800049643)
			(0.11879999999999999, -0.0025899146220176826)
			(0.12374999999999999, -0.002680062827759779)
			(0.12869999999999998, -0.0026945120063746712)
			(0.13365, -0.0027760207614579892)
			(0.1386, -0.002871398196728887)
			(0.14354999999999998, -0.0029429068477727868)
			(0.1485, -0.003002711097429499)
			(0.15344999999999998, -0.003070191665350964)
			(0.15839999999999999, -0.003145627068124828)
			(0.16335, -0.003209725648960384)
			(0.16829999999999998, -0.0032738538122199035)
			(0.17325, -0.0033153410594313182)
			(0.17819999999999997, -0.003330877385595289)
			(0.18314999999999998, -0.0033432507181931494)
			(0.1881, -0.003385614431099033)
			(0.19304999999999997, -0.003400876619210723)
			(0.19799999999999998, -0.0034233472320557433)
			(0.20295, -0.003443490767539802)
			(0.20789999999999997, -0.003494900902923847)
			(0.21284999999999998, -0.003538500663120418)
			(0.2178, -0.0035743090827491317)
			(0.22274999999999998, -0.003574864913396532)
			(0.22769999999999999, -0.0035857551294397705)
			(0.23264999999999997, -0.00361654753257827)
			(0.23759999999999998, -0.003639382758714911)
			(0.24255, -0.0036370680049761187)
			(0.24749999999999997, -0.0036342394640126985)
			(0.25244999999999995, -0.003691940171424679)
			(0.25739999999999996, -0.0036907314236067396)
			(0.26234999999999997, -0.003749961868791213)
			(0.2673, -0.003735103487259118)
			(0.27225, -0.0037394512365636103)
			(0.2772, -0.003754984458331747)
			(0.28214999999999996, -0.003771413916707309)
			(0.28709999999999997, -0.003755729605050747)
			(0.29205, -0.0037995474255634694)
			(0.297, -0.0038222975298865224)
			(0.30195, -0.0038196233953880694)
			(0.30689999999999995, -0.0038350920073624306)
			(0.31184999999999996, -0.0037734174964455525)
			(0.31679999999999997, -0.003730538430223908)
			(0.32175, -0.003738697051876134)
			(0.3267, -0.003729465442972329)
			(0.33164999999999994, -0.0037217271230090047)
			(0.33659999999999995, -0.0036900825881338007)
			(0.34154999999999996, -0.003694241371181189)
			(0.3465, -0.0036500670786000977)
			(0.35145, -0.003685280205514958)
			(0.35639999999999994, -0.0037025104944431054)
			(0.36134999999999995, -0.0036928798560812524)
			(0.36629999999999996, -0.003689102248615666)
			(0.37124999999999997, -0.003715171216705759)
			(0.3762, -0.003748136888303023)
			(0.38115, -0.0038049586204555444)
			(0.38609999999999994, -0.0037471274864421156)
			(0.39104999999999995, -0.003750551586631755)
			(0.39599999999999996, -0.0037398318501282737)
			(0.40095, -0.003760398192567056)
			(0.4059, -0.0037598896658829653)
			(0.41084999999999994, -0.003758258190224115)
			(0.41579999999999995, -0.0037445883383148194)
			(0.42074999999999996, -0.003763286271774742)
			(0.42569999999999997, -0.003714968651531603)
			(0.43065, -0.0036830845682318146)
			(0.4356, -0.0037091127081449317)
			(0.44054999999999994, -0.003731192052777622)
			(0.44549999999999995, -0.003710717887509276)
			(0.45044999999999996, -0.0036872274142986973)
			(0.45539999999999997, -0.0036480068199358697)
			(0.46035, -0.0036378197475896564)
			(0.46529999999999994, -0.0035977122693911143)
			(0.47024999999999995, -0.0035925373071792463)
			(0.47519999999999996, -0.0035890042180524214)
			(0.48014999999999997, -0.003554070177763821)
			(0.4851, -0.0035761372542531937)
			(0.49004999999999993, -0.003588483887112765)
			(0.49499999999999994, -0.0036006845582881803)
			(0.49994999999999995, -0.003636552386917569)
			(0.5048999999999999, -0.0036227385302409444)
			(0.5098499999999999, -0.003648051569224247)
			(0.5147999999999999, -0.003680480257091281)
			(0.5197499999999999, -0.0037053785956961934)
			(0.5246999999999999, -0.0037142670820068616)
			(0.52965, -0.003684030967705758)
			(0.5346, -0.0036836709182394758)
			(0.53955, -0.0037210679659281077)
			(0.5445, -0.0037205623823548576)
			(0.54945, -0.0037008628727521783)
			(0.5544, -0.003629067839450194)
			(0.5593499999999999, -0.0036349960737710467)
			(0.5642999999999999, -0.003672960537209429)
			(0.5692499999999999, -0.0037071985767095947)
			(0.5741999999999999, -0.0036688042677241297)
			(0.5791499999999999, -0.003637206074755508)
			(0.5841, -0.0035846755223885593)
			(0.58905, -0.003559140251980862)
			(0.594, -0.0035465705813467812)
			(0.59895, -0.0034623454172862016)
			(0.6039, -0.003448648119442421)
			(0.6088499999999999, -0.003354492739187217)
			(0.6137999999999999, -0.0033795847585329115)
			(0.6187499999999999, -0.0034165421046778717)
			(0.6236999999999999, -0.0033421657994760264)
			(0.6286499999999999, -0.003328156929180092)
			(0.6335999999999999, -0.003363400823811365)
			(0.63855, -0.0032713878419421325)
			(0.6435, -0.0033272421697770173)
			(0.64845, -0.0033992593661814553)
			(0.6534, -0.003406425460955626)
			(0.65835, -0.003481769647990392)
			(0.6632999999999999, -0.003529941285218777)
			(0.6682499999999999, -0.003543362312586438)
			(0.6731999999999999, -0.003581690334082023)
			(0.6781499999999999, -0.0035222841142860425)
			(0.6830999999999999, -0.003533422268458928)
			(0.6880499999999999, -0.0035274061599106696)
			(0.693, -0.003449643317850236)
			(0.69795, -0.003433599170445574)
			(0.7029, -0.003538201606992977)
			(0.70785, -0.003590577626180405)
			(0.7127999999999999, -0.003642643087486863)
			(0.7177499999999999, -0.003541283356463241)
			(0.7226999999999999, -0.0034743067241591306)
			(0.7276499999999999, -0.0033886411848733228)
			(0.7325999999999999, -0.003226106133771263)
			(0.7375499999999999, -0.00322019647666983)
			(0.7424999999999999, -0.003223948866693631)
			(0.74745, -0.0032135394169376754)
			(0.7524, -0.003120146935357697)
			(0.75735, -0.00310450170112465)
			(0.7623, -0.003137510808110499)
			(0.7672499999999999, -0.0032072029369343545)
			(0.7721999999999999, -0.003119930308931083)
			(0.7771499999999999, -0.0029614303088610545)
			(0.7820999999999999, -0.002783750977891464)
			(0.7870499999999999, -0.002717043598358695)
			(0.7919999999999999, -0.002593632784269569)
			(0.7969499999999999, -0.00254920206880732)
			(0.8019, -0.002349492047898737)
			(0.80685, -0.0022890701688064135)
			(0.8118, -0.0022277484634888327)
			(0.81675, -0.002205883940823687)
			(0.8216999999999999, -0.0019949316242106187)
			(0.8266499999999999, -0.0016971188466778598)
			(0.8315999999999999, -0.0015498755451282113)
			(0.8365499999999999, -0.0014738034660973164)
			(0.8414999999999999, -0.0011980964961367609)
			(0.8464499999999999, -0.0012081604844498594)
			(0.8513999999999999, -0.0013095172548899063)
			(0.85635, -0.0014060957667686265)
			(0.8613, -0.0013573459177752905)
			(0.86625, -0.0009891411754666018)
			(0.8712, -0.0006602937479002481)
			(0.8761499999999999, -1.1112966267760693e-05)
			(0.8810999999999999, -0.00014318654285824938)
			(0.8860499999999999, -9.767109534566449e-05)
			(0.8909999999999999, 0.00011352588280697129)
			(0.8959499999999999, 0.00015544386226109863)
			(0.9008999999999999, 0.0006549787435786422)
			(0.9058499999999999, 0.0008147151733237937)
			(0.9107999999999999, 0.0009379578299769923)
			(0.91575, 0.0012951345236053684)
			(0.9207, 0.001630192109210773)
			(0.9256499999999999, 0.0018488571454509733)
			(0.9305999999999999, 0.0020433413567686894)
			(0.9355499999999999, 0.0026024838038768375)
			(0.9404999999999999, 0.0024305780675086225)
			(0.9454499999999999, 0.002504907965441493)
			(0.9503999999999999, 0.002357604008451233)
			(0.9553499999999999, 0.0031389067096455766)
			(0.9602999999999999, 0.0033012602997371633)
			(0.9652499999999999, 0.0036917251083591423)
			(0.9702, 0.002124000154899992)
			(0.97515, 0.005185213556803558)
			(0.9800999999999999, 0.007090830381406204)
			(0.9850499999999999, 0.0052254042201290195)
		};
		\addlegendentry{$m^f_j(p)$}
		\addplot[dashed, green] coordinates {
			(0.0, -0.001309094178422412)
			(0.0049499999999999995, -0.0017449604632587858)
			(0.009899999999999999, -0.002049878185831828)
			(0.014849999999999999, -0.0023163507462686563)
			(0.019799999999999998, -0.0025595755118589096)
			(0.024749999999999998, -0.0027866526079869604)
			(0.029699999999999997, -0.003002424329305754)
			(0.03465, -0.003207952243721696)
			(0.039599999999999996, -0.003404403062280157)
			(0.04454999999999999, -0.0035913404742096504)
			(0.049499999999999995, -0.0037710811206355843)
			(0.05445, -0.003937729565034671)
			(0.059399999999999994, -0.004106769750739332)
			(0.06434999999999999, -0.004271511998301126)
			(0.0693, -0.004433403928266439)
			(0.07425, -0.0045932762395363815)
			(0.07919999999999999, -0.00475155977557186)
			(0.08414999999999999, -0.00490745584725537)
			(0.08909999999999998, -0.005061285122164048)
			(0.09405, -0.005212794253125686)
			(0.09899999999999999, -0.005361760476400529)
			(0.10394999999999999, -0.005509026169882457)
			(0.1089, -0.005648477814938684)
			(0.11384999999999999, -0.005791396412556053)
			(0.11879999999999999, -0.005932615332581736)
			(0.12374999999999999, -0.006071570975056689)
			(0.12869999999999998, -0.006208654275940707)
			(0.13365, -0.00634479495412844)
			(0.1386, -0.006479481430219146)
			(0.14354999999999998, -0.006612105800464037)
			(0.1485, -0.006743260443407235)
			(0.15344999999999998, -0.006873235211267606)
			(0.15839999999999999, -0.00700203329397875)
			(0.16335, -0.007124885775350274)
			(0.16829999999999998, -0.0072511715241280455)
			(0.17325, -0.007376378514780101)
			(0.17819999999999997, -0.007500212282398453)
			(0.18314999999999998, -0.007622652882510339)
			(0.1881, -0.007744152961331375)
			(0.19304999999999997, -0.007864679389312977)
			(0.19799999999999998, -0.007984627601585728)
			(0.20295, -0.008104357516828722)
			(0.20789999999999997, -0.008223790767686904)
			(0.21284999999999998, -0.008342909366321637)
			(0.2178, -0.008456861823723648)
			(0.22274999999999998, -0.008574797801635992)
			(0.22769999999999999, -0.008692141497298689)
			(0.23264999999999997, -0.008804008024851153)
			(0.23759999999999998, -0.008920473423658155)
			(0.24255, -0.009041548006295906)
			(0.24749999999999997, -0.009157644837602324)
			(0.25244999999999995, -0.009273505847953214)
			(0.25739999999999996, -0.009389193738292749)
			(0.26234999999999997, -0.009504782327586206)
			(0.2673, -0.009620684838622186)
			(0.27225, -0.009732047502047501)
			(0.2772, -0.00984311788953009)
			(0.28214999999999996, -0.009963286742319402)
			(0.28709999999999997, -0.010074314293675118)
			(0.29205, -0.010195276452427729)
			(0.297, -0.010311865178066704)
			(0.30195, -0.010428768004554513)
			(0.30689999999999995, -0.010545589736238533)
			(0.31184999999999996, -0.010662788045047647)
			(0.31679999999999997, -0.010780722222222221)
			(0.32175, -0.010899060357456783)
			(0.3267, -0.011013087636470936)
			(0.33164999999999994, -0.011131729488703924)
			(0.33659999999999995, -0.011250751123090746)
			(0.34154999999999996, -0.011370098672299333)
			(0.3465, -0.011489818485861964)
			(0.35145, -0.011610214767156863)
			(0.35639999999999994, -0.011731267983945665)
			(0.36134999999999995, -0.01185251773490977)
			(0.36629999999999996, -0.011974212292254628)
			(0.37124999999999997, -0.012096390960809103)
			(0.3762, -0.01221912392481682)
			(0.38115, -0.01233763242375602)
			(0.38609999999999994, -0.012461518122977345)
			(0.39104999999999995, -0.012586122349102773)
			(0.39599999999999996, -0.012711013157894737)
			(0.40095, -0.012831326923076924)
			(0.4059, -0.012962642474916389)
			(0.41084999999999994, -0.013089821922428331)
			(0.41579999999999995, -0.013217661904761905)
			(0.42074999999999996, -0.013345754373927958)
			(0.42569999999999997, -0.013474738062283735)
			(0.43065, -0.013604463874345548)
			(0.4356, -0.01372955403027103)
			(0.44054999999999994, -0.013860709161931817)
			(0.44549999999999995, -0.013987505730659027)
			(0.45044999999999996, -0.014125601229211858)
			(0.45539999999999997, -0.014259133892739875)
			(0.46035, -0.01439351914580265)
			(0.46529999999999994, -0.014528756967670012)
			(0.47024999999999995, -0.014665075768942238)
			(0.47519999999999996, -0.014802699734948884)
			(0.48014999999999997, -0.01494114640672783)
			(0.4851, -0.015080661906599768)
			(0.49004999999999993, -0.015215712504869498)
			(0.49499999999999994, -0.015357259244689224)
			(0.49994999999999995, -0.015499685339690109)
			(0.5048999999999999, -0.015643208667736757)
			(0.5098499999999999, -0.01578793190109445)
			(0.5147999999999999, -0.015927635284486284)
			(0.5197499999999999, -0.016080434836574266)
			(0.5246999999999999, -0.016229019648829432)
			(0.52965, -0.016378874524714825)
			(0.5346, -0.016530074295473957)
			(0.53955, -0.016682522227017695)
			(0.5445, -0.01682994025294374)
			(0.54945, -0.01698476543209877)
			(0.5544, -0.017141129291127953)
			(0.5593499999999999, -0.01729880117222723)
			(0.5642999999999999, -0.017457593707250342)
			(0.5692499999999999, -0.01761813699261993)
			(0.5741999999999999, -0.017780489500699954)
			(0.5791499999999999, -0.017944050047214354)
			(0.5841, -0.01810893454371715)
			(0.58905, -0.01827540232108317)
			(0.594, -0.018443859520313268)
			(0.59895, -0.018606936602278357)
			(0.6039, -0.01877841925777332)
			(0.6088499999999999, -0.01895168054850178)
			(0.6137999999999999, -0.01912671502057613)
			(0.6187499999999999, -0.019303876498176133)
			(0.6236999999999999, -0.019482303590285112)
			(0.6286499999999999, -0.019663309256286787)
			(0.6335999999999999, -0.019846542841648593)
			(0.63855, -0.020032472787245743)
			(0.6435, -0.02022188573021182)
			(0.64845, -0.020414288863764837)
			(0.6534, -0.02060192034383954)
			(0.65835, -0.02080003720930233)
			(0.6632999999999999, -0.021000985250737465)
			(0.6682499999999999, -0.021205137125748504)
			(0.6731999999999999, -0.021411707598784194)
			(0.6781499999999999, -0.0216215450617284)
			(0.6830999999999999, -0.021835094043887147)
			(0.6880499999999999, -0.022052487261146496)
			(0.693, -0.022273471844660196)
			(0.69795, -0.02249742039473684)
			(0.7029, -0.02271602473262032)
			(0.70785, -0.022945651937457512)
			(0.7127999999999999, -0.023178165975103738)
			(0.7177499999999999, -0.023414174524982404)
			(0.7226999999999999, -0.02365278581661891)
			(0.7276499999999999, -0.023896175784099197)
			(0.7325999999999999, -0.024144611441307577)
			(0.7375499999999999, -0.02439701968205905)
			(0.7424999999999999, -0.024653443672839503)
			(0.74745, -0.024913837922895355)
			(0.7524, -0.025178902086677366)
			(0.75735, -0.02543869476268413)
			(0.7623, -0.025715577276524647)
			(0.7672499999999999, -0.025999598976109217)
			(0.7721999999999999, -0.026289101133391456)
			(0.7771499999999999, -0.026583550802139037)
			(0.7820999999999999, -0.02688518687329079)
			(0.7870499999999999, -0.0271955)
			(0.7919999999999999, -0.027513052531041072)
			(0.7969499999999999, -0.02783981409001957)
			(0.8019, -0.028174683049147442)
			(0.80685, -0.028520050411522634)
			(0.8118, -0.02886048734177215)
			(0.81675, -0.029226117009750814)
			(0.8216999999999999, -0.02960555122494432)
			(0.8266499999999999, -0.029994479954180985)
			(0.8315999999999999, -0.030395024764150944)
			(0.8365499999999999, -0.030809317132442286)
			(0.8414999999999999, -0.031237279448621553)
			(0.8464499999999999, -0.031679878395860284)
			(0.8513999999999999, -0.032138528074866314)
			(0.85635, -0.03261623651452282)
			(0.8613, -0.03311662750716332)
			(0.86625, -0.03362225667655786)
			(0.8712, -0.03417266563944531)
			(0.8761499999999999, -0.03474871314102564)
			(0.8810999999999999, -0.03534770784641069)
			(0.8860499999999999, -0.035971890243902446)
			(0.8909999999999999, -0.03662445901639344)
			(0.8959499999999999, -0.03730378625954199)
			(0.9008999999999999, -0.03801472144288578)
			(0.9058499999999999, -0.03876779746835443)
			(0.9107999999999999, -0.03957042538975502)
			(0.91575, -0.04041970990566038)
			(0.9207, -0.04128661)
			(0.9256499999999999, -0.042267080000000005)
			(0.9305999999999999, -0.043316391428571424)
			(0.9355499999999999, -0.044453932307692304)
			(0.9404999999999999, -0.04570719)
			(0.9454499999999999, -0.04706993454545454)
			(0.9503999999999999, -0.048582416)
			(0.9553499999999999, -0.050255484444444436)
			(0.9602999999999999, -0.052156770000000005)
			(0.9652499999999999, -0.05434625142857143)
			(0.9702, -0.056979113333333324)
			(0.97515, -0.06011860317460317)
			(0.9800999999999999, -0.0643110792079208)
			(0.9850499999999999, -0.06986772368421053)
			
		};
		\addlegendentry{$m_j(p)$}
		\addplot [] coordinates {(1,0.075) (1,-0.075)};
		\addplot [] coordinates {(0,0.075) (1,0.075)};
		%				\node[anchor=south, text=blue] at (axis cs:0.4,0.53) {$f_1(U)$};
		%				\node[anchor=south, text=orange] at (axis cs:0.4,0.1) {$f_2(U)$};
	\end{axis}
\end{tikzpicture}

%% file: empirical_bounds_fin_cris.tex
\begin{tikzpicture}
	\begin{axis}[
		xlabel={$p$},
		ylabel={},
		xtick={0.2, 0.4, 0.6, 0.8, 1},
		ytick={-0.35, -0.175, 0, 0.375, 0.65},
		xmin=0, xmax=1.05,
		ymin=-0.4, ymax=0.8,
		axis lines=middle,
		width=0.75\textwidth,
		height=0.5\textwidth,
		xlabel style={below right},
		ylabel style={above left},
		%				axis x line=middle, % Add this line
		axis x line shift=0.4, % Add this line
		legend style={
			at={(-0.15,0.5)}, % position relative to axis
			anchor=east          % anchor legend at its right side
		}
		]
		\addplot[blue] coordinates {
			(0.0, 0.013954305555555555)
			(0.0049499999999999995, 0.01571735856573705)
			(0.009899999999999999, 0.016818048000000002)
			(0.014849999999999999, 0.017600457831325304)
			(0.019799999999999998, 0.0182844314516129)
			(0.024749999999999998, 0.019539341463414635)
			(0.029699999999999997, 0.02014237959183673)
			(0.03465, 0.020657040983606555)
			(0.039599999999999996, 0.02116435390946502)
			(0.04454999999999999, 0.022135556016597505)
			(0.049499999999999995, 0.0225721125)
			(0.05445, 0.023001184100418407)
			(0.059399999999999994, 0.023425663865546217)
			(0.06434999999999999, 0.024272601694915255)
			(0.0693, 0.024689710638297873)
			(0.07425, 0.025108752136752135)
			(0.07919999999999999, 0.025511012875536483)
			(0.08414999999999999, 0.026320112554112547)
			(0.08909999999999998, 0.026724956521739125)
			(0.09405, 0.027130432314410483)
			(0.09899999999999999, 0.027519469298245613)
			(0.10394999999999999, 0.028304132743362832)
			(0.1089, 0.028697248888888888)
			(0.11384999999999999, 0.029084522321428573)
			(0.11879999999999999, 0.029456448430493273)
			(0.12374999999999999, 0.03020997737556561)
			(0.12869999999999998, 0.030590281818181818)
			(0.13365, 0.030968127853881273)
			(0.1386, 0.03134722935779816)
			(0.14354999999999998, 0.03210952314814815)
			(0.1485, 0.03249278604651163)
			(0.15344999999999998, 0.03287928037383178)
			(0.15839999999999999, 0.03325755399061033)
			(0.16335, 0.03400205687203792)
			(0.16829999999999998, 0.03436931904761904)
			(0.17325, 0.0347376985645933)
			(0.17819999999999997, 0.03510886057692308)
			(0.18314999999999998, 0.035850917475728154)
			(0.1881, 0.03621864390243903)
			(0.19304999999999997, 0.03658603431372549)
			(0.19799999999999998, 0.03694329556650246)
			(0.20295, 0.037655557213930345)
			(0.20789999999999997, 0.038013705)
			(0.21284999999999998, 0.03836688442211055)
			(0.2178, 0.03871999494949495)
			(0.22274999999999998, 0.03943367346938776)
			(0.22769999999999999, 0.03979421025641025)
			(0.23264999999999997, 0.04015626804123711)
			(0.23759999999999998, 0.04052073575129534)
			(0.24255, 0.04123787958115183)
			(0.24749999999999997, 0.041590757894736846)
			(0.25244999999999995, 0.04194255555555555)
			(0.25739999999999996, 0.04228973404255319)
			(0.26234999999999997, 0.04299378494623656)
			(0.2673, 0.043349718918918916)
			(0.27225, 0.04370441304347827)
			(0.2772, 0.044042475409836065)
			(0.28214999999999996, 0.04472199447513812)
			(0.28709999999999997, 0.045063844444444445)
			(0.29205, 0.04540903910614525)
			(0.297, 0.04575547191011235)
			(0.30195, 0.046443744318181816)
			(0.30689999999999995, 0.046790331428571426)
			(0.31184999999999996, 0.04714072413793104)
			(0.31679999999999997, 0.04749515606936416)
			(0.32175, 0.048213409356725134)
			(0.3267, 0.04857856470588235)
			(0.33164999999999994, 0.04894004142011834)
			(0.33659999999999995, 0.04930567261904762)
			(0.34154999999999996, 0.05004480722891566)
			(0.3465, 0.050415593939393946)
			(0.35145, 0.05079082317073171)
			(0.35639999999999994, 0.051170588957055216)
			(0.36134999999999995, 0.05194151552795031)
			(0.36629999999999996, 0.05233218125)
			(0.37124999999999997, 0.0527253144654088)
			(0.3762, 0.053111329113924056)
			(0.38115, 0.05388842307692308)
			(0.38609999999999994, 0.05428234193548387)
			(0.39104999999999995, 0.05468098701298701)
			(0.39599999999999996, 0.05508404575163398)
			(0.40095, 0.05589952317880795)
			(0.4059, 0.056310633333333325)
			(0.41084999999999994, 0.05672685906040268)
			(0.41579999999999995, 0.057145466216216224)
			(0.42074999999999996, 0.05798029452054793)
			(0.42569999999999997, 0.058398648275862074)
			(0.43065, 0.05882145833333333)
			(0.4356, 0.059246209790209794)
			(0.44054999999999994, 0.06010387234042553)
			(0.44549999999999995, 0.06052798571428572)
			(0.45044999999999996, 0.06095230215827338)
			(0.45539999999999997, 0.061380565217391304)
			(0.46035, 0.062206088235294124)
			(0.46529999999999994, 0.06262667407407407)
			(0.47024999999999995, 0.0630519776119403)
			(0.47519999999999996, 0.06348175187969926)
			(0.48014999999999997, 0.06391723484848484)
			(0.4851, 0.06480702307692307)
			(0.49004999999999993, 0.06526071317829456)
			(0.49499999999999994, 0.0657146796875)
			(0.49994999999999995, 0.06617483464566931)
			(0.5048999999999999, 0.06710573600000001)
			(0.5098499999999999, 0.06757601612903227)
			(0.5147999999999999, 0.06804923577235772)
			(0.5197499999999999, 0.06852963114754099)
			(0.5246999999999999, 0.06951059166666666)
			(0.52965, 0.07001299159663867)
			(0.5346, 0.07051989830508475)
			(0.53955, 0.07103511965811966)
			(0.5445, 0.07208689565217392)
			(0.54945, 0.07261916666666667)
			(0.5544, 0.07315781415929205)
			(0.5593499999999999, 0.07370511607142857)
			(0.5642999999999999, 0.07481717272727273)
			(0.5692499999999999, 0.07538509174311928)
			(0.5741999999999999, 0.07596168518518519)
			(0.5791499999999999, 0.07654203738317758)
			(0.5841, 0.0777246761904762)
			(0.58905, 0.0783279423076923)
			(0.594, 0.07893382524271846)
			(0.59895, 0.07954817647058825)
			(0.6039, 0.08080516000000001)
			(0.6088499999999999, 0.08144947474747476)
			(0.6137999999999999, 0.08210083673469389)
			(0.6187499999999999, 0.08276004123711342)
			(0.6236999999999999, 0.0841105894736842)
			(0.6286499999999999, 0.08480181914893617)
			(0.6335999999999999, 0.0855042258064516)
			(0.63855, 0.08621901086956521)
			(0.6435, 0.08769108888888888)
			(0.64845, 0.08844208988764046)
			(0.6534, 0.08920052272727273)
			(0.65835, 0.08997397701149426)
			(0.6632999999999999, 0.09156308235294118)
			(0.6682499999999999, 0.0923550238095238)
			(0.6731999999999999, 0.09316363855421686)
			(0.6781499999999999, 0.09398519512195122)
			(0.6830999999999999, 0.0956819875)
			(0.6880499999999999, 0.09655602531645568)
			(0.693, 0.09745039743589744)
			(0.69795, 0.09835993506493505)
			(0.7029, 0.10023017333333334)
			(0.70785, 0.10119494594594594)
			(0.7127999999999999, 0.10217931506849316)
			(0.7177499999999999, 0.10319059722222222)
			(0.7226999999999999, 0.10527498571428573)
			(0.7276499999999999, 0.10632201449275364)
			(0.7325999999999999, 0.10738317647058823)
			(0.7375499999999999, 0.10845476119402984)
			(0.7424999999999999, 0.11066430769230769)
			(0.74745, 0.11180137500000001)
			(0.7524, 0.11297171428571429)
			(0.75735, 0.11415703225806453)
			(0.7623, 0.11658466666666668)
			(0.7672499999999999, 0.11785993220338983)
			(0.7721999999999999, 0.11916374137931034)
			(0.7771499999999999, 0.12049217543859649)
			(0.7820999999999999, 0.12325465454545455)
			(0.7870499999999999, 0.12468370370370369)
			(0.7919999999999999, 0.1261555471698113)
			(0.7969499999999999, 0.12765119230769228)
			(0.8019, 0.13080626)
			(0.80685, 0.13246222448979592)
			(0.8118, 0.13418652083333335)
			(0.81675, 0.13597704255319149)
			(0.8216999999999999, 0.13950264444444446)
			(0.8266499999999999, 0.14136281818181817)
			(0.8315999999999999, 0.1433049534883721)
			(0.8365499999999999, 0.14533376190476188)
			(0.8414999999999999, 0.149662625)
			(0.8464499999999999, 0.15195746153846154)
			(0.8513999999999999, 0.15432271052631574)
			(0.85635, 0.15679964864864862)
			(0.8613, 0.1619372285714286)
			(0.86625, 0.16471526470588235)
			(0.8712, 0.1676248787878788)
			(0.8761499999999999, 0.17069690625)
			(0.8810999999999999, 0.1773375333333333)
			(0.8860499999999999, 0.18093855172413792)
			(0.8909999999999999, 0.1846958214285714)
			(0.8959499999999999, 0.18859662962962961)
			(0.9008999999999999, 0.19707632)
			(0.9058499999999999, 0.20166183333333332)
			(0.9107999999999999, 0.20660517391304348)
			(0.91575, 0.21195259090909088)
			(0.9207, 0.2238071)
			(0.9256499999999999, 0.2300917368421053)
			(0.9305999999999999, 0.2370698333333333)
			(0.9355499999999999, 0.2445643529411765)
			(0.9404999999999999, 0.26153699999999996)
			(0.9454499999999999, 0.2714842857142857)
			(0.9503999999999999, 0.28285946153846153)
			(0.9553499999999999, 0.29588583333333335)
			(0.9602999999999999, 0.31114318181818185)
			(0.9652499999999999, 0.3507886666666667)
			(0.9702, 0.37759175)
			(0.97515, 0.41143642857142854)
			(0.9800999999999999, 0.4556198333333333)
			(0.9850499999999999, 0.5941335000000001)
		};
		\addlegendentry{$M_j(p)$}
		\addplot[dashdotted, red] coordinates {
			(0.0, 0.013727813046830229)
			(0.0049499999999999995, 0.01475723232350181)
			(0.009899999999999999, 0.015385168420541611)
			(0.014849999999999999, 0.01586619568892892)
			(0.019799999999999998, 0.016404785932469815)
			(0.024749999999999998, 0.01794971852936185)
			(0.029699999999999997, 0.018862822201495426)
			(0.03465, 0.019401704337019966)
			(0.039599999999999996, 0.01995452326302094)
			(0.04454999999999999, 0.021318353078334032)
			(0.049499999999999995, 0.021880135235686594)
			(0.05445, 0.022348801014786673)
			(0.059399999999999994, 0.02302596026001756)
			(0.06434999999999999, 0.024034196199526608)
			(0.0693, 0.0245622828101742)
			(0.07425, 0.024724873578337633)
			(0.07919999999999999, 0.02518109340118875)
			(0.08414999999999999, 0.02619341146657759)
			(0.08909999999999998, 0.02670661556201517)
			(0.09405, 0.027254372073770215)
			(0.09899999999999999, 0.02780826077651497)
			(0.10394999999999999, 0.028639563806488275)
			(0.1089, 0.02920199191065421)
			(0.11384999999999999, 0.02969921963586549)
			(0.11879999999999999, 0.030440542680142145)
			(0.12374999999999999, 0.03135418655169336)
			(0.12869999999999998, 0.03191041422554472)
			(0.13365, 0.032462624353197564)
			(0.1386, 0.0329472372288551)
			(0.14354999999999998, 0.03370299698616662)
			(0.1485, 0.0342289501519135)
			(0.15344999999999998, 0.03464218289933059)
			(0.15839999999999999, 0.03517096784027806)
			(0.16335, 0.036315248754122884)
			(0.16829999999999998, 0.036552562332871345)
			(0.17325, 0.03678676128370771)
			(0.17819999999999997, 0.03734090738911958)
			(0.18314999999999998, 0.03798441697773174)
			(0.1881, 0.03848069991472873)
			(0.19304999999999997, 0.0388847909705599)
			(0.19799999999999998, 0.04034742420589536)
			(0.20295, 0.0413503215732774)
			(0.20789999999999997, 0.04181672726885596)
			(0.21284999999999998, 0.042082344208545897)
			(0.2178, 0.04234957693980158)
			(0.22274999999999998, 0.043601811761072605)
			(0.22769999999999999, 0.043952319867707604)
			(0.23264999999999997, 0.044288870141987215)
			(0.23759999999999998, 0.04503672621910037)
			(0.24255, 0.045770160819548114)
			(0.24749999999999997, 0.046299553518479614)
			(0.25244999999999995, 0.04673777050435419)
			(0.25739999999999996, 0.047448819279949574)
			(0.26234999999999997, 0.04796613703430919)
			(0.2673, 0.04880640342380307)
			(0.27225, 0.04952325663549249)
			(0.2772, 0.04992654260524211)
			(0.28214999999999996, 0.05048737766407733)
			(0.28709999999999997, 0.051002637380227554)
			(0.29205, 0.0515788176332208)
			(0.297, 0.05204201885587934)
			(0.30195, 0.053154816538490364)
			(0.30689999999999995, 0.05372732580302184)
			(0.31184999999999996, 0.05473095417889059)
			(0.31679999999999997, 0.05514128339379747)
			(0.32175, 0.05615140268882857)
			(0.3267, 0.05629535995630416)
			(0.33164999999999994, 0.05715321270221801)
			(0.33659999999999995, 0.057744493545733556)
			(0.34154999999999996, 0.05855669909623374)
			(0.3465, 0.05926503052867778)
			(0.35145, 0.059522320331329474)
			(0.35639999999999994, 0.059846846256016016)
			(0.36134999999999995, 0.06044257503779214)
			(0.36629999999999996, 0.060842880136108356)
			(0.37124999999999997, 0.06136259481770404)
			(0.3762, 0.06190683230453522)
			(0.38115, 0.0638148581460448)
			(0.38609999999999994, 0.06397943733457455)
			(0.39104999999999995, 0.06547579184308143)
			(0.39599999999999996, 0.06594075373727569)
			(0.40095, 0.06682915339815523)
			(0.4059, 0.06712384103605509)
			(0.41084999999999994, 0.06788305905539327)
			(0.41579999999999995, 0.06839985732587155)
			(0.42074999999999996, 0.06961407942257182)
			(0.42569999999999997, 0.0699748376296481)
			(0.43065, 0.07044301794745318)
			(0.4356, 0.07072502927731729)
			(0.44054999999999994, 0.07139610089342746)
			(0.44549999999999995, 0.07219814163706227)
			(0.45044999999999996, 0.07284850671453216)
			(0.45539999999999997, 0.0728576660720532)
			(0.46035, 0.0737256422752472)
			(0.46529999999999994, 0.07448670598274543)
			(0.47024999999999995, 0.07453417316671948)
			(0.47519999999999996, 0.07495835475716205)
			(0.48014999999999997, 0.07566774432576583)
			(0.4851, 0.07636486169135108)
			(0.49004999999999993, 0.07686761215769758)
			(0.49499999999999994, 0.07740777731137244)
			(0.49994999999999995, 0.078045036786728)
			(0.5048999999999999, 0.07911382291891521)
			(0.5098499999999999, 0.0795696727983039)
			(0.5147999999999999, 0.08007275708500883)
			(0.5197499999999999, 0.08063348850442933)
			(0.5246999999999999, 0.08188619401165338)
			(0.52965, 0.0819997289131335)
			(0.5346, 0.0826202355935799)
			(0.53955, 0.08323841469611147)
			(0.5445, 0.08480174869635092)
			(0.54945, 0.08546083511867938)
			(0.5544, 0.08638377589094309)
			(0.5593499999999999, 0.08680207614648548)
			(0.5642999999999999, 0.08783876744916873)
			(0.5692499999999999, 0.0885199146568234)
			(0.5741999999999999, 0.08887038441530569)
			(0.5791499999999999, 0.0898407931332509)
			(0.5841, 0.09117968297042049)
			(0.58905, 0.09174912148788707)
			(0.594, 0.09276833820426139)
			(0.59895, 0.09289600221932506)
			(0.6039, 0.09402521744232097)
			(0.6088499999999999, 0.09485216730155148)
			(0.6137999999999999, 0.0956779716987017)
			(0.6187499999999999, 0.09676050386471237)
			(0.6236999999999999, 0.09820302171096804)
			(0.6286499999999999, 0.09919262871546446)
			(0.6335999999999999, 0.10086631161822755)
			(0.63855, 0.10188760349396966)
			(0.6435, 0.10327315464289631)
			(0.64845, 0.10400574407335599)
			(0.6534, 0.10494141581641094)
			(0.65835, 0.1061440873135016)
			(0.6632999999999999, 0.10827321750137131)
			(0.6682499999999999, 0.1099129060214646)
			(0.6731999999999999, 0.11019073968518975)
			(0.6781499999999999, 0.11115102908467757)
			(0.6830999999999999, 0.11287318041423404)
			(0.6880499999999999, 0.11387865088101387)
			(0.693, 0.11454652136437957)
			(0.69795, 0.1152735959874624)
			(0.7029, 0.11745756299678847)
			(0.70785, 0.11790421659607442)
			(0.7127999999999999, 0.11931843343530879)
			(0.7177499999999999, 0.12040151754673542)
			(0.7226999999999999, 0.12312897936135833)
			(0.7276499999999999, 0.12363859342102546)
			(0.7325999999999999, 0.12582928621106362)
			(0.7375499999999999, 0.12714316422612476)
			(0.7424999999999999, 0.12962415664030313)
			(0.74745, 0.13101325255936574)
			(0.7524, 0.13144475734733296)
			(0.75735, 0.1328597577315813)
			(0.7623, 0.13420699916109707)
			(0.7672499999999999, 0.13650740919077642)
			(0.7721999999999999, 0.13783893186475127)
			(0.7771499999999999, 0.13880483854203227)
			(0.7820999999999999, 0.14347702687066546)
			(0.7870499999999999, 0.14540679178341345)
			(0.7919999999999999, 0.14696735361335533)
			(0.7969499999999999, 0.14920269193778946)
			(0.8019, 0.1534127855938573)
			(0.80685, 0.15504047638690813)
			(0.8118, 0.15689582314404152)
			(0.81675, 0.15958142769687855)
			(0.8216999999999999, 0.16305364006308934)
			(0.8266499999999999, 0.1644032781803619)
			(0.8315999999999999, 0.1667542390872506)
			(0.8365499999999999, 0.17033890220979514)
			(0.8414999999999999, 0.17290995586540828)
			(0.8464499999999999, 0.17745461615787386)
			(0.8513999999999999, 0.1794001961329595)
			(0.85635, 0.18228927649576748)
			(0.8613, 0.18754539297374698)
			(0.86625, 0.19076365425862676)
			(0.8712, 0.19208564876057466)
			(0.8761499999999999, 0.1923423638206237)
			(0.8810999999999999, 0.20218317316526693)
			(0.8860499999999999, 0.2055826351819588)
			(0.8909999999999999, 0.2099693013733141)
			(0.8959499999999999, 0.21351939366976647)
			(0.9008999999999999, 0.22447744116548826)
			(0.9058499999999999, 0.23019353568070144)
			(0.9107999999999999, 0.23286788685217444)
			(0.91575, 0.23662571478407274)
			(0.9207, 0.24910916894862858)
			(0.9256499999999999, 0.25257663387823315)
			(0.9305999999999999, 0.25752988078736516)
			(0.9355499999999999, 0.2678325532194221)
			(0.9404999999999999, 0.2759561121692381)
			(0.9454499999999999, 0.2823256797399948)
			(0.9503999999999999, 0.28690450517159916)
			(0.9553499999999999, 0.2989090837027894)
			(0.9602999999999999, 0.3154822129213017)
			(0.9652499999999999, 0.3492011495671069)
			(0.9702, 0.379799971042957)
			(0.97515, 0.40368064926416114)
			(0.9800999999999999, 0.4311052154647103)
			(0.9850499999999999, 0.5407087431235826)
		};
		\addlegendentry{$M^f_j(p)$}
		\addplot[dash pattern=on 8pt off 2pt] coordinates {
			(0.0, 0.013954305555555555)
			(0.0049499999999999995, 0.014719071713147411)
			(0.009899999999999999, 0.015087860000000002)
			(0.014849999999999999, 0.016190441767068277)
			(0.019799999999999998, 0.0166595)
			(0.024749999999999998, 0.01765202845528455)
			(0.029699999999999997, 0.018023530612244897)
			(0.03465, 0.01852950819672131)
			(0.039599999999999996, 0.01882701234567901)
			(0.04454999999999999, 0.01964803734439834)
			(0.049499999999999995, 0.020363341666666666)
			(0.05445, 0.020664761506276146)
			(0.059399999999999994, 0.020219894957983188)
			(0.06434999999999999, 0.021221504237288132)
			(0.0693, 0.021514757446808506)
			(0.07425, 0.02173680769230769)
			(0.07919999999999999, 0.021652776824034336)
			(0.08414999999999999, 0.022603744588744583)
			(0.08909999999999998, 0.022803443478260868)
			(0.09405, 0.023138187772925763)
			(0.09899999999999999, 0.02346843421052631)
			(0.10394999999999999, 0.02447092477876106)
			(0.1089, 0.024690346666666668)
			(0.11384999999999999, 0.024961008928571428)
			(0.11879999999999999, 0.025422829596412552)
			(0.12374999999999999, 0.0277328778280543)
			(0.12869999999999998, 0.028087536363636364)
			(0.13365, 0.028339543378995434)
			(0.1386, 0.028500321100917433)
			(0.14354999999999998, 0.02889662037037037)
			(0.1485, 0.028986060465116276)
			(0.15344999999999998, 0.029177686915887846)
			(0.15839999999999999, 0.0296901455399061)
			(0.16335, 0.030368407582938392)
			(0.16829999999999998, 0.0307057)
			(0.17325, 0.028852947368421047)
			(0.17819999999999997, 0.029155)
			(0.18314999999999998, 0.030028529126213593)
			(0.1881, 0.03047535121951219)
			(0.19304999999999997, 0.030694240196078432)
			(0.19799999999999998, 0.031108778325123154)
			(0.20295, 0.03162679104477612)
			(0.20789999999999997, 0.031935855)
			(0.21284999999999998, 0.032268934673366834)
			(0.2178, 0.03254733333333333)
			(0.22274999999999998, 0.0330886887755102)
			(0.22769999999999999, 0.033336517948717954)
			(0.23264999999999997, 0.03359457216494845)
			(0.23759999999999998, 0.03386219689119171)
			(0.24255, 0.03438794764397905)
			(0.24749999999999997, 0.03464242631578947)
			(0.25244999999999995, 0.03488463492063491)
			(0.25739999999999996, 0.03512431914893617)
			(0.26234999999999997, 0.03539991935483871)
			(0.2673, 0.03574711351351351)
			(0.27225, 0.03590936956521739)
			(0.2772, 0.036184213114754095)
			(0.28214999999999996, 0.036948928176795584)
			(0.28709999999999997, 0.03718623888888889)
			(0.29205, 0.037523240223463684)
			(0.297, 0.03756655056179775)
			(0.30195, 0.03781332954545455)
			(0.30689999999999995, 0.03783418857142857)
			(0.31184999999999996, 0.0379552816091954)
			(0.31679999999999997, 0.038115491329479764)
			(0.32175, 0.03863241520467837)
			(0.3267, 0.03899635294117647)
			(0.33164999999999994, 0.039184775147929)
			(0.33659999999999995, 0.03936041666666667)
			(0.34154999999999996, 0.039944481927710836)
			(0.3465, 0.04025397575757575)
			(0.35145, 0.04007674390243902)
			(0.35639999999999994, 0.04020148466257669)
			(0.36134999999999995, 0.03990608695652174)
			(0.36629999999999996, 0.03998803125)
			(0.37124999999999997, 0.040287578616352196)
			(0.3762, 0.04055953164556962)
			(0.38115, 0.04115339102564102)
			(0.38609999999999994, 0.04174468387096774)
			(0.39104999999999995, 0.0420563961038961)
			(0.39599999999999996, 0.04258440522875817)
			(0.40095, 0.043471046357615895)
			(0.4059, 0.0436474)
			(0.41084999999999994, 0.0440531610738255)
			(0.41579999999999995, 0.04474306081081081)
			(0.42074999999999996, 0.04556855479452055)
			(0.42569999999999997, 0.04618022068965517)
			(0.43065, 0.04659722222222222)
			(0.4356, 0.04710355944055944)
			(0.44054999999999994, 0.04722236170212766)
			(0.44549999999999995, 0.04758561428571428)
			(0.45044999999999996, 0.04703869784172661)
			(0.45539999999999997, 0.04747003623188406)
			(0.46035, 0.04806677205882354)
			(0.46529999999999994, 0.04835341481481482)
			(0.47024999999999995, 0.048616402985074615)
			(0.47519999999999996, 0.04871293233082707)
			(0.48014999999999997, 0.049035840909090904)
			(0.4851, 0.04966711538461538)
			(0.49004999999999993, 0.049690302325581394)
			(0.49499999999999994, 0.0501653125)
			(0.49994999999999995, 0.050615330708661414)
			(0.5048999999999999, 0.051821087999999994)
			(0.5098499999999999, 0.05228115322580645)
			(0.5147999999999999, 0.05284320325203252)
			(0.5197499999999999, 0.053344090163934425)
			(0.5246999999999999, 0.05398345833333333)
			(0.52965, 0.054225)
			(0.5346, 0.05461073728813559)
			(0.53955, 0.05501528205128206)
			(0.5445, 0.05573253913043478)
			(0.54945, 0.05604605263157894)
			(0.5544, 0.05649469911504425)
			(0.5593499999999999, 0.056985294642857155)
			(0.5642999999999999, 0.05750997272727273)
			(0.5692499999999999, 0.058661284403669724)
			(0.5741999999999999, 0.05903631481481481)
			(0.5791499999999999, 0.059792859813084104)
			(0.5841, 0.06083852380952381)
			(0.58905, 0.06095497115384615)
			(0.594, 0.0617045922330097)
			(0.59895, 0.06123441176470589)
			(0.6039, 0.06222917999999999)
			(0.6088499999999999, 0.06271041414141415)
			(0.6137999999999999, 0.0624623163265306)
			(0.6187499999999999, 0.06212499999999999)
			(0.6236999999999999, 0.06346733684210525)
			(0.6286499999999999, 0.06394298936170212)
			(0.6335999999999999, 0.06420601075268817)
			(0.63855, 0.06505834782608697)
			(0.6435, 0.06608091111111111)
			(0.64845, 0.06664449438202247)
			(0.6534, 0.06710697727272728)
			(0.65835, 0.06764890804597701)
			(0.6632999999999999, 0.06986701176470587)
			(0.6682499999999999, 0.07071019047619048)
			(0.6731999999999999, 0.07147446987951808)
			(0.6781499999999999, 0.07202560975609755)
			(0.6830999999999999, 0.07441367499999998)
			(0.6880499999999999, 0.07499059493670886)
			(0.693, 0.07560052564102564)
			(0.69795, 0.07625715584415585)
			(0.7029, 0.07794029333333333)
			(0.70785, 0.07909041891891891)
			(0.7127999999999999, 0.07948847945205478)
			(0.7177499999999999, 0.079789)
			(0.7226999999999999, 0.08114435714285714)
			(0.7276499999999999, 0.0816128115942029)
			(0.7325999999999999, 0.08326698529411763)
			(0.7375499999999999, 0.08409071641791044)
			(0.7424999999999999, 0.08628726153846153)
			(0.74745, 0.08734703125)
			(0.7524, 0.08895353968253966)
			(0.75735, 0.08896948387096774)
			(0.7623, 0.0908691)
			(0.7672499999999999, 0.09300130508474574)
			(0.7721999999999999, 0.09423063793103446)
			(0.7771499999999999, 0.09530433333333334)
			(0.7820999999999999, 0.09592101818181818)
			(0.7870499999999999, 0.09604988888888888)
			(0.7919999999999999, 0.0965098113207547)
			(0.7969499999999999, 0.09809038461538461)
			(0.8019, 0.0974928)
			(0.80685, 0.09870914285714284)
			(0.8118, 0.10088447916666667)
			(0.81675, 0.10197359574468083)
			(0.8216999999999999, 0.10405108888888888)
			(0.8266499999999999, 0.10614879545454546)
			(0.8315999999999999, 0.10558467441860465)
			(0.8365499999999999, 0.10782697619047618)
			(0.8414999999999999, 0.10928550000000001)
			(0.8464499999999999, 0.11150905128205128)
			(0.8513999999999999, 0.11397992105263158)
			(0.85635, 0.11379199999999999)
			(0.8613, 0.11523402857142856)
			(0.86625, 0.1169145882352941)
			(0.8712, 0.11895248484848485)
			(0.8761499999999999, 0.12046659375)
			(0.8810999999999999, 0.12414560000000001)
			(0.8860499999999999, 0.12810093103448275)
			(0.8909999999999999, 0.13229632142857145)
			(0.8959499999999999, 0.13332955555555553)
			(0.9008999999999999, 0.13770028)
			(0.9058499999999999, 0.146220875)
			(0.9107999999999999, 0.14968178260869566)
			(0.91575, 0.1498338636363636)
			(0.9207, 0.15816840000000001)
			(0.9256499999999999, 0.15931599999999999)
			(0.9305999999999999, 0.16506322222222222)
			(0.9355499999999999, 0.171305)
			(0.9404999999999999, 0.18882900000000002)
			(0.9454499999999999, 0.1942020714285714)
			(0.9503999999999999, 0.20569169230769233)
			(0.9553499999999999, 0.2192124166666667)
			(0.9602999999999999, 0.23391936363636365)
			(0.9652499999999999, 0.27189655555555553)
			(0.9702, 0.29535300000000003)
			(0.97515, 0.3200784285714286)
			(0.9800999999999999, 0.3560105)
			(0.9850499999999999, 0.50117225)
		};
		\addlegendentry{$\mathrm{MES}_p(X_j,S)$}
		\addplot[dotted] coordinates {
			(0.0, 0.010356796956488914)
			(0.0049499999999999995, 0.008777674539058984)
			(0.009899999999999999, 0.00815527122302095)
			(0.014849999999999999, 0.007883652862286926)
			(0.019799999999999998, 0.007517961967228786)
			(0.024749999999999998, 0.008347578330292793)
			(0.029699999999999997, 0.007197147263266474)
			(0.03465, 0.006912063540730242)
			(0.039599999999999996, 0.0068079625718736715)
			(0.04454999999999999, 0.007012562100940121)
			(0.049499999999999995, 0.007045976942835337)
			(0.05445, 0.006245426279741015)
			(0.059399999999999994, 0.006836224615865988)
			(0.06434999999999999, 0.006788612734091683)
			(0.0693, 0.006520450493302329)
			(0.07425, 0.006464145011409802)
			(0.07919999999999999, 0.006157308162606055)
			(0.08414999999999999, 0.0057513407290723465)
			(0.08909999999999998, 0.0058777914719674905)
			(0.09405, 0.00593594443197846)
			(0.09899999999999999, 0.005818223164986269)
			(0.10394999999999999, 0.005595911469257637)
			(0.1089, 0.005537961591171173)
			(0.11384999999999999, 0.005617761194622766)
			(0.11879999999999999, 0.006278762547883026)
			(0.12374999999999999, 0.005970371639583715)
			(0.12869999999999998, 0.005836190188521468)
			(0.13365, 0.005938168365432598)
			(0.1386, 0.006155593477093822)
			(0.14354999999999998, 0.0055968029365502485)
			(0.1485, 0.005358402672279588)
			(0.15344999999999998, 0.0055003957126129495)
			(0.15839999999999999, 0.00537725417585188)
			(0.16335, 0.005655973443050596)
			(0.16829999999999998, 0.005474976321249565)
			(0.17325, 0.00508378902819446)
			(0.17819999999999997, 0.004838780014598171)
			(0.18314999999999998, 0.00430238031215696)
			(0.1881, 0.004381490355154479)
			(0.19304999999999997, 0.0060525086295721555)
			(0.19799999999999998, 0.005917297813650589)
			(0.20295, 0.006266811498735834)
			(0.20789999999999997, 0.006286260865602317)
			(0.21284999999999998, 0.005731780739307614)
			(0.2178, 0.005235918075070312)
			(0.22274999999999998, 0.005640511285167704)
			(0.22769999999999999, 0.005972488640788532)
			(0.23264999999999997, 0.005591427185502099)
			(0.23759999999999998, 0.00506198748569922)
			(0.24255, 0.004953659040871262)
			(0.24749999999999997, 0.004882425423616836)
			(0.25244999999999995, 0.004581080065195435)
			(0.25739999999999996, 0.00496200189536471)
			(0.26234999999999997, 0.004607272343091079)
			(0.2673, 0.005215921503734796)
			(0.27225, 0.0055285123973891435)
			(0.2772, 0.005335357618763767)
			(0.28214999999999996, 0.004674036526031212)
			(0.28709999999999997, 0.0047336833693873365)
			(0.29205, 0.005200693160753836)
			(0.297, 0.006614495842077175)
			(0.30195, 0.005801545045675515)
			(0.30689999999999995, 0.006950517491329788)
			(0.31184999999999996, 0.006977738915852708)
			(0.31679999999999997, 0.006784073850997695)
			(0.32175, 0.007184446297726564)
			(0.3267, 0.0069432219814235694)
			(0.33164999999999994, 0.008128255449374254)
			(0.33659999999999995, 0.008713225253245818)
			(0.34154999999999996, 0.007992938174691757)
			(0.3465, 0.007963864606756754)
			(0.35145, 0.007822507547684885)
			(0.35639999999999994, 0.0083841316727431)
			(0.36134999999999995, 0.008333716715887002)
			(0.36629999999999996, 0.007911668425810744)
			(0.37124999999999997, 0.007796817608645728)
			(0.3762, 0.007763381324380743)
			(0.38115, 0.00941201559551473)
			(0.38609999999999994, 0.009431520571233564)
			(0.39104999999999995, 0.011084736138431552)
			(0.39599999999999996, 0.011101165843568463)
			(0.40095, 0.011510481690947951)
			(0.4059, 0.011926956331784365)
			(0.41084999999999994, 0.011335602647851035)
			(0.41579999999999995, 0.01161640090819936)
			(0.42074999999999996, 0.011764290928957411)
			(0.42569999999999997, 0.011965130147861805)
			(0.43065, 0.012135506540357935)
			(0.4356, 0.012105705341711365)
			(0.44054999999999994, 0.013116633593499292)
			(0.44549999999999995, 0.012131120874637597)
			(0.45044999999999996, 0.012223258000308044)
			(0.45539999999999997, 0.011715421779979161)
			(0.46035, 0.011096972938839492)
			(0.46529999999999994, 0.011137368464522174)
			(0.47024999999999995, 0.010273790696192364)
			(0.47519999999999996, 0.01008772620575395)
			(0.48014999999999997, 0.010386884808279345)
			(0.4851, 0.009774810504806488)
			(0.49004999999999993, 0.009828974968936937)
			(0.49499999999999994, 0.009376259865620662)
			(0.49994999999999995, 0.00980732950346618)
			(0.5048999999999999, 0.009753916349153288)
			(0.5098499999999999, 0.009957492970216017)
			(0.5147999999999999, 0.009563006469867345)
			(0.5197499999999999, 0.009871082260932503)
			(0.5246999999999999, 0.010008189417946099)
			(0.52965, 0.009684994661503871)
			(0.5346, 0.009755705609224648)
			(0.53955, 0.010102747705266578)
			(0.5445, 0.011513876239119354)
			(0.54945, 0.011279864438577848)
			(0.5544, 0.011745532761079253)
			(0.5593499999999999, 0.011442333333814031)
			(0.5642999999999999, 0.01083893545835614)
			(0.5692499999999999, 0.010967050178405737)
			(0.5741999999999999, 0.010706327993266438)
			(0.5791499999999999, 0.011205209317019306)
			(0.5841, 0.010240591497622017)
			(0.58905, 0.010877971729110212)
			(0.594, 0.01036344313697791)
			(0.59895, 0.01027194186418686)
			(0.6039, 0.011446033367700774)
			(0.6088499999999999, 0.010092640901169243)
			(0.6137999999999999, 0.009597122672329551)
			(0.6187499999999999, 0.010162232628154822)
			(0.6236999999999999, 0.012107472653428776)
			(0.6286499999999999, 0.011775841401260938)
			(0.6335999999999999, 0.012245432443006266)
			(0.63855, 0.012002251444632305)
			(0.6435, 0.014417491609941045)
			(0.64845, 0.01437836932025814)
			(0.6534, 0.014494714810865329)
			(0.65835, 0.01562567695302195)
			(0.6632999999999999, 0.01692578731725531)
			(0.6682499999999999, 0.016433688305657636)
			(0.6731999999999999, 0.01620628510333067)
			(0.6781499999999999, 0.017660926184724916)
			(0.6830999999999999, 0.017095085495260384)
			(0.6880499999999999, 0.016926957109085254)
			(0.693, 0.016758555576957372)
			(0.69795, 0.017195116364881143)
			(0.7029, 0.01836178479018654)
			(0.70785, 0.018164076052999023)
			(0.7127999999999999, 0.017903970543864303)
			(0.7177499999999999, 0.017066471820736912)
			(0.7226999999999999, 0.01729386592386883)
			(0.7276499999999999, 0.019662619247791013)
			(0.7325999999999999, 0.018042789864283478)
			(0.7375499999999999, 0.017505605957085158)
			(0.7424999999999999, 0.017667248793259625)
			(0.74745, 0.018398265979629173)
			(0.7524, 0.01696348414586753)
			(0.75735, 0.016734755341418037)
			(0.7623, 0.014633304971250522)
			(0.7672499999999999, 0.016253952113320425)
			(0.7721999999999999, 0.01616225446913461)
			(0.7771499999999999, 0.016265878616236566)
			(0.7820999999999999, 0.01933426645575051)
			(0.7870499999999999, 0.020191619948933655)
			(0.7919999999999999, 0.022693568386366048)
			(0.7969499999999999, 0.023241794423618515)
			(0.8019, 0.024827921440747355)
			(0.80685, 0.024714888667989)
			(0.8118, 0.022953256317885345)
			(0.81675, 0.025123456311978678)
			(0.8216999999999999, 0.023100972112663595)
			(0.8266499999999999, 0.022133307176981653)
			(0.8315999999999999, 0.025273926754344384)
			(0.8365499999999999, 0.025322158197069754)
			(0.8414999999999999, 0.021332287888568908)
			(0.8464499999999999, 0.024582932313892886)
			(0.8513999999999999, 0.022309814351241984)
			(0.85635, 0.020953208719949535)
			(0.8613, 0.023127019977938732)
			(0.86625, 0.021923473804140142)
			(0.8712, 0.02035038702857791)
			(0.8761499999999999, 0.017660818276267067)
			(0.8810999999999999, 0.026441384796464844)
			(0.8860499999999999, 0.029572211914738)
			(0.8909999999999999, 0.033811436500293855)
			(0.8959499999999999, 0.02984725819479537)
			(0.9008999999999999, 0.029637429998805843)
			(0.9058499999999999, 0.03586503594270558)
			(0.9107999999999999, 0.035882960714299125)
			(0.91575, 0.035406438316149674)
			(0.9207, 0.03690242597033729)
			(0.9256499999999999, 0.03327546482343577)
			(0.9305999999999999, 0.040083754287326565)
			(0.9355499999999999, 0.04177830540414595)
			(0.9404999999999999, 0.06920042295644682)
			(0.9454499999999999, 0.06639491077876833)
			(0.9503999999999999, 0.06286808167650045)
			(0.9553499999999999, 0.06684202906040655)
			(0.9602999999999999, 0.07682101475948908)
			(0.9652499999999999, 0.07879969684527031)
			(0.9702, 0.08175437352824813)
			(0.97515, 0.10685786574521036)
			(0.9800999999999999, 0.1035623621479959)
			(0.9850499999999999, 0.10500880945177109)
		};
		\addlegendentry{$m^f_j(p)$}
		\addplot[dashed, green] coordinates {
			(0.0, 0.01206416205533597)
			(0.0049499999999999995, 0.00614780876494024)
			(0.009899999999999999, 0.004374520000000001)
			(0.014849999999999999, 0.0027136506024096384)
			(0.019799999999999998, 0.001953266129032258)
			(0.024749999999999998, 0.0006999105691056912)
			(0.029699999999999997, 0.0001285673469387762)
			(0.03465, -0.00042977459016393287)
			(0.039599999999999996, -0.000987390946502056)
			(0.04454999999999999, -0.00206803734439834)
			(0.049499999999999995, -0.0026039166666666667)
			(0.05445, -0.003131995815899582)
			(0.059399999999999994, -0.0036589159663865545)
			(0.06434999999999999, -0.004683733050847458)
			(0.0693, -0.005170314893617022)
			(0.07425, -0.005638931623931624)
			(0.07919999999999999, -0.006111197424892704)
			(0.08414999999999999, -0.006972831168831169)
			(0.08909999999999998, -0.007389939130434783)
			(0.09405, -0.007806336244541484)
			(0.09899999999999999, -0.008222258771929825)
			(0.10394999999999999, -0.009026)
			(0.1089, -0.00941888888888889)
			(0.11384999999999999, -0.009799040178571427)
			(0.11879999999999999, -0.010169923766816143)
			(0.12374999999999999, -0.010905285067873303)
			(0.12869999999999998, -0.011269945454545453)
			(0.13365, -0.011635095890410957)
			(0.1386, -0.011998027522935778)
			(0.14354999999999998, -0.01272849074074074)
			(0.1485, -0.01307920930232558)
			(0.15344999999999998, -0.013430411214953272)
			(0.15839999999999999, -0.013775924882629107)
			(0.16335, -0.014463436018957344)
			(0.16829999999999998, -0.01480895238095238)
			(0.17325, -0.015156607655502391)
			(0.17819999999999997, -0.01550666346153846)
			(0.18314999999999998, -0.016207223300970876)
			(0.1881, -0.016530341463414634)
			(0.19304999999999997, -0.016854980392156865)
			(0.19799999999999998, -0.01718266009852217)
			(0.20295, -0.01783895024875622)
			(0.20789999999999997, -0.018170055000000004)
			(0.21284999999999998, -0.018495914572864324)
			(0.2178, -0.018822085858585864)
			(0.22274999999999998, -0.01946847448979592)
			(0.22769999999999999, -0.019791097435897433)
			(0.23264999999999997, -0.020110840206185567)
			(0.23759999999999998, -0.02042925906735751)
			(0.24255, -0.021075931937172776)
			(0.24749999999999997, -0.02139465789473684)
			(0.25244999999999995, -0.021709285714285716)
			(0.25739999999999996, -0.022026313829787234)
			(0.26234999999999997, -0.02265718279569892)
			(0.2673, -0.022972016216216213)
			(0.27225, -0.02328253260869565)
			(0.2772, -0.023590251366120217)
			(0.28214999999999996, -0.02418502762430939)
			(0.28709999999999997, -0.024482538888888888)
			(0.29205, -0.024783201117318434)
			(0.297, -0.02508443820224719)
			(0.30195, -0.025690238636363636)
			(0.30689999999999995, -0.02599370285714286)
			(0.31184999999999996, -0.026297086206896553)
			(0.31679999999999997, -0.02660304046242775)
			(0.32175, -0.0272196081871345)
			(0.3267, -0.02753146470588235)
			(0.33164999999999994, -0.027843721893491123)
			(0.33659999999999995, -0.028158505952380956)
			(0.34154999999999996, -0.028768090361445786)
			(0.3465, -0.029075230303030306)
			(0.35145, -0.029384835365853658)
			(0.35639999999999994, -0.029693036809815946)
			(0.36134999999999995, -0.030310037267080744)
			(0.36629999999999996, -0.030622875)
			(0.37124999999999997, -0.030937974842767297)
			(0.3762, -0.031254892405063295)
			(0.38115, -0.03189417307692308)
			(0.38609999999999994, -0.03221709032258064)
			(0.39104999999999995, -0.032540681818181816)
			(0.39599999999999996, -0.03286459477124183)
			(0.40095, -0.03352106622516556)
			(0.4059, -0.03385300666666667)
			(0.41084999999999994, -0.034187067114093965)
			(0.41579999999999995, -0.03451931081081081)
			(0.42074999999999996, -0.03519017123287671)
			(0.42569999999999997, -0.03552847586206896)
			(0.43065, -0.03586626388888889)
			(0.4356, -0.036207384615384614)
			(0.44054999999999994, -0.036898865248226956)
			(0.44549999999999995, -0.037247142857142854)
			(0.45044999999999996, -0.037599654676258996)
			(0.45539999999999997, -0.03795478260869565)
			(0.46035, -0.03866820588235294)
			(0.46529999999999994, -0.039030481481481485)
			(0.47024999999999995, -0.03939785820895522)
			(0.47519999999999996, -0.03976720300751879)
			(0.48014999999999997, -0.0401417803030303)
			(0.4851, -0.04090633076923077)
			(0.49004999999999993, -0.04129606976744186)
			(0.49499999999999994, -0.041687375)
			(0.49994999999999995, -0.04208418897637796)
			(0.5048999999999999, -0.042873968)
			(0.5098499999999999, -0.04327741129032258)
			(0.5147999999999999, -0.04368032520325203)
			(0.5197499999999999, -0.044088270491803284)
			(0.5246999999999999, -0.04492366666666667)
			(0.52965, -0.04535068907563026)
			(0.5346, -0.04578277966101695)
			(0.53955, -0.04622047008547009)
			(0.5445, -0.04711552173913044)
			(0.54945, -0.04754506140350877)
			(0.5544, -0.04797951327433628)
			(0.5593499999999999, -0.04841440178571428)
			(0.5642999999999999, -0.0492725)
			(0.5692499999999999, -0.04970694495412844)
			(0.5741999999999999, -0.050144175925925925)
			(0.5791499999999999, -0.05058775700934579)
			(0.5841, -0.05147900952380952)
			(0.58905, -0.051923740384615384)
			(0.594, -0.05237244660194174)
			(0.59895, -0.052829362745098044)
			(0.6039, -0.05375626000000001)
			(0.6088499999999999, -0.054228676767676766)
			(0.6137999999999999, -0.054709489795918366)
			(0.6187499999999999, -0.05519959793814433)
			(0.6236999999999999, -0.05620375789473684)
			(0.6286499999999999, -0.056713744680851066)
			(0.6335999999999999, -0.057214150537634414)
			(0.63855, -0.05772120652173913)
			(0.6435, -0.05876192222222223)
			(0.64845, -0.059297325842696626)
			(0.6534, -0.059844772727272726)
			(0.65835, -0.06040465517241379)
			(0.6632999999999999, -0.06154258823529411)
			(0.6682499999999999, -0.06212659523809524)
			(0.6731999999999999, -0.0627243734939759)
			(0.6781499999999999, -0.06332024390243902)
			(0.6830999999999999, -0.0645553625)
			(0.6880499999999999, -0.06519307594936707)
			(0.693, -0.06584711538461538)
			(0.69795, -0.06651774025974026)
			(0.7029, -0.06789654666666667)
			(0.70785, -0.0685943918918919)
			(0.7127999999999999, -0.06930491780821918)
			(0.7177499999999999, -0.070034)
			(0.7226999999999999, -0.0715362857142857)
			(0.7276499999999999, -0.07230984057971016)
			(0.7325999999999999, -0.07305095588235294)
			(0.7375499999999999, -0.07380016417910448)
			(0.7424999999999999, -0.0753574923076923)
			(0.74745, -0.07617046875)
			(0.7524, -0.07698430158730157)
			(0.75735, -0.07780970967741936)
			(0.7623, -0.07947115)
			(0.7672499999999999, -0.08030649152542374)
			(0.7721999999999999, -0.0811661724137931)
			(0.7771499999999999, -0.08204854385964912)
			(0.7820999999999999, -0.08389683636363636)
			(0.7870499999999999, -0.08486624074074074)
			(0.7919999999999999, -0.08585864150943395)
			(0.7969499999999999, -0.0868564230769231)
			(0.8019, -0.08894512)
			(0.80685, -0.09002689795918367)
			(0.8118, -0.09109560416666666)
			(0.81675, -0.09219268085106382)
			(0.8216999999999999, -0.09445355555555557)
			(0.8266499999999999, -0.09563513636363638)
			(0.8315999999999999, -0.096868)
			(0.8365499999999999, -0.09814764285714286)
			(0.8414999999999999, -0.10079064999999998)
			(0.8464499999999999, -0.1021521282051282)
			(0.8513999999999999, -0.10351884210526316)
			(0.85635, -0.1049574054054054)
			(0.8613, -0.10804180000000001)
			(0.86625, -0.10969961764705881)
			(0.8712, -0.11144330303030302)
			(0.8761499999999999, -0.11325537499999999)
			(0.8810999999999999, -0.11721849999999998)
			(0.8860499999999999, -0.11940344827586205)
			(0.8909999999999999, -0.12159457142857143)
			(0.8959499999999999, -0.12387040740740742)
			(0.9008999999999999, -0.12888824)
			(0.9058499999999999, -0.13169316666666667)
			(0.9107999999999999, -0.13454378260869562)
			(0.91575, -0.13762331818181817)
			(0.9207, -0.14459165000000002)
			(0.9256499999999999, -0.14859026315789473)
			(0.9305999999999999, -0.1527693888888889)
			(0.9355499999999999, -0.15741770588235293)
			(0.9404999999999999, -0.168205)
			(0.9454499999999999, -0.17464642857142856)
			(0.9503999999999999, -0.18192876923076923)
			(0.9553499999999999, -0.19020299999999998)
			(0.9602999999999999, -0.19916736363636361)
			(0.9652499999999999, -0.2208983333333333)
			(0.9702, -0.23533125)
			(0.97515, -0.250635)
			(0.9800999999999999, -0.269885)
			(0.9850499999999999, -0.33257025)
		};
		\addlegendentry{$m_j(p)$}
		\addplot [] coordinates {(1,0.7) (1,-0.4)};
		\addplot [] coordinates {(0,0.7) (1,0.7)};
		%				\node[anchor=south, text=blue] at (axis cs:0.4,0.53) {$f_1(U)$};
		%				\node[anchor=south, text=orange] at (axis cs:0.4,0.1) {$f_2(U)$};
	\end{axis}
\end{tikzpicture}

%% file: empirical_bounds_pandemic.tex
\begin{tikzpicture}
	\begin{axis}[
		xlabel={$p$},
		ylabel={},
		xtick={0.2, 0.4, 0.6, 0.8, 1},
		ytick={-0.11, -0.05, 0, 0.0475, 0.085},
		xmin=0, xmax=1.05,
		ymin=-0.12, ymax=0.10,
		axis lines=middle,
		width=0.75\textwidth,
		height=0.5\textwidth,
		xlabel style={below right},
		ylabel style={above left},
		%				axis x line=middle, % Add this line
		axis x line shift=0.12, % Add this line
		legend style={
			at={(-0.2,0.5)}, % position relative to axis
			anchor=east          % anchor legend at its right side
		}
		]
		\addplot[blue] coordinates {
			(0.0, -0.00392251931330472)
			(0.00485, -0.003146060344827586)
			(0.0097, -0.0025151515151515154)
			(0.01455, -0.002132560869565217)
			(0.0194, -0.001772746724890829)
			(0.02425, -0.0014152938596491225)
			(0.0291, -0.0010739647577092514)
			(0.03395, -0.0007325929203539824)
			(0.0388, -8.090178571428559e-05)
			(0.04365, 0.0002311479820627798)
			(0.0485, 0.0005423963963963959)
			(0.05335, 0.0008547918552036193)
			(0.0582, 0.0011453954545454547)
			(0.06305, 0.0014359817351598176)
			(0.0679, 0.0017288394495412844)
			(0.07275000000000001, 0.0020196175115207377)
			(0.0776, 0.0025941720930232563)
			(0.08245, 0.0028834953271028047)
			(0.0873, 0.003172943661971832)
			(0.09215000000000001, 0.0034463018867924543)
			(0.097, 0.0037110473933649295)
			(0.10185, 0.003974757142857143)
			(0.1067, 0.004237813397129187)
			(0.11155000000000001, 0.00449966826923077)
			(0.1164, 0.00502018932038835)
			(0.12125, 0.005282443902439025)
			(0.1261, 0.005535490196078433)
			(0.13095, 0.005787581280788178)
			(0.1358, 0.006042158415841584)
			(0.14065, 0.006298597014925372)
			(0.14550000000000002, 0.006556009999999998)
			(0.15035, 0.007060863636363637)
			(0.1552, 0.007301538071065989)
			(0.16005, 0.007541030612244897)
			(0.1649, 0.007777379487179486)
			(0.16975, 0.008011582474226805)
			(0.1746, 0.0082469689119171)
			(0.17945, 0.008476958333333335)
			(0.18430000000000002, 0.00870824607329843)
			(0.18915, 0.009164116402116403)
			(0.194, 0.009382744680851065)
			(0.19885, 0.009593556149732622)
			(0.2037, 0.009803451612903226)
			(0.20855, 0.010012913513513515)
			(0.2134, 0.010222119565217392)
			(0.21825, 0.010428792349726776)
			(0.22310000000000002, 0.010634917582417584)
			(0.22795, 0.011012494444444444)
			(0.2328, 0.011199541899441341)
			(0.23765, 0.011387955056179774)
			(0.2425, 0.011574412429378532)
			(0.24735000000000001, 0.01176296590909091)
			(0.2522, 0.011949337142857144)
			(0.25705, 0.012137482758620692)
			(0.2619, 0.012509116279069768)
			(0.26675, 0.012697391812865496)
			(0.2716, 0.012886111764705882)
			(0.27645000000000003, 0.013074781065088758)
			(0.2813, 0.01326432738095238)
			(0.28615, 0.013455820359281438)
			(0.29100000000000004, 0.013644698795180722)
			(0.29585, 0.013835339393939394)
			(0.3007, 0.014215570552147238)
			(0.30555, 0.01440630864197531)
			(0.3104, 0.014594018633540373)
			(0.31525000000000003, 0.0147840625)
			(0.3201, 0.014976477987421384)
			(0.32495, 0.01516996835443038)
			(0.3298, 0.015365222929936305)
			(0.33465, 0.01556172435897436)
			(0.3395, 0.01595918831168831)
			(0.34435, 0.01615492156862745)
			(0.3492, 0.01634729605263158)
			(0.35405000000000003, 0.016541960264900662)
			(0.3589, 0.016737979999999996)
			(0.36375, 0.016932161073825506)
			(0.36860000000000004, 0.017128527027027024)
			(0.37345, 0.017522020547945206)
			(0.3783, 0.017712931034482758)
			(0.38315, 0.01790534027777778)
			(0.388, 0.01809520979020979)
			(0.39285000000000003, 0.018286683098591546)
			(0.3977, 0.018480489361702122)
			(0.40255, 0.01867537857142857)
			(0.4074, 0.01887173381294964)
			(0.41225, 0.019270233576642336)
			(0.4171, 0.019472595588235293)
			(0.42195, 0.019677444444444442)
			(0.4268, 0.019883059701492536)
			(0.43165000000000003, 0.020088556390977442)
			(0.4365, 0.020294939393939393)
			(0.44135, 0.020502664122137405)
			(0.44620000000000004, 0.020710161538461537)
			(0.45105, 0.021125015625)
			(0.4559, 0.021329055118110233)
			(0.46075, 0.021535523809523806)
			(0.4656, 0.021743647999999994)
			(0.47045000000000003, 0.021954314516129027)
			(0.4753, 0.02216534146341463)
			(0.48015, 0.022376770491803276)
			(0.485, 0.022794341666666666)
			(0.48985, 0.02300357142857143)
			(0.49470000000000003, 0.02321448305084746)
			(0.49955, 0.023428905982905982)
			(0.5044, 0.023643112068965515)
			(0.50925, 0.023857399999999997)
			(0.5141, 0.02407516666666666)
			(0.51895, 0.0242967610619469)
			(0.5238, 0.024743882882882883)
			(0.5286500000000001, 0.024970163636363635)
			(0.5335, 0.025199247706422016)
			(0.53835, 0.0254314537037037)
			(0.5432, 0.025666336448598125)
			(0.54805, 0.025905386792452828)
			(0.5529000000000001, 0.026143942857142853)
			(0.55775, 0.026386874999999997)
			(0.5626, 0.02688504901960784)
			(0.56745, 0.027135683168316833)
			(0.5723, 0.027388359999999997)
			(0.57715, 0.027637606060606057)
			(0.5820000000000001, 0.02788047959183673)
			(0.58685, 0.028127649484536086)
			(0.5917, 0.028379291666666667)
			(0.59655, 0.028634389473684213)
			(0.6014, 0.02915551612903226)
			(0.6062500000000001, 0.02942143478260869)
			(0.6111, 0.029689956043956037)
			(0.61595, 0.029962977777777774)
			(0.6208, 0.03024066292134832)
			(0.62565, 0.0305195)
			(0.6305000000000001, 0.030804735632183902)
			(0.63535, 0.03137876470588235)
			(0.6402, 0.031666297619047615)
			(0.64505, 0.03195715662650602)
			(0.6499, 0.032252878048780485)
			(0.65475, 0.032545135802469136)
			(0.6596, 0.032840825)
			(0.66445, 0.03314284810126582)
			(0.6693, 0.03344275641025641)
			(0.67415, 0.03405863157894737)
			(0.679, 0.03436890666666667)
			(0.6838500000000001, 0.03468414864864864)
			(0.6887, 0.0349996301369863)
			(0.69355, 0.035322513888888886)
			(0.6984, 0.03565280281690141)
			(0.70325, 0.03599057142857143)
			(0.7081000000000001, 0.03633550724637682)
			(0.71295, 0.03699613432835821)
			(0.7178, 0.0373389696969697)
			(0.72265, 0.03768884615384615)
			(0.7275, 0.03804509375)
			(0.7323500000000001, 0.03840973015873016)
			(0.7372000000000001, 0.03878606451612903)
			(0.74205, 0.03917367213114754)
			(0.7469, 0.03995428813559322)
			(0.75175, 0.04036279310344827)
			(0.7566, 0.04078331578947368)
			(0.7614500000000001, 0.041218821428571424)
			(0.7663, 0.04166672727272727)
			(0.77115, 0.04212138888888889)
			(0.776, 0.042581999999999995)
			(0.78085, 0.04305686538461539)
			(0.7857000000000001, 0.04404706)
			(0.79055, 0.044560469387755104)
			(0.7954, 0.04509266666666667)
			(0.80025, 0.04563208510638298)
			(0.8051, 0.046191108695652176)
			(0.8099500000000001, 0.04675586666666666)
			(0.8148, 0.04734345454545454)
			(0.81965, 0.04795713953488372)
			(0.8245, 0.049248)
			(0.82935, 0.049936825)
			(0.8342, 0.05142102631578947)
			(0.8390500000000001, 0.05142102631578947)
			(0.8439, 0.05217518918918918)
			(0.84875, 0.05296163888888889)
			(0.8536, 0.05374937142857142)
			(0.85845, 0.0554441515151515)
			(0.8633000000000001, 0.05633134375)
			(0.86815, 0.05720219354838709)
			(0.873, 0.058059966666666664)
			(0.87785, 0.05896558620689654)
			(0.8827, 0.059933321428571426)
			(0.8875500000000001, 0.060935629629629624)
			(0.8924000000000001, 0.061989846153846145)
			(0.89725, 0.06422658333333334)
			(0.9021, 0.06517613043478261)
			(0.90695, 0.06613131818181818)
			(0.9118, 0.0671602857142857)
			(0.9166500000000001, 0.06824214999999999)
			(0.9215, 0.06940942105263158)
			(0.92635, 0.0706965)
			(0.9312, 0.07211864705882354)
			(0.93605, 0.07532566666666665)
			(0.9409000000000001, 0.07720692857142857)
			(0.94575, 0.07883307692307692)
			(0.9506, 0.08062575)
			(0.95545, 0.08259054545454546)
			(0.9603, 0.084574)
			(0.9651500000000001, 0.08620422222222222)
		};
		\addlegendentry{$M_j(p)$}
		\addplot[dashdotted, red] coordinates {
			(0.0, -0.0042086370733541755)
			(0.00485, -0.004791960998341367)
			(0.0097, -0.004165842858780914)
			(0.01455, -0.0035341732260346986)
			(0.0194, -0.003167790205521746)
			(0.02425, -0.0027508848344947787)
			(0.0291, -0.002512150788449889)
			(0.03395, -0.002116301931150291)
			(0.0388, -0.0013503697519556723)
			(0.04365, -0.001065667401596774)
			(0.0485, -0.0008540580740635044)
			(0.05335, -0.0005540983587851213)
			(0.0582, -0.00026304005768161933)
			(0.06305, 8.85556315075637e-05)
			(0.0679, 0.00031874001855880215)
			(0.07275000000000001, 0.0004821890160853522)
			(0.0776, 0.0010148822724844473)
			(0.08245, 0.0008295826300576115)
			(0.0873, 0.0011855780583411712)
			(0.09215000000000001, 0.001229758507570604)
			(0.097, 0.0014524206432540892)
			(0.10185, 0.0015574926990074383)
			(0.1067, 0.0015649634889971487)
			(0.11155000000000001, 0.0019261145715061055)
			(0.1164, 0.00243702583716436)
			(0.12125, 0.002643220847881974)
			(0.1261, 0.00294537312508274)
			(0.13095, 0.003197428677003126)
			(0.1358, 0.003028544056430178)
			(0.14065, 0.0032383421359527863)
			(0.14550000000000002, 0.0029108101462732637)
			(0.15035, 0.003437615145696939)
			(0.1552, 0.003732856650919084)
			(0.16005, 0.0038174313821080948)
			(0.1649, 0.004034528928771177)
			(0.16975, 0.004186670282987462)
			(0.1746, 0.004265271470983585)
			(0.17945, 0.004437686240871988)
			(0.18430000000000002, 0.004269102735474982)
			(0.18915, 0.00448954513478899)
			(0.194, 0.00470665679250257)
			(0.19885, 0.004825112808710072)
			(0.2037, 0.004889795757542277)
			(0.20855, 0.004933629362445365)
			(0.2134, 0.0052511609609267075)
			(0.21825, 0.005358401799343055)
			(0.22310000000000002, 0.0056119537922710225)
			(0.22795, 0.005680436448103638)
			(0.2328, 0.005814601801345462)
			(0.23765, 0.005897282421291553)
			(0.2425, 0.005895224093180471)
			(0.24735000000000001, 0.006342343889435245)
			(0.2522, 0.006397061659022585)
			(0.25705, 0.006619263253881782)
			(0.2619, 0.006415646959199143)
			(0.26675, 0.006420436519057121)
			(0.2716, 0.006757194414534201)
			(0.27645000000000003, 0.00690981904010379)
			(0.2813, 0.007004126682969528)
			(0.28615, 0.0072108985044700155)
			(0.29100000000000004, 0.007224149805436858)
			(0.29585, 0.0075350084451974)
			(0.3007, 0.008112429275711602)
			(0.30555, 0.008044595710676256)
			(0.3104, 0.008145661252052605)
			(0.31525000000000003, 0.008421265731637656)
			(0.3201, 0.00855765955990695)
			(0.32495, 0.008501434667856527)
			(0.3298, 0.008741571719306807)
			(0.33465, 0.008981554133686028)
			(0.3395, 0.009513678164556841)
			(0.34435, 0.009766200615542718)
			(0.3492, 0.010230395807883017)
			(0.35405000000000003, 0.01031852717788141)
			(0.3589, 0.010626848244914893)
			(0.36375, 0.010804621961400519)
			(0.36860000000000004, 0.010832461976656877)
			(0.37345, 0.011350093226122382)
			(0.3783, 0.011491224339359124)
			(0.38315, 0.011761676053954735)
			(0.388, 0.011811637786224028)
			(0.39285000000000003, 0.011870502179061354)
			(0.3977, 0.011660848083850381)
			(0.40255, 0.011768935600054549)
			(0.4074, 0.011944304218577829)
			(0.41225, 0.011895256870809064)
			(0.4171, 0.011954255072473855)
			(0.42195, 0.012068052231566484)
			(0.4268, 0.012105463851131537)
			(0.43165000000000003, 0.01228764921166907)
			(0.4365, 0.012342992820113197)
			(0.44135, 0.012711466665855596)
			(0.44620000000000004, 0.012678990559383416)
			(0.45105, 0.013092347622975493)
			(0.4559, 0.013441561423817012)
			(0.46075, 0.013531931674663608)
			(0.4656, 0.013723821295380335)
			(0.47045000000000003, 0.013734430333856835)
			(0.4753, 0.01351970413291835)
			(0.48015, 0.013907608243366293)
			(0.485, 0.014325595141457759)
			(0.48985, 0.014395942376358276)
			(0.49470000000000003, 0.01482765354561806)
			(0.49955, 0.014802511349411398)
			(0.5044, 0.014830716506103485)
			(0.50925, 0.015311557815331014)
			(0.5141, 0.015569840832123687)
			(0.51895, 0.015587592176190176)
			(0.5238, 0.015836443104939795)
			(0.5286500000000001, 0.01588678743428775)
			(0.5335, 0.015329567310560125)
			(0.53835, 0.015524363352241903)
			(0.5432, 0.015568336783488967)
			(0.54805, 0.015611225119667064)
			(0.5529000000000001, 0.015734646212972755)
			(0.55775, 0.015731149957335634)
			(0.5626, 0.015552770642647573)
			(0.56745, 0.015560082417494896)
			(0.5723, 0.015888871221959342)
			(0.57715, 0.01633506352598787)
			(0.5820000000000001, 0.016313566501001516)
			(0.58685, 0.01615716122594356)
			(0.5917, 0.016134125782806607)
			(0.59655, 0.016128022430162474)
			(0.6014, 0.016085411997902813)
			(0.6062500000000001, 0.016254325281423423)
			(0.6111, 0.01646037953335351)
			(0.61595, 0.016301856709340534)
			(0.6208, 0.016093128593062205)
			(0.62565, 0.01616752005407253)
			(0.6305000000000001, 0.01617591710024068)
			(0.63535, 0.01647162029374564)
			(0.6402, 0.016352144316072487)
			(0.64505, 0.016340101924299782)
			(0.6499, 0.016660081111290262)
			(0.65475, 0.017068959250606808)
			(0.6596, 0.016858102840541754)
			(0.66445, 0.017356470479822007)
			(0.6693, 0.01768348051690151)
			(0.67415, 0.01761934051273855)
			(0.679, 0.017512441589916546)
			(0.6838500000000001, 0.017750951825397546)
			(0.6887, 0.01815385521457907)
			(0.69355, 0.01818166847104396)
			(0.6984, 0.018541347389433444)
			(0.70325, 0.0190814898525419)
			(0.7081000000000001, 0.019405693221867623)
			(0.71295, 0.0197571377109464)
			(0.7178, 0.01986733408090087)
			(0.72265, 0.020267461996779472)
			(0.7275, 0.0204802802326519)
			(0.7323500000000001, 0.02055977018766522)
			(0.7372000000000001, 0.020822810890566527)
			(0.74205, 0.02086373218562886)
			(0.7469, 0.02127167395981545)
			(0.75175, 0.021612998764127218)
			(0.7566, 0.021652438450712467)
			(0.7614500000000001, 0.02216876728260851)
			(0.7663, 0.022568832313454548)
			(0.77115, 0.02281313334716921)
			(0.776, 0.024309669140040263)
			(0.78085, 0.024310793313557642)
			(0.7857000000000001, 0.024629125517113902)
			(0.79055, 0.02457324110544656)
			(0.7954, 0.024509749214764637)
			(0.80025, 0.02500728328274607)
			(0.8051, 0.024293636083319203)
			(0.8099500000000001, 0.024821897567801315)
			(0.8148, 0.02522097830564576)
			(0.81965, 0.025477076511037523)
			(0.8245, 0.025437962312424373)
			(0.82935, 0.02545924800106981)
			(0.8342, 0.024998152054130456)
			(0.8390500000000001, 0.025392087230963253)
			(0.8439, 0.025189642062116845)
			(0.84875, 0.026466142251986624)
			(0.8536, 0.027009880931269086)
			(0.85845, 0.02832373028696739)
			(0.8633000000000001, 0.028685209058199355)
			(0.86815, 0.02903174033820302)
			(0.873, 0.03053912040464026)
			(0.87785, 0.031246065898772192)
			(0.8827, 0.03159170862032578)
			(0.8875500000000001, 0.03241648110648577)
			(0.8924000000000001, 0.03311122628656585)
			(0.89725, 0.03375222829625261)
			(0.9021, 0.03423054185105018)
			(0.90695, 0.03448977920901398)
			(0.9118, 0.0365843777592371)
			(0.9166500000000001, 0.038196602846587914)
			(0.9215, 0.04088879746876859)
			(0.92635, 0.04205800516131916)
			(0.9312, 0.04352147919033641)
			(0.93605, 0.04448446250103977)
			(0.9409000000000001, 0.04631077873851412)
			(0.94575, 0.049021049468456375)
			(0.9506, 0.049688846016865525)
			(0.95545, 0.050572135408998624)
			(0.9603, 0.05198554414083599)
			(0.9651500000000001, 0.05160984575899508)
		};
		\addlegendentry{$M^f_j(p)$}
		\addplot[dash pattern=on 8pt off 2pt] coordinates {
			(0.0, -0.004954639484978539)
			(0.00485, -0.00484651724137931)
			(0.0097, -0.005067679653679652)
			(0.01455, -0.004753895652173912)
			(0.0194, -0.0047995633187772925)
			(0.02425, -0.0046565)
			(0.0291, -0.004489237885462556)
			(0.03395, -0.00435020353982301)
			(0.0388, -0.004233508928571428)
			(0.04365, -0.0042927668161434965)
			(0.0485, -0.004290536036036035)
			(0.05335, -0.004406479638009049)
			(0.0582, -0.004390713636363636)
			(0.06305, -0.004378095890410958)
			(0.0679, -0.004234422018348624)
			(0.07275000000000001, -0.004222059907834101)
			(0.0776, -0.00422566976744186)
			(0.08245, -0.004206859813084111)
			(0.0873, -0.004183915492957746)
			(0.09215000000000001, -0.00428013679245283)
			(0.097, -0.004226511848341233)
			(0.10185, -0.0042072523809523815)
			(0.1067, -0.003987057416267942)
			(0.11155000000000001, -0.004058091346153847)
			(0.1164, -0.004659936893203883)
			(0.12125, -0.004646336585365854)
			(0.1261, -0.004362774509803921)
			(0.13095, -0.004290674876847291)
			(0.1358, -0.004070618811881189)
			(0.14065, -0.0040037860696517415)
			(0.14550000000000002, -0.0040282)
			(0.15035, -0.003970055555555556)
			(0.1552, -0.0040405634517766496)
			(0.16005, -0.004053938775510205)
			(0.1649, -0.004162497435897436)
			(0.16975, -0.004160860824742268)
			(0.1746, -0.004313979274611399)
			(0.17945, -0.0043409114583333325)
			(0.18430000000000002, -0.0033999685863874336)
			(0.18915, -0.003346402116402116)
			(0.194, -0.003253627659574468)
			(0.19885, -0.003091823529411764)
			(0.2037, -0.003132698924731182)
			(0.20855, -0.0029029351351351337)
			(0.2134, -0.0021095489130434773)
			(0.21825, -0.0021696557377049166)
			(0.22310000000000002, -0.001959862637362636)
			(0.22795, -0.0019929055555555552)
			(0.2328, -0.0018991843575418988)
			(0.23765, -0.0016534662921348312)
			(0.2425, -0.0017628587570621464)
			(0.24735000000000001, -0.0022225795454545454)
			(0.2522, -0.0019825714285714286)
			(0.25705, -0.0021188908045977005)
			(0.2619, -0.0022300232558139527)
			(0.26675, -0.0023502807017543852)
			(0.2716, -0.0023949823529411753)
			(0.27645000000000003, -0.0026933964497041405)
			(0.2813, -0.0023974345238095226)
			(0.28615, -0.002406011976047903)
			(0.29100000000000004, -0.00250706626506024)
			(0.29585, -0.00250950909090909)
			(0.3007, -0.0023162147239263802)
			(0.30555, -0.002044469135802469)
			(0.3104, -0.0018857453416149066)
			(0.31525000000000003, -0.0015032937499999993)
			(0.3201, -0.0014396666666666666)
			(0.32495, -0.0010768227848101267)
			(0.3298, -0.0005452738853503179)
			(0.33465, -4.733333333333299e-05)
			(0.3395, 0.00047647402597402616)
			(0.34435, 0.00047763398692810444)
			(0.3492, 0.00018128289473684354)
			(0.35405000000000003, 0.000507086092715232)
			(0.3589, 0.0003999066666666673)
			(0.36375, 0.00046017449664429615)
			(0.36860000000000004, 0.0005308108108108113)
			(0.37345, 6.760273972602467e-06)
			(0.3783, 6.915172413793148e-05)
			(0.38315, 4.206250000000024e-05)
			(0.388, 9.957342657342658e-05)
			(0.39285000000000003, -1.4239436619718086e-05)
			(0.3977, 9.130496453900747e-05)
			(0.40255, -1.573571428571436e-05)
			(0.4074, -4.5669064748201585e-05)
			(0.41225, -3.4817518248175564e-05)
			(0.4171, -0.00013974999999999976)
			(0.42195, -0.00014078518518518494)
			(0.4268, -0.0003796641791044772)
			(0.43165000000000003, -4.039849624060163e-05)
			(0.4365, -0.00012973484848484808)
			(0.44135, -0.0002534427480916029)
			(0.44620000000000004, -8.356153846153813e-05)
			(0.45105, 0.0004207187500000004)
			(0.4559, 0.0005941968503937015)
			(0.46075, 0.0007821031746031757)
			(0.4656, 0.000945848)
			(0.47045000000000003, 0.0014483951612903228)
			(0.4753, 0.0015353414634146335)
			(0.48015, 0.0019182868852459028)
			(0.485, 0.0017610000000000006)
			(0.48985, 0.0020922521008403367)
			(0.49470000000000003, 0.0020616525423728807)
			(0.49955, 0.0020782393162393164)
			(0.5044, 0.0022089655172413793)
			(0.50925, 0.002139834782608696)
			(0.5141, 0.0016557894736842105)
			(0.51895, 0.0015673097345132755)
			(0.5238, 0.002178243243243245)
			(0.5286500000000001, 0.001433881818181818)
			(0.5335, 0.0018106972477064226)
			(0.53835, 0.0018708518518518523)
			(0.5432, 0.0018629813084112157)
			(0.54805, 0.0023448301886792454)
			(0.5529000000000001, 0.002177476190476192)
			(0.55775, 0.0028605865384615397)
			(0.5626, 0.0028229019607843133)
			(0.56745, 0.0028693960396039614)
			(0.5723, 0.0024374500000000025)
			(0.57715, 0.0023189090909090924)
			(0.5820000000000001, 0.002333418367346941)
			(0.58685, 0.0025891134020618583)
			(0.5917, 0.0018879375000000005)
			(0.59655, 0.0020962105263157894)
			(0.6014, 0.0028228924731182785)
			(0.6062500000000001, 0.0027230434782608694)
			(0.6111, 0.0027870769230769225)
			(0.61595, 0.0025765222222222206)
			(0.6208, 0.0025935280898876398)
			(0.62565, 0.0024571249999999992)
			(0.6305000000000001, 0.0022465517241379313)
			(0.63535, 0.0024971764705882344)
			(0.6402, 0.0021483690476190483)
			(0.64505, 0.002888963855421687)
			(0.6499, 0.0039397073170731696)
			(0.65475, 0.00339458024691358)
			(0.6596, 0.0030988375)
			(0.66445, 0.0031399240506329116)
			(0.6693, 0.003551294871794871)
			(0.67415, 0.004631144736842104)
			(0.679, 0.004590119999999998)
			(0.6838500000000001, 0.0038944864864864865)
			(0.6887, 0.004646520547945208)
			(0.69355, 0.004462597222222222)
			(0.6984, 0.004551985915492958)
			(0.70325, 0.004401685714285713)
			(0.7081000000000001, 0.004477478260869566)
			(0.71295, 0.004395761194029851)
			(0.7178, 0.004716954545454546)
			(0.72265, 0.0046889384615384595)
			(0.7275, 0.004394750000000001)
			(0.7323500000000001, 0.005413920634920634)
			(0.7372000000000001, 0.00581390322580645)
			(0.74205, 0.005146590163934427)
			(0.7469, 0.005621661016949153)
			(0.75175, 0.005390741379310344)
			(0.7566, 0.0055847894736842105)
			(0.7614500000000001, 0.005472517857142859)
			(0.7663, 0.0049314909090909095)
			(0.77115, 0.005134129629629628)
			(0.776, 0.0031539622641509435)
			(0.78085, 0.002980192307692309)
			(0.7857000000000001, 0.003637599999999999)
			(0.79055, 0.0037051632653061235)
			(0.7954, 0.003345791666666667)
			(0.80025, 0.00237468085106383)
			(0.8051, 0.004141456521739128)
			(0.8099500000000001, 0.004504488888888888)
			(0.8148, 0.0032656590909090903)
			(0.81965, 0.0026097674418604656)
			(0.8245, 0.0023200243902439036)
			(0.82935, 0.001999475000000003)
			(0.8342, 0.0024562051282051287)
			(0.8390500000000001, 0.002089447368421053)
			(0.8439, 0.0016353783783783801)
			(0.84875, 0.0007495277777777788)
			(0.8536, -0.0013651999999999974)
			(0.85845, -0.00014790909090908996)
			(0.8633000000000001, -0.00014599999999999856)
			(0.86815, -2.1612903225802998e-05)
			(0.873, 0.0006421333333333334)
			(0.87785, -1.6413793103447462e-05)
			(0.8827, -8.371428571428419e-05)
			(0.8875500000000001, -0.0028755925925925923)
			(0.8924000000000001, -0.0033161923076923073)
			(0.89725, -0.006349999999999999)
			(0.9021, -0.006812739130434781)
			(0.90695, -0.0073550454545454535)
			(0.9118, -0.009375238095238094)
			(0.9166500000000001, -0.012981799999999998)
			(0.9215, -0.01598936842105263)
			(0.92635, -0.018313999999999997)
			(0.9312, -0.01800235294117647)
			(0.93605, -0.032034466666666664)
			(0.9409000000000001, -0.03156928571428571)
			(0.94575, -0.03583330769230769)
			(0.9506, -0.04235166666666667)
			(0.95545, -0.048868636363636364)
			(0.9603, -0.0505617)
			(0.9651500000000001, -0.05398944444444445)
		};
		\addlegendentry{$\mathrm{MES}_p(X_j,S)$}
		\addplot[dotted] coordinates {
			(0.0, -0.005235687673893938)
			(0.00485, -0.005219112748379773)
			(0.0097, -0.005568912059238703)
			(0.01455, -0.005847975407241981)
			(0.0194, -0.00611070128350507)
			(0.02425, -0.00653022540202337)
			(0.0291, -0.006581866895432717)
			(0.03395, -0.006701707096496809)
			(0.0388, -0.0073702208716471)
			(0.04365, -0.007524853245113695)
			(0.0485, -0.007923289332264119)
			(0.05335, -0.008029117216737018)
			(0.0582, -0.008163891902516205)
			(0.06305, -0.008392450969673346)
			(0.0679, -0.008645879280411845)
			(0.07275000000000001, -0.008773390127631356)
			(0.0776, -0.008873091619656538)
			(0.08245, -0.008301131796505936)
			(0.0873, -0.008266977563398442)
			(0.09215000000000001, -0.008373377106178175)
			(0.097, -0.008550265156895193)
			(0.10185, -0.008654673977227483)
			(0.1067, -0.008825018125302904)
			(0.11155000000000001, -0.009042027088957977)
			(0.1164, -0.009387824084762582)
			(0.12125, -0.009512821385192025)
			(0.1261, -0.009267127178824129)
			(0.13095, -0.00950731417143207)
			(0.1358, -0.0093048340449384)
			(0.14065, -0.009353389138915622)
			(0.14550000000000002, -0.009496653729269177)
			(0.15035, -0.01001708697301523)
			(0.1552, -0.010101810420367292)
			(0.16005, -0.0102624905030255)
			(0.1649, -0.0103550254048542)
			(0.16975, -0.01041075909778212)
			(0.1746, -0.010495372184563081)
			(0.17945, -0.010644567151798998)
			(0.18430000000000002, -0.011232507566521702)
			(0.18915, -0.011536381977657243)
			(0.194, -0.011557047092678609)
			(0.19885, -0.01174835154415355)
			(0.2037, -0.011891165322598343)
			(0.20855, -0.012016821806394093)
			(0.2134, -0.012177045308894996)
			(0.21825, -0.012226895464767892)
			(0.22310000000000002, -0.012365550673543672)
			(0.22795, -0.012032717546780603)
			(0.2328, -0.012468118577998104)
			(0.23765, -0.012805508323849595)
			(0.2425, -0.01289189339994835)
			(0.24735000000000001, -0.012943010658070115)
			(0.2522, -0.01291094491657557)
			(0.25705, -0.013044674082782226)
			(0.2619, -0.013758826319483277)
			(0.26675, -0.013974004560827752)
			(0.2716, -0.01423846174096141)
			(0.27645000000000003, -0.01433428311406221)
			(0.2813, -0.014632582850091839)
			(0.28615, -0.01506634487611768)
			(0.29100000000000004, -0.01521252321600845)
			(0.29585, -0.015297172433724844)
			(0.3007, -0.015645931826368926)
			(0.30555, -0.015931260098726988)
			(0.3104, -0.016179817890808388)
			(0.31525000000000003, -0.016319476520498647)
			(0.3201, -0.016447660039383197)
			(0.32495, -0.016813763569601063)
			(0.3298, -0.016965191300317015)
			(0.33465, -0.017347760660964343)
			(0.3395, -0.017795631590177077)
			(0.34435, -0.01768901129322008)
			(0.3492, -0.01788890331352638)
			(0.35405000000000003, -0.017967949343205076)
			(0.3589, -0.018545961365407784)
			(0.36375, -0.018902486267649128)
			(0.36860000000000004, -0.019025712985487442)
			(0.37345, -0.019458500352360814)
			(0.3783, -0.01954967341394236)
			(0.38315, -0.01949938064334307)
			(0.388, -0.019642613144257193)
			(0.39285000000000003, -0.019837227217584873)
			(0.3977, -0.019769337985106158)
			(0.40255, -0.020192538030818996)
			(0.4074, -0.020320945062004145)
			(0.41225, -0.020678807946338175)
			(0.4171, -0.021225122823861967)
			(0.42195, -0.021273251088556197)
			(0.4268, -0.021583088721739016)
			(0.43165000000000003, -0.021705660450612597)
			(0.4365, -0.02202914216044653)
			(0.44135, -0.022138771404677918)
			(0.44620000000000004, -0.022482038388282395)
			(0.45105, -0.023021027174006694)
			(0.4559, -0.02314101650788595)
			(0.46075, -0.0231889552414228)
			(0.4656, -0.023522318043878453)
			(0.47045000000000003, -0.023975762508161514)
			(0.4753, -0.024378582103513614)
			(0.48015, -0.024956886567030875)
			(0.485, -0.02532543636573543)
			(0.48985, -0.025914686592202626)
			(0.49470000000000003, -0.026038093699065477)
			(0.49955, -0.0260194569817982)
			(0.5044, -0.026269136357561167)
			(0.50925, -0.026478510340871498)
			(0.5141, -0.026529450453597052)
			(0.51895, -0.026701809699510498)
			(0.5238, -0.02736682368264939)
			(0.5286500000000001, -0.027157494291230967)
			(0.5335, -0.027546328068051365)
			(0.53835, -0.027810072401794835)
			(0.5432, -0.027916505139618936)
			(0.54805, -0.028587591625851744)
			(0.5529000000000001, -0.028719016208913505)
			(0.55775, -0.028654574303071965)
			(0.5626, -0.029636987882704698)
			(0.56745, -0.03002212346002259)
			(0.5723, -0.030031507929268727)
			(0.57715, -0.02995896961982154)
			(0.5820000000000001, -0.03020233600111834)
			(0.58685, -0.030752437322827435)
			(0.5917, -0.031303457712840356)
			(0.59655, -0.03164761204314494)
			(0.6014, -0.03190108302525987)
			(0.6062500000000001, -0.03205273381830924)
			(0.6111, -0.03236493213402945)
			(0.61595, -0.032775848307522636)
			(0.6208, -0.032687570234283835)
			(0.62565, -0.0327502386594095)
			(0.6305000000000001, -0.0334458106432365)
			(0.63535, -0.03367493953275052)
			(0.6402, -0.03397885655317438)
			(0.64505, -0.03411062425418561)
			(0.6499, -0.034629628293545586)
			(0.65475, -0.035190264784701365)
			(0.6596, -0.03596586336478706)
			(0.66445, -0.03629915780298789)
			(0.6693, -0.03675313745217136)
			(0.67415, -0.038235297078803535)
			(0.679, -0.038518605224847964)
			(0.6838500000000001, -0.03876299433817559)
			(0.6887, -0.03887156788906545)
			(0.69355, -0.03894138073782608)
			(0.6984, -0.04064708213866752)
			(0.70325, -0.04063790399852254)
			(0.7081000000000001, -0.04031801850898125)
			(0.71295, -0.040855354948075434)
			(0.7178, -0.04117393118087441)
			(0.72265, -0.0414250786253239)
			(0.7275, -0.041889551674449146)
			(0.7323500000000001, -0.042186551624965395)
			(0.7372000000000001, -0.04218238502399433)
			(0.74205, -0.042589809972176346)
			(0.7469, -0.04371138623959929)
			(0.75175, -0.04371952906803166)
			(0.7566, -0.04491375268005123)
			(0.7614500000000001, -0.0460591327619847)
			(0.7663, -0.046285699966141604)
			(0.77115, -0.046720251581077815)
			(0.776, -0.046702435587261046)
			(0.78085, -0.04711302358056223)
			(0.7857000000000001, -0.04807243990443419)
			(0.79055, -0.048761794966103805)
			(0.7954, -0.04927884038388867)
			(0.80025, -0.048880234471643344)
			(0.8051, -0.049657599721994024)
			(0.8099500000000001, -0.04966717650090774)
			(0.8148, -0.05032163936461651)
			(0.81965, -0.05068012317984637)
			(0.8245, -0.0511266974799682)
			(0.82935, -0.051665186610859434)
			(0.8342, -0.053023918872433184)
			(0.8390500000000001, -0.05346481699443725)
			(0.8439, -0.05503476309353755)
			(0.84875, -0.05519016009794364)
			(0.8536, -0.05520414700327111)
			(0.85845, -0.0574904751113608)
			(0.8633000000000001, -0.05846722625099093)
			(0.86815, -0.05911956630891521)
			(0.873, -0.06060007280853335)
			(0.87785, -0.06296903865032434)
			(0.8827, -0.06336371950504137)
			(0.8875500000000001, -0.06347538143884461)
			(0.8924000000000001, -0.06425990690463362)
			(0.89725, -0.06495862482445701)
			(0.9021, -0.065945352001493)
			(0.90695, -0.06777395779009733)
			(0.9118, -0.07143993561318662)
			(0.9166500000000001, -0.07668502311866625)
			(0.9215, -0.08133137913820347)
			(0.92635, -0.07539298590315549)
			(0.9312, -0.07558424328831809)
			(0.93605, -0.08124193767333174)
			(0.9409000000000001, -0.08169433471211467)
			(0.94575, -0.0843351958738047)
			(0.9506, -0.08562481077159907)
			(0.95545, -0.08808942872190263)
			(0.9603, -0.09192135257753134)
			(0.9651500000000001, -0.09880532359432015)
		};
		\addlegendentry{$m^f_j(p)$}
		\addplot[dashed, green] coordinates {
			(0.0, -0.004855388888888889)
			(0.00485, -0.005800487068965518)
			(0.0097, -0.006211038961038962)
			(0.01455, -0.006609521739130434)
			(0.0194, -0.007005449781659388)
			(0.02425, -0.007383315789473685)
			(0.0291, -0.007761837004405287)
			(0.03395, -0.008129353982300885)
			(0.0388, -0.008847772321428572)
			(0.04365, -0.009168865470852017)
			(0.0485, -0.009475990990990992)
			(0.05335, -0.009778239819004525)
			(0.0582, -0.010077536363636362)
			(0.06305, -0.010347242009132422)
			(0.0679, -0.010615325688073395)
			(0.07275000000000001, -0.010885612903225808)
			(0.0776, -0.011418325581395347)
			(0.08245, -0.011686934579439253)
			(0.0873, -0.011955525821596242)
			(0.09215000000000001, -0.01222193396226415)
			(0.097, -0.012489156398104264)
			(0.10185, -0.012750471428571428)
			(0.1067, -0.012980038277511962)
			(0.11155000000000001, -0.01321104326923077)
			(0.1164, -0.013661621359223301)
			(0.12125, -0.013883721951219511)
			(0.1261, -0.014107647058823529)
			(0.13095, -0.014332162561576354)
			(0.1358, -0.014548336633663367)
			(0.14065, -0.014755313432835818)
			(0.14550000000000002, -0.014958359999999999)
			(0.15035, -0.015367575757575755)
			(0.1552, -0.015566715736040607)
			(0.16005, -0.01576612244897959)
			(0.1649, -0.01576612244897959)
			(0.16975, -0.016152752577319587)
			(0.1746, -0.016348854922279794)
			(0.17945, -0.016546796875000002)
			(0.18430000000000002, -0.016745120418848167)
			(0.18915, -0.017143783068783065)
			(0.194, -0.017345489361702125)
			(0.19885, -0.017544754010695187)
			(0.2037, -0.017745209677419355)
			(0.20855, -0.017943913513513515)
			(0.2134, -0.018144097826086957)
			(0.21825, -0.01834343169398907)
			(0.22310000000000002, -0.018544604395604392)
			(0.22795, -0.018948422222222223)
			(0.2328, -0.0191498938547486)
			(0.23765, -0.019350646067415725)
			(0.2425, -0.0195525988700565)
			(0.24735000000000001, -0.01975683522727273)
			(0.2522, -0.019962651428571426)
			(0.25705, -0.02017056896551724)
			(0.2619, -0.020586610465116274)
			(0.26675, -0.020795169590643273)
			(0.2716, -0.021006158823529412)
			(0.27645000000000003, -0.02121855621301775)
			(0.2813, -0.021431744047619047)
			(0.28615, -0.02164611976047904)
			(0.29100000000000004, -0.021862277108433734)
			(0.29585, -0.022080672727272727)
			(0.3007, -0.022500061349693245)
			(0.30555, -0.022712234567901234)
			(0.3104, -0.022926298136645963)
			(0.31525000000000003, -0.023142425)
			(0.3201, -0.02335741509433962)
			(0.32495, -0.023573525316455693)
			(0.3298, -0.023788382165605092)
			(0.33465, -0.024004461538461537)
			(0.3395, -0.024437837662337664)
			(0.34435, -0.024655666666666666)
			(0.3492, -0.02487432236842105)
			(0.35405000000000003, -0.02509009933774834)
			(0.3589, -0.025307533333333333)
			(0.36375, -0.02552587919463087)
			(0.36860000000000004, -0.025742527027027028)
			(0.37345, -0.026177239726027396)
			(0.3783, -0.02639710344827586)
			(0.38315, -0.026616868055555555)
			(0.388, -0.026838790209790207)
			(0.39285000000000003, -0.027062908450704224)
			(0.3977, -0.027288113475177303)
			(0.40255, -0.027515249999999998)
			(0.4074, -0.02774408633093525)
			(0.41225, -0.02820834306569343)
			(0.4171, -0.028444470588235293)
			(0.42195, -0.02868358518518518)
			(0.4268, -0.02891788805970149)
			(0.43165000000000003, -0.029149360902255637)
			(0.4365, -0.029382090909090906)
			(0.44135, -0.029614496183206107)
			(0.44620000000000004, -0.0298492)
			(0.45105, -0.03032915625)
			(0.4559, -0.030570543307086615)
			(0.46075, -0.03081553968253968)
			(0.4656, -0.031063032)
			(0.47045000000000003, -0.031313540322580644)
			(0.4753, -0.031566926829268295)
			(0.48015, -0.03182395901639344)
			(0.485, -0.03233941666666667)
			(0.48985, -0.03260304201680673)
			(0.49470000000000003, -0.03287086440677966)
			(0.49955, -0.03313968376068376)
			(0.5044, -0.033409224137931036)
			(0.50925, -0.033683356521739136)
			(0.5141, -0.033960368421052635)
			(0.51895, -0.03424098230088496)
			(0.5238, -0.03479727927927928)
			(0.5286500000000001, -0.03507723636363636)
			(0.5335, -0.03535887155963303)
			(0.53835, -0.03564478703703703)
			(0.5432, -0.035934121495327105)
			(0.54805, -0.036227952830188676)
			(0.5529000000000001, -0.03651898095238095)
			(0.55775, -0.03681232692307693)
			(0.5626, -0.03740287254901961)
			(0.56745, -0.03770236633663366)
			(0.5723, -0.03800491)
			(0.57715, -0.03830925252525253)
			(0.5820000000000001, -0.038616673469387755)
			(0.58685, -0.03892972164948454)
			(0.5917, -0.03924908333333333)
			(0.59655, -0.03957191578947369)
			(0.6014, -0.04023559139784946)
			(0.6062500000000001, -0.04057467391304348)
			(0.6111, -0.04092061538461538)
			(0.61595, -0.04127255555555556)
			(0.6208, -0.041624)
			(0.62565, -0.04198154545454546)
			(0.6305000000000001, -0.04233106896551724)
			(0.63535, -0.04304768235294117)
			(0.6402, -0.04341497619047619)
			(0.64505, -0.04378309638554217)
			(0.6499, -0.04415792682926829)
			(0.65475, -0.044541530864197536)
			(0.6596, -0.044923450000000004)
			(0.66445, -0.045303544303797474)
			(0.6693, -0.04568961538461539)
			(0.67415, -0.04648705263157895)
			(0.679, -0.04689894666666667)
			(0.6838500000000001, -0.04731906756756756)
			(0.6887, -0.04775065753424658)
			(0.69355, -0.048194208333333335)
			(0.6984, -0.04863801408450705)
			(0.70325, -0.0490928)
			(0.7081000000000001, -0.04955060869565217)
			(0.71295, -0.050496761194029856)
			(0.7178, -0.050978303030303034)
			(0.72265, -0.051473830769230774)
			(0.7275, -0.05198125)
			(0.7323500000000001, -0.0524986507936508)
			(0.7372000000000001, -0.053027887096774196)
			(0.74205, -0.05357404918032787)
			(0.7469, -0.05469991525423729)
			(0.75175, -0.0552835)
			(0.7566, -0.05587424561403509)
			(0.7614500000000001, -0.056486017857142856)
			(0.7663, -0.057106890909090914)
			(0.77115, -0.05774833333333334)
			(0.776, -0.05840241509433962)
			(0.78085, -0.05907146153846154)
			(0.7857000000000001, -0.060340620000000005)
			(0.79055, -0.06099081632653061)
			(0.7954, -0.061658395833333345)
			(0.80025, -0.06234374468085107)
			(0.8051, -0.06304602173913045)
			(0.8099500000000001, -0.06373731111111111)
			(0.8148, -0.06442427272727272)
			(0.81965, -0.06510316279069768)
			(0.8245, -0.06653234146341463)
			(0.82935, -0.06726019999999999)
			(0.8342, -0.06801923076923076)
			(0.8390500000000001, -0.06879481578947368)
			(0.8439, -0.06958281081081082)
			(0.84875, -0.07039477777777778)
			(0.8536, -0.0711882)
			(0.85845, -0.07279330303030304)
			(0.8633000000000001, -0.07364615625)
			(0.86815, -0.07454967741935484)
			(0.873, -0.07551336666666668)
			(0.87785, -0.07651937931034482)
			(0.8827, -0.07751142857142856)
			(0.8875500000000001, -0.07856685185185183)
			(0.8924000000000001, -0.07969584615384616)
			(0.89725, -0.08211916666666667)
			(0.9021, -0.0834431304347826)
			(0.90695, -0.08485349999999998)
			(0.9118, -0.08628561904761904)
			(0.9166500000000001, -0.08766145)
			(0.9215, -0.08915305263157894)
			(0.92635, -0.09079061111111111)
			(0.9312, -0.09261282352941176)
			(0.93605, -0.09670939999999999)
			(0.9409000000000001, -0.09915342857142855)
			(0.94575, -0.10192846153846154)
			(0.9506, -0.10471441666666666)
			(0.95545, -0.10797336363636365)
			(0.9603, -0.1118039)
			(0.9651500000000001, -0.11642588888888888)
		};
		\addlegendentry{$m_j(p)$}
		\addplot [] coordinates {(1,0.09) (1,-0.15)};
		\addplot [] coordinates {(0,0.09) (1,0.09)};
		%				\node[anchor=south, text=blue] at (axis cs:0.4,0.53) {$f_1(U)$};
		%				\node[anchor=south, text=orange] at (axis cs:0.4,0.1) {$f_2(U)$};
	\end{axis}
\end{tikzpicture}